\documentclass[letter, 11pt]{article}
\usepackage{natbib}
\usepackage{hyperref}
\usepackage{amssymb,amsmath,xcolor,graphicx,xspace,float,colortbl,rotating,dsfont} % 
\usepackage{textcomp}
\usepackage{appendix}  %% 100
\usepackage{bm}  %% 100
\usepackage{boxedminipage}  %% 100
\usepackage{color}  %% 100
\usepackage{endnotes}  %% 100
\usepackage[onehalfspacing]{setspace}  %% 100
\usepackage{tabulary}  %% 100  
\usepackage{varioref}  %% 100
\usepackage{wrapfig}  %% 100  
\usepackage[linesnumbered,ruled]{algorithm2e}
\usepackage{tikz}
\usepackage{subcaption}
\usepackage{float}
\usepackage{tabularx} 
\restylefloat{table}
\usepackage{makecell}

\usepackage[margin=1in]{geometry}

\title{Keeping in Place After the Storm—Emergency Assistance and Evictions\footnote{\href{https://islahbilal.github.io/evictions.pdf}{Link to latest version.}}\thanks{We are grateful to the organizers of the 2026 Africa Meeting of the Econometric Society and the 2026 Africa Business School household finance conference for the opportunity to present the paper. We would like to specifically thank Andres Almazan for discussing the paper alongside Fatima Zahra Filali Adib, Michael Boutros, Taha Choukhmane, Niclas Moneke, Adair Morse and Amine Ouazad for their comments and feedback.}}
\author{Bilal Islah\footnote{Africa Business School, University Mohammed VI Polytechnic, Morocco. e-mail: \href{mailto:bilal.islah@um6p.ma}{bilal.islah@um6p.ma}} \hspace{5em} Ahmed Zoulati\footnote{Africa Business School, University Mohammed VI Polytechnic, Morocco. e-mail: \href{mailto:ahmed.zoulati@um6p.ma}{ahmed.zoulati@um6p.ma}}}

\date{November 2026}

\begin{document}

\maketitle

\begin{abstract}
We offer evidence that federal emergency assistance (FEMA) in the days following natural disasters mitigate evictions in comparison to similar emergency scenarios where FEMA aid is not provided. We find an approximate 10.9\% increase in overall evictions after hurricane natural disaster events driven in large part by areas in close proximity of the hurricane path that do not receive FEMA rental assistance. Furthermore, we also show that FEMA aid acts as a liquidity buffer to other forms of emergency credit, specifically we find that overall transactions volumes decrease during hurricane events in locations that do receive FEMA aid. This effect largely reverses in areas that do not receive FEMA aid. Overall, this suggests that the availability of emergency liquidity during natural disaster events is indeed a binding constraint with real household financial consequences, in particular through our documented channel of evictions and in usage of high-cost credit.
\end{abstract}

\doublespacing

\break

\section{Introduction}

The loss of structures and, in particular, housing, is often among the stark lasting visuals in the wake of natural disasters. The resulting financial loss and displacement highlight the importance of both the individual and government role in pre-disaster mitigation and post-disaster response. At the micro level, the loss born by households can be large and has been documented in various forms in the literature.

Historically, the destruction of housing in such events has led to rebuilt neighborhoods that differ significantly from their original form and community composition. The Boston fire of 1871 \citep{hornbeck2017} and the San Francisco 1906 earthquake \citep{SIODLA201548} provide specific historical moments of migration and reconstruction where entire neighborhoods were rebuilt but also with a notable change in the former population to which \citep{BOUSTAN2020103257} find systematic evidence of out-migration following disaster events in the United States. In contrast, we study instances of assistance in preserving housing access. That is, we examine the existence of federal assistance households to maintain their ability to make housing payments due to potential disruptions and damages directly related to the natural disaster.

Prior to the creation of the Federal Emergency Management Agency (FEMA) under the Carter administration in 1979, many natural disasters that have marked the history of the United States have required federal emergency funding through ad-hoc legislation in the aftermath of their occurrence. One form of disaster relief has been the creation of temporary emergency housing in response to damage and destruction of housing property. This has further expanded to include rental assistance, starting initially under the Stafford Act of 1988 and then further expanded under the Individuals and Households Program in 2002\footnote{\citep{crs_r47015}}. The goal of such programs to recognize the disruption in household financial ability to cope with uninsured or under-insured emergency expenses.  

In the same vein, we study the availability of emergency rental assistance through FEMA in the wake of hurricane disaster events and their impact on rental evictions in the state of Florida. We find that notably the rate of eviction is mitigated via the availability of FEMA rental assistance, and furthermore we document, in the case of payday loans, that the supply of emergency assistance corresponds with a substitution away from higher cost credit. 

\subsection{Literature Review}

The household financial impacts of natural disasters have been increasingly studied, with significant attention given to household resilience and recovery mechanisms. The intersection of disasters and eviction risks has also garnered attention. \citep{brennan2022perfect} explore the heightened vulnerability of renters in disaster-affected areas, showing that disasters exacerbate eviction rates, particularly for vulnerable populations. \citep{raymond2022preventing} examine the legal and policy dimensions of eviction laws, emphasizing the gaps in tenant protections in states such as Florida. These findings are particularly relevant to our work, underscoring the cascading effects of disasters on eviction, financial distress, and reliance on high-cost credit. Finally, \citep{Aron-Dine} studies household relocation decisions in relation to hurricane disaster risk.

Furthermore, the consequences of hurricane events has notably received attention in the literature. \citep{gallagher2017household} analyze the financial consequences of hurricane Katrina, highlighting the crucial role of federal aid in stabilizing household finances. \citep{deryugina2018economic} corroborate these findings, using tax return data to show long-term income declines and wealth disruptions. \citep{del2024household} extend this analysis to hurricane Harvey, examining household decision-making in the aftermath of a disaster. They identify shifts in financial behavior, including increased reliance on savings and external credit.  Similarly, \citep{collier2024credit} demonstrates that FEMA assistance mitigates financial distress, using credit files data they show a reduction in delinquencies and defaults in affected regions. Our study complements this result by focusing on evictions and on substitution effects with higher cost credit, particularly how households resort to alternative financial lenders like payday loans when formal aid is absent or insufficient.

The literature on payday loans has highlighted the dual nature of high-cost credit, as both a financial lifeline and a debt trap. We similarly contrast its availability in the context of natural disasters.\citep{bhutta2015payday} describe how payday loans exacerbate financial strain for vulnerable borrowers, while \citep{morse2011payday} offers a contrasting view, emphasizing their utility during emergencies. \citep{gathergood2019payday} shows that payday loans increase financial distress, especially for low-income borrowers, but regulatory interventions can mitigate harmful effects. In the same vein, \citep{dobridge2016better} finds mixed effects, while access to high-cost credit aids short-term liquidity, it negatively affects long-term financial outcomes. Extending this debate, our findings suggest that payday loans act as a substitute for formal disaster relief, especially in regions excluded from FEMA’s coverage. This substitution underscores the dual-edged nature of payday lending in disaster recovery contexts, offering immediate relief but also amplifying financial fragility as we observe a difference in subsequent default rates.

Beyond individual outcomes, the broader economic implications of disasters have been extensively analyzed. \citep{bernstein2019disaster} highlight the long-term economic vulnerabilities of disaster-prone areas, particularly the impact of rising sea levels on property prices.
The interplay between financial aid and alternative credit sources offers another dimension to understanding recovery dynamics. \citep{collier2024credit} analyze demand for disaster recovery loans, illustrating how credit availability shapes household recovery trajectories. \citep{lane2024adapting} underscores the importance of guaranteed credit programs in building resilience, particularly for lower-income households. \citep{malmin2023access} explores the role of federal credit in shaping future wealth trajectories, emphasizing the need for equitable and targeted recovery programs. 

Finally, recent studies on disaster aid distribution and its long-term effects highlight systemic inequalities. \citep{billings2022let} document the inequities in financial aid distribution during Hurricane Harvey, revealing gaps in support for vulnerable populations. \citep{bufe2021financial} explores how financial shocks impact lower-income households, identifying key factors that contribute to financial resilience, such as savings, credit access, and social support networks. \citep{ratnadiwakara2020areas} finds that flood-prone areas increasingly attract lower-income and less creditworthy populations due to declining property values and housing costs.

Using Florida as a case study, our research extends prior work by integrating multiple granular datasets to examine the dynamics of financial distress through eviction risks and high-cost credit during the recovery following disasters. Our granular spatial and temporal analysis reveals how gaps in disaster aid exacerbate loss of housing via increased evictions, driving households to substitute formal aid with high-cost credit options. This approach reinforces the importance of inclusive disaster recovery programs that address the nuanced needs of vulnerable populations, ultimately potentially mitigating long-term financial harm and fostering resilience.

The rest of the paper is as follows. In section 2, we describe the various data used and our spatial-temporal approach. In section 3 ,we present our main result linking eviction outcomes and access to emergency financial assistance. And then in section 4, we further extend this analysis to understand the interdependence with access to high-cost credit via payday loans.

\section{Data and Methodology}

\subsection{Data}
Our data comes from several administrative sources. Our access to evictions data comes from the Evictions Lab at Princeton, which sources their data via county level court records on evictions. Our access to payday loans data comes via a freedom of information request to the state of Florida's Office of Financial Regulation. We also utilize several federal level datasets. Specifically we retrieve FEMA's (Federal Emergency Management Agency) Web Disaster Declarations and Housing Assistance Program Data for Renters, as well as publicly available data from NOAA (the National Oceanic and Atmospheric Administration) and the NHC (National Hurricane Center). Finally, estimates of urban and rural population, housing units, and characteristics at the census level come from the American Community Survey.

\subsubsection{Eviction Data}

Princeton's Eviction Lab data provides Census-Tract level information at a weekly level, including eviction, filing, and at-risk outcome variables \citep{evictionlab}. Covering the period from 2003 to 2017. Table \ref{tab:evict-stat} shows summary statistics. The evict variable describes the number of evictions aggregated at the weekly-census tract level. The filing variable describes the number of court filings for eviction and the atrisk variable adjusts this for the number of individuals living in the household at risk of being evicted.

\begin{table}[H]
\centering
\caption{Summary Statistics: Evictions dataset, measures per weekly-census tract.}
\label{tab:evict-stat}
\begin{tabular}{lccc}
\hline
Statistic  & \texttt{evict} & \texttt{filing} & \texttt{atrisk} \\
\hline
25th Perc. & 1 & 1 & 1 \\
Median    & 1 & 1 & 1 \\
Mean     & 1.219 & 1.770 & 1.676 \\
75th Perc. & 1 & 2 & 2 \\
Std. Dev.  & 1.26 & 1.59 & 1.49 \\
\hline
\multicolumn{2}{c}{\footnotesize Observations:  498793} & \multicolumn{2}{c}{\footnotesize Unique Census Tracts: 3980}    \\
\hline\hline
\end{tabular}
\label{tab:summary_stats}
\end{table}

\subsubsection{Payday Loan Data}

Our dataset on payday loans includes loan-level transactions from 2002 to 2018 in the state of Florida, totaling 101 million daily loan-level observations. We are able to observe the loan date, due date, transaction amount, and any related fees. We also observe the date of repayment or whether the loan is outstanding. Furthermore, each loan observation has information on the retailer and borrower's ZIP code.

\begin{table}[H]
\centering
\caption{Summary Statistics: Payday Loan Activity, measures per ZIP--day.}
\label{tab:payday-stat}
\begin{tabular}{lccc}
\hline
Statistic  & \texttt{Amount} & \texttt{Transaction\_Volume} & \texttt{Default} \\
\hline
25th Perc. & 300 & 14 & 0 \\
Median     & 500 & 30 & 1 \\
Mean       & 394.67 & 46.55 & 1.49 \\
75th Perc. & 500 & 59 & 2 \\
Std. Dev.  & 132.73 & 55.79 & 5.26 \\
\hline
\multicolumn{2}{c}{\footnotesize Observations: 2{,}052{,}435} &
\multicolumn{2}{c}{\footnotesize Unique ZIP Codes: 549} \\
\hline\hline
\end{tabular}
\end{table}

\subsubsection{Hurricane Forecast Cone}

The National Hurricane Center (NHC) provides authoritative information on the timing, location, and progression of tropical storms and hurricanes affecting the United States. In our study, this data is critical for reconstructing the geographic footprint of storms that impacted Florida between 2004 and 2018. However, the forecast cones are only available as far back as 2008. 

\begin{figure}[H]
    \centering
    \includegraphics[scale=0.5]{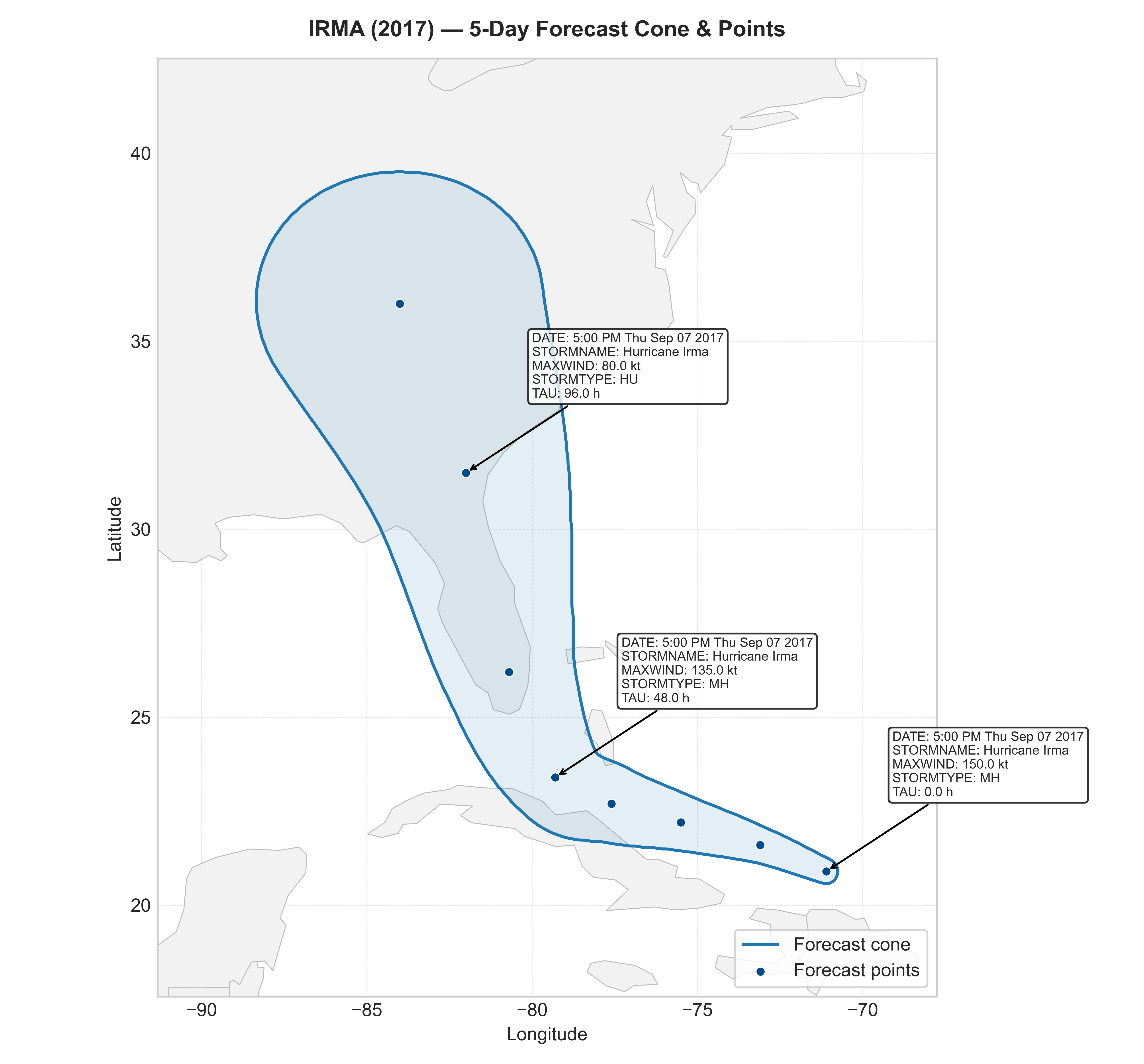}
    \caption{5-Day Track Forecast Cone}
    \label{fig:cone}
\end{figure}

For storm events prior to 2008, we relied on NHC forecast cone archives,\footnote{\url{https://www.nhc.noaa.gov/aboutcone.shtml} provides details on how the forecast cone is constructed.} which record the storm’s position and projected path at successive forecast times. From these records, we extracted the initial forecast points that represent the best estimate of the storm’s current location, thereby tracing the realized path of each hurricane. Figure \ref{fig:cone} shows an example of a forecast cone. The first point represents the actual location of the hurricane, and as we move away from the initial point, the cone covers a larger area with increasing uncertainty, a method of construction chosen by the NHC based on the prediction error over the past five years.

For storm events prior to 2008, we complemented NHC records with storm track data from NOAA’s Historical Hurricane Tracks database. By combining these sources, we were able to build a consistent dataset of storm paths across the full study period. This provided the foundation for identifying geographic areas affected by each hurricane and for distinguishing treated from control communities in the empirical analysis.

\subsubsection{FEMA Data}

Since 1979, the Federal Emergency Management Agency (FEMA) has served as the federal government’s lead agency in responding to and facilitating recovery from most significant crises. To provide insight into FEMA’s Housing Assistance Program for renters, the Individual Assistance (IA) reporting team created a dataset beginning with disaster declaration DR1439 (2002). This dataset includes aggregated, non-personally identifiable information such as the number of applications, inspections, the extent of damages, aid provided, and other relevant metrics, broken down by state, county, and ZIP code where registrations were made. FEMA’s Housing Assistance Program for renters, a component of the Individuals and Households Program (IHP), offers critical financial support to those impacted by federally declared disasters. Assistance includes temporary rental aid, Other Needs Assistance (ONA) for uninsured expenses (e.g., personal property or childcare), and Direct Housing Assistance (e.g., transportable housing units or repaired multi-family homes) for cases where rental housing is unavailable.

Before an individual can receive FEMA aid, they must meet specific general eligibility criteria. Applicants are required by FEMA to confirm that the disaster-damaged home is their primary residence that their disaster-related needs are not already covered by another source, such as insurance or other programs. If applicants have insurance, FEMA will require proof of the settlement or a letter explaining why coverage was denied before determining the assistance they are eligible to receive.  

We utilize two FEMA datasets: the FEMA Web Disaster Declarations and the Housing Assistance Program Data on Renters. The first dataset includes information on all declared hurricanes and tropical storms affecting Florida, including the disaster name and its corresponding Disaster Number (a sequentially assigned identifier for declared disasters). The Disaster Number is essential for linking to the second dataset, which provides detailed information about housing assistance. We then identify ZIP codes where at least one individual has received rental disaster assistance.

Table \ref{tab:disasters} presents a list of all major disasters in Florida that were approved for the Individuals and Households Program (IHP). Table \ref{tab:program-renters} provides summary statistics of key variables for the Housing Assistance Program on Renters.

\begin{table}[H]
\centering
\caption{Major Disasters and Declaration Types}
\label{tab:disasters}
\begin{tabular}{@{}lll@{}}
\hline
Declaration Date & Disaster Name        & \textbf{Declaration Type} \\ \hline
2004-08-13  & Tropical Storm Bonnie and Hurricane Charley  & Major Disaster            \\
2004-09-04  & Hurricane Frances                            & Major Disaster            \\
2004-09-16  & Hurricane Ivan                               & Major Disaster            \\
2004-09-26  & Hurricane Jeanne                             & Major Disaster            \\
2005-07-10  & Hurricane Dennis                             & Major Disaster            \\
2005-08-28  & Hurricane Katrina                            & Major Disaster            \\
2005-10-24  & Hurricane Wilma                              & Major Disaster            \\
2008-08-24  & Tropical Storm Fay                           & Major Disaster            \\
2012-07-03  & Tropical Storm Debby                         & Major Disaster            \\
2016-09-28  & Hurricane Hermine                            & Major Disaster            \\
2017-09-10  & Hurricane Irma                               & Major Disaster            \\
\hline
\end{tabular}
\end{table}

\begin{table}[H]
\begin{center}
\caption{This table reports summary statistics on FEMA Assistance by disaster. Registered for FEMA Aid denotes the number of  households with valid FEMA registrations in declared counties during the official registration period. Approved for FEMA Aid refers to applicants approved for FEMA Individual and Household Program (IHP) assistance. Renters Approved for Rental Assistance counts renter households that received rental assistance. Total Rental Assistance Disbursed is the total dollar amount of rental assistance paid for each disaster, and Average Rental Assistance per Renter is calculated as total rental assistance disbursed divided by the number of renters approved for rental assistance.}
\label{tab:program-renters}
\resizebox{\textwidth}{!}{
\begin{tabular}{@{}lcccccc@{}}
\hline
Disaster Name & \makecell{Registered for\\FEMA Aid} & \makecell{Approved for\\FEMA Aid} & \makecell{Renters Approved\\for Rental Assistance} & \makecell{Total Rental\\Assistance Disbursed (\$)} & \makecell{Average Rental Assistance\\per Renter (\$)} \\
\hline
\makecell{Tropical Storm Bonnie and \\ Hurricane Charley} & 50,136 & 28,945 & 19,338 & 26,273,487 & 1,359 \\
Hurricane Frances & 114,540 & 66,784 & 33,418 & 46,208,593 & 1,383 \\
Hurricane Ivan & 35,961 & 21,203 & 12,463 & 11,684,323 & 938 \\
Hurricane Jeanne & 89,278 & 57,075 & 30,751 & 39,083,876 & 1,271 \\
Hurricane Dennis & 9,131 & 4,238 & 1,640 & 1,006,428 & 614 \\
Hurricane Wilma & 150,501 & 57,542 & 23,834 & 23,939,083 & 1,004 \\
Tropical Storm Fay & 5,463 & 1,622 & 879 & 2,098,470 & 2,387 \\
Tropical Storm Debby & 4,455 & 1,752 & 1,093 & 2,771,489 & 2,536 \\
Hurricane Hermine & 1,643 & 585 & 378 & 797,293 & 2,109 \\
Hurricane Irma & 1,420,062 & 444,735 & 240,131 & 304,094,846 & 1,266 \\
\hline
\end{tabular}
}
\end{center}
\end{table}

\begin{table}[H]\centering
\caption{This table shows FEMA application outcome counts over the sample and their ineligibility reasons. FEMA disaster assistance eligibility is determined in two stages: first, a Presidential disaster declaration must designate specific counties as eligible for Individual Assistance; then, households within those areas must demonstrate disaster-related damage to their primary residence, unmet needs not covered by insurance, and basic eligibility criteria (e.g., identity and occupancy verification). FEMA emphasizes that aid is limited to uninsured or underinsured losses and is based on verified damage and necessity rather than income alone.}
\begin{tabularx}{\textwidth}{lXrr}
\hline
\textbf{Category} & \textbf{Description} & \textbf{Count} & \textbf{Share (\%)} \\
\hline
\multicolumn{4}{l}{\textit{Overall Applications}} \\
\hline
-- & Total Applicants & 4,516,155 & 100.00 \\
-- & Eligible Applicants & 2,540,160 & 56.22 \\
-- & Ineligible Applicants & 1,975,995 & 43.78 \\
\hline
\multicolumn{4}{l}{\textit{Reasons for Ineligibility (conditional on being ineligible)}} \\
\hline
1 & No disaster-related damages or needs & 700,340 & 35.45 \\
2 & Applicant has insurance & 666,162 & 33.72 \\
3 & No response / application withdrawn & 349,126 & 17.67 \\
4 & Missing required documentation & 134,532 & 6.81 \\
-- & Other & 125,835 & 6.37 \\
\hline
\end{tabularx}
\end{table}

\begin{table}[H]
\begin{center}
\caption{This table reports the distribution of FEMA rental assistance across renter income groups, aggregated across all disaster events in the sample. Total Rental Assistance Amount represents the cumulative rental assistance disbursed to eligible renters within each income category. Percentage denotes each group’s share of total rental assistance. The Overall Average Rental Assistance Amount per Renter is calculated across all storms combined.}
\label{tab:ra_eligible_stats}
\renewcommand{\arraystretch}{1.2}
%\resizebox{\textwidth}{!}{
\begin{tabular}{@{}lrr@{}}
\hline
Income Group & Total Rental Assistance Amount (\$) & Percentage (\%) \\
\hline
$<$ \$15{,}000 & 128{,}929{,}816.82 & 28.15 \\
\$15{,}000--\$30{,}000 & 161{,}242{,}766.89 & 35.21 \\
\$30{,}001--\$60{,}000 & 87{,}797{,}451.98 & 19.17 \\
\$60{,}001--\$120{,}000 & 18{,}358{,}446.00 & 4.01 \\
\$120{,}001--\$175{,}000 & 1{,}314{,}990.00 & 0.29 \\
$>$ \$175{,}000 & 2{,}669{,}687.00 & 0.58 \\
0 & 57{,}643{,}167.26 & 12.59 \\
\hline
\hline
\multicolumn{3}{l}{Overall Average Rental Assistance Amount per Renter (\$): 1{,}258} \\
\hline
\end{tabular}
%}
\end{center}
\caption{This table uses individual demographic information, in particular self-declared income when applying for FEMA assistance. We report the total rental assistance per income bracket across disasters in the sample and the percentage of individuals in each income bracket. A value of 0 indicates applicants who either refused to provide their income, or in some cases reported they were self-employed.}
\end{table}

\subsection{Methodology}

We analyze our eviction and payday loan datasets to measure the impact of hurricanes and tropical storms on households. Our focus is on renters, and since our eviction data is at the census tract level, we weighted the outcome variables (evictions, filings, and at-risk individuals) based on the number of renters in each census tract. For the payday loan dataset, we aggregate defaults and number of transactions that we observe at the ZIP code level per day.

We use two models to measure the impact of natural disasters on households. First, we plot event studys to visually analyze the different outcome variables for both the treated and control groups. Then, we used a difference-in-differences model to confirm our results from the event study. Finally, we also then study the existence of federal aid assistance in an interacted difference-in-differences setting to measure the role of FEMA aid.

In the following three sections, we will explain how the treated and control groups were constructed and provide a detailed explanation of how the event study and difference-in-differences models were conducted.

\subsubsection{Construction of treatment and control group}

To assess the impact of hurricanes on households, we classify communities into treatment and control groups based on their geographic proximity to the storm path. The treated group consists of ZIP codes whose boundaries intersect the realized hurricane track, extended by a buffer. This buffer reflects the forecast uncertainty at the initial advisory position, as defined by the National Hurricane Center (NHC) when constructing its cones of uncertainty. It captures not only areas directly along the storm’s path but also nearby regions likely to experience significant effects. In this way, the treated group represents communities most plausibly exposed to wind, rainfall, and other storm-related disruptions. 

By contrast, the control group includes ZIP codes in Florida that did not intersect the buffered storm path. These areas lie outside the spatial footprint of the hurricane and thus provide a counterfactual for how outcomes would have evolved in the absence of storm exposure. This distinction ensures that observed differences between the treated and control groups can be more directly attributed to hurricane impacts rather than broader regional dynamics.  

\begin{figure}[H]
    \centering
    \includegraphics[width=0.5\linewidth]{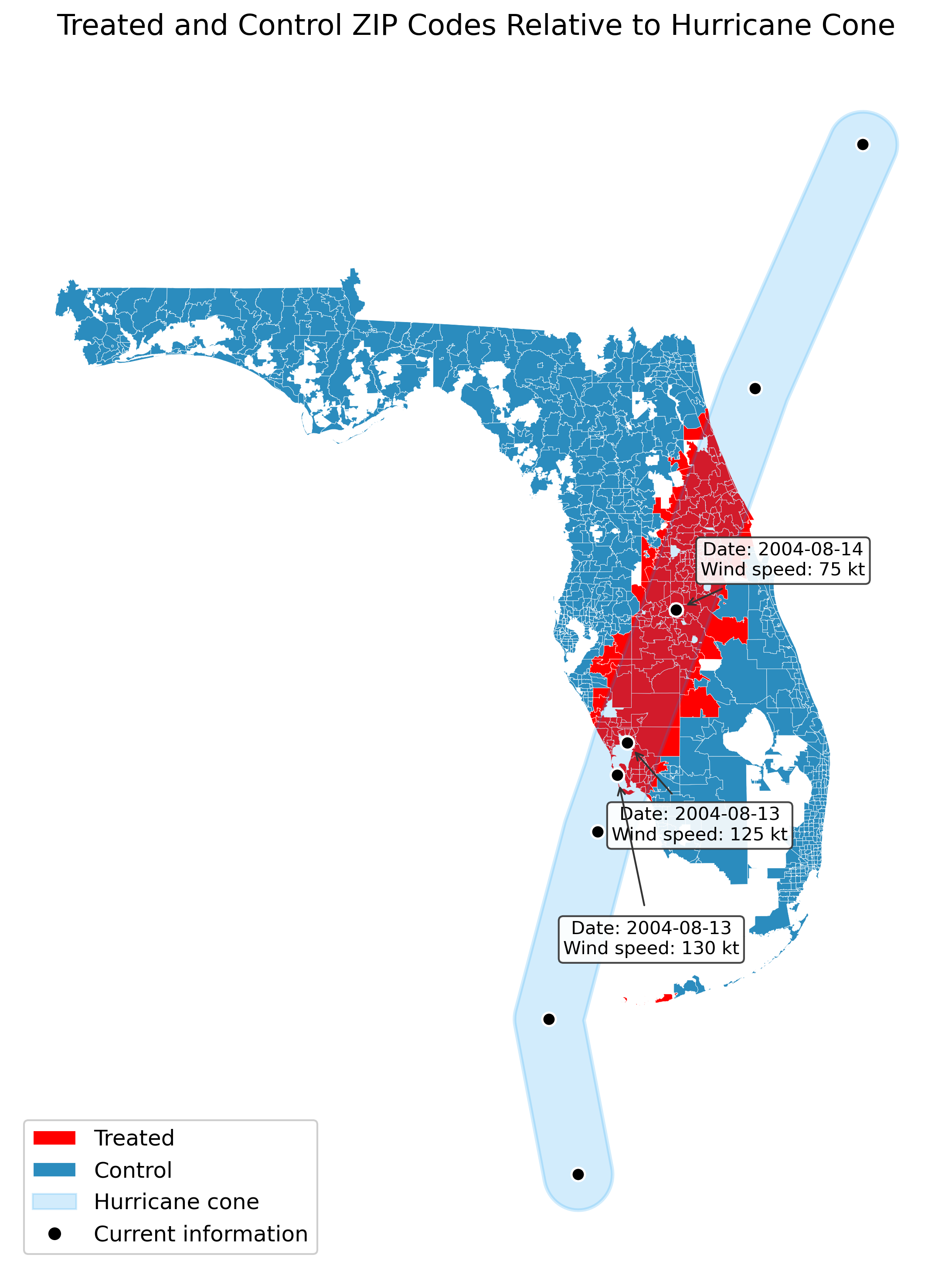}
    \caption{Treatment and Control ZIP Codes for Hurricane Charley (2004).
            The figure shows the hurricane’s observed track (black points), the buffer zone used to define exposure (light blue), and ZIP Codes classified as treated (red) or control (blue).}
    \label{fig:charley_path}
\end{figure}

Figure \ref{fig:charley_path} illustrates our spatial treatment assignment strategy using Hurricane Charley (2004) as an example. The map shows how ZIP Code Tabulation Areas (ZCTAs) intersecting the storm’s path are classified as treated, while those outside the exposure zone serve as controls.

\subsubsection{Event Study Framework}

To construct the data for the event study, we treat each hurricane and tropical storm as a separate event study. We merge the constructed treated and control group dataset of storm affected geographies with the evictions dataset at the ZIP code level for six months before and six months after the event. For the treated group, we use the exact date of the storm path for each ZIP code and merge with the observed week in the eviction dataset. For the control group, whose geographies were constructed to remain outside the storm’s path, we assign a pseudo-date: when the storm event occurred on a single day, we assign that same day to all control ZIP codes, whereas for multi-day events, we randomly distribute the storm dates across control ZIP codes (e.g., if the storm lasted two days, some control ZIP codes are assigned the first day while others are assigned the second). Since our evictions dataset is at the weekly level, we calculate the relative event date in weeks and also in months to the storm date observed. We follow the same approach with payday loans, calculating the relative event date at the daily level since our payday loan data is recorded daily.

The estimating equation for our event study model is given by :

\begin{equation} 
y_{h,i,\tau} = \left( \sum_{j = -m}^{n} \beta_j \cdot \large\mathds{1} \{\tau=j\}\cdot D_{h,i} \right) + \alpha_{h,i} + \delta_{h,\tau} + \epsilon_{h,i,\tau}
\end{equation}

In this equation, \( y_{h,i,\tau} \) represents the outcome variable for unit \( i \) at relative time \( \tau \) for hurricane event \( h \). The term \( \large\mathds{1}\{\tau=j\}\cdot D_{h,i} \) is an indicator variable that equals 1 if the hurricane event \( h \) occurred at relative time to impact \( j \) in a treated ZIP code \( i \), the unit of observation, and 0 otherwise. The coefficients \( \beta_j \) capture the dynamic effects of the event over time. For \( \tau > 0 \), these coefficients show the post-event impact, while for \( \tau  < 0 \), they provide a measure for pre-event trends. The summation term thus captures the event study terms over all pre- and post-event periods.

The model also includes \( \alpha_{h,i} \), which denotes event-by-unit fixed effects to account for event-specific and time-invariant heterogeneity across units, and \( \delta_{h,\tau} \), which represents event-by-time fixed effects to control for event-specific time-specific shocks common to all units, ensuring that all the identification comes from within an event. Together, these terms ensure the model effectively isolates the impact of the event while accounting for unobserved heterogeneity.

As in \citep{cengiz} and \cite{dube2024minimum}, this event study regression aligns events based on event time rather than calendar time, making it comparable to a scenario where all events occur simultaneously. This approach aims to correct well documented challenges in the standard difference-in-differences approach \citep{Baker}. 

To properly isolate both treated and control groups and to prevent overlap, we only consider a ZIP code as part of the treated group when it is the first event affecting each ZIP code during the event window. Essentially, we exclude ZIP codes that experience treatment of any other storm events during the construction of the event window of the control group. Similarly, for the control group, we only consider the never-treated during the event window.

In our setup, we conduct multiple event studies focusing on evictions, filings, transaction volumes, and default rates for both treated and control groups. Considering the eviction process spans potentially several months from initial notice, to filing, to eviction, we measured the effect spanning six months. For the payday loans data, we aggregated daily observations at the weekly level to account for within week seasonality.

\subsubsection{Difference-in-Differences Framework}

We also a difference-in-differences analysis comparing the treated and control groups in pre and post-event framework. The model is as follows:

\begin{equation}
y_{h,i,\tau} = \beta \cdot \large\mathds{1}\{\tau\geq0\}\cdot D_{h,i} + \alpha_{h,i} + \delta_{h,\tau} + \epsilon_{h,i,\tau} 
\end{equation}

where $y_{h,i,\tau}$ denotes the dependent variable, representing (evictions, filings, transaction volume, defaults) for observation of hurricane event $h$, unit $i$, at relative time $\tau$. $D_{h,i}$ is an indicator variable for the treatment group, equals one for the treated ZIP code and zero otherwise for hurricane event $h$. The indicator variable $\large\mathds{1}\{\tau\geq0\}$ represents the post-treatment period. By constructing the panel based on relative time, we ensure a framework in which events are aligned to occur simultaneously. And $\alpha_{h,i}, \delta_{h,\tau}$ denotes the event specific fixed effects for unit and time (ZIP code and week/month), respectively.

Finally, to shed light on the role of FEMA aid, we interact the availability of FEMA rental assistance at the ZIP code level to determine its impact via a triple difference-in-differences approach. Given the possible endogeneity of individuals receiving FEMA assistance, we also implement an instrumental variable design for FEMA eligibility, leveraging the fact that FEMA assistance is made at the county level\footnote{\url{https://www.fema.gov/node/how-do-i-know-if-my-area-eligible-assistance?}} and then assessed individually per application. This means that individuals applying in ZIP codes that fall within eligible county boundaries can apply to be assessed and those outside are ineligible by default.

Essentially, our instrumental variable approach mimics a fuzzy regression discontinuity at county boundaries that are declared FEMA eligible or not. ZIP codes that fall within eligible county boundaries have a positive probability of at least one individual receiving FEMA assistance, whereas ZIP codes outside the eligibility county boundary will never receive FEMA assistance for a given disaster event. \autoref{fig:bin-scatter} shows the decrease in the proportion of ZIP codes with recipients of FEMA assistance as a function of distance from the path of the storm. Hence, we instrument for FEMA eligibility in the first stage of a two-stage least squares adaptation of our difference-in-differences framework in the following way:

\begin{equation}
FEMA_{h,i} = \pi_1 Z_{h,i} + \pi_2 Z_{h,i} \cdot X_{h,i} + \mu_{h,i}
\end{equation}

For each hurricane event we estimate FEMA eligibility for each ZIP code taking into account that eligibility varies at county borders ($Z_{h,i}$ is an indicator for a ZIP code being within an eligible FEMA county for a given disaster $h$) and that the individual's damage assessment likely varies with distance from the storm path ($X_{h,i}$ represents of each ZIP code from the storm path in nautical miles).

The second stage includes the instrumented FEMA variable in a triple interaction as follows:

\begin{equation}
y_{h,i,\tau} = \beta \cdot \large\mathds{1}\{\tau\geq0\}\cdot D_{h,i} \cdot FEMA_{h,i} + \alpha_{h,i} + \delta_{h,\tau} + \epsilon_{h,i,\tau} 
\end{equation}

In the next section, we present the findings from the evictions and payday loan analysis.

\section{Results}

\subsection{Evictions}

\subsubsection{Hurricanes}

\begin{figure}[H]
  \centering
  \subfloat[Eviction: ZIP Code - Monthly]{\includegraphics[width=.49\linewidth]{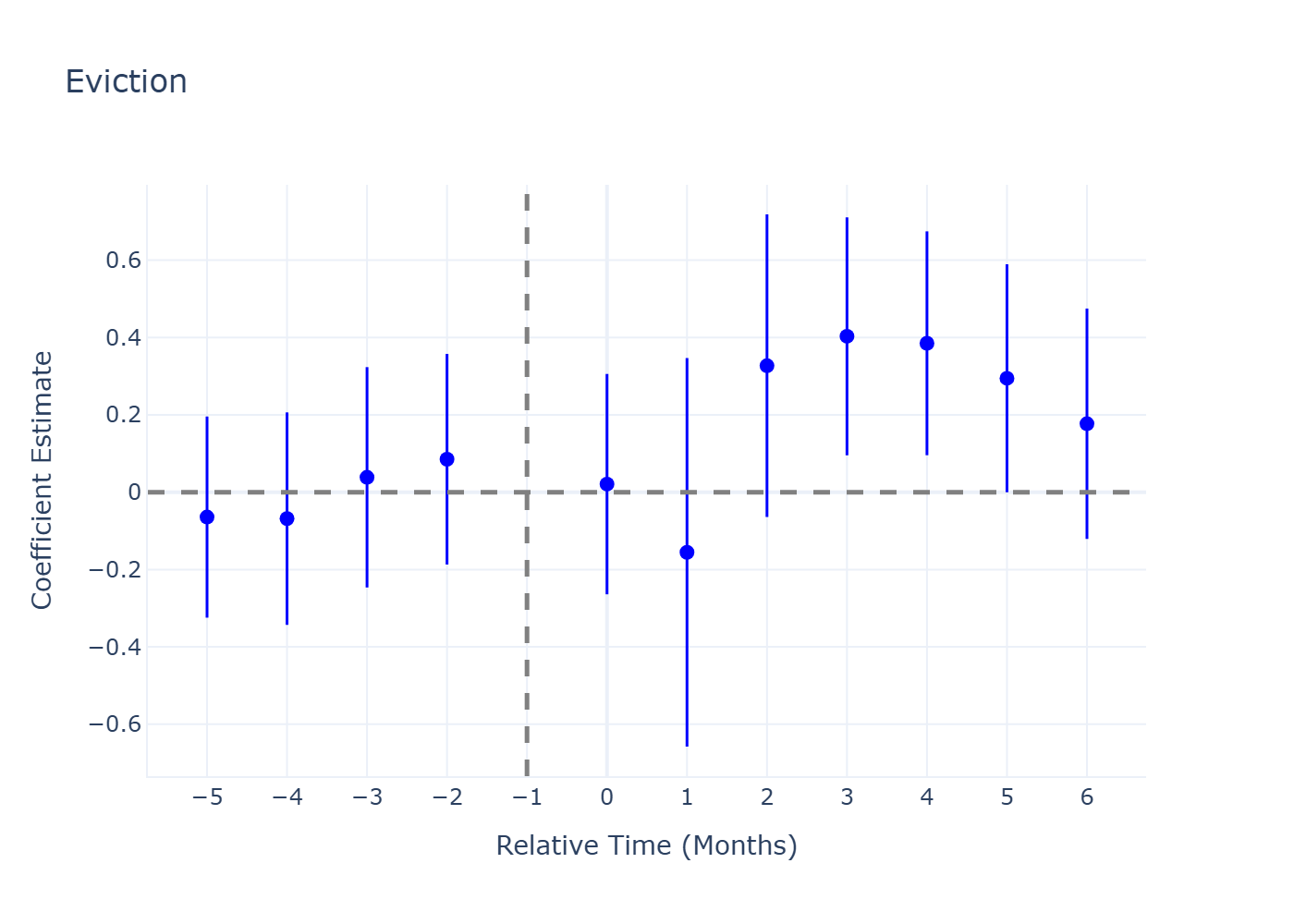}}\hfill
  \subfloat[Filing: ZIP Code - Monthly] {\includegraphics[width=.49\linewidth]{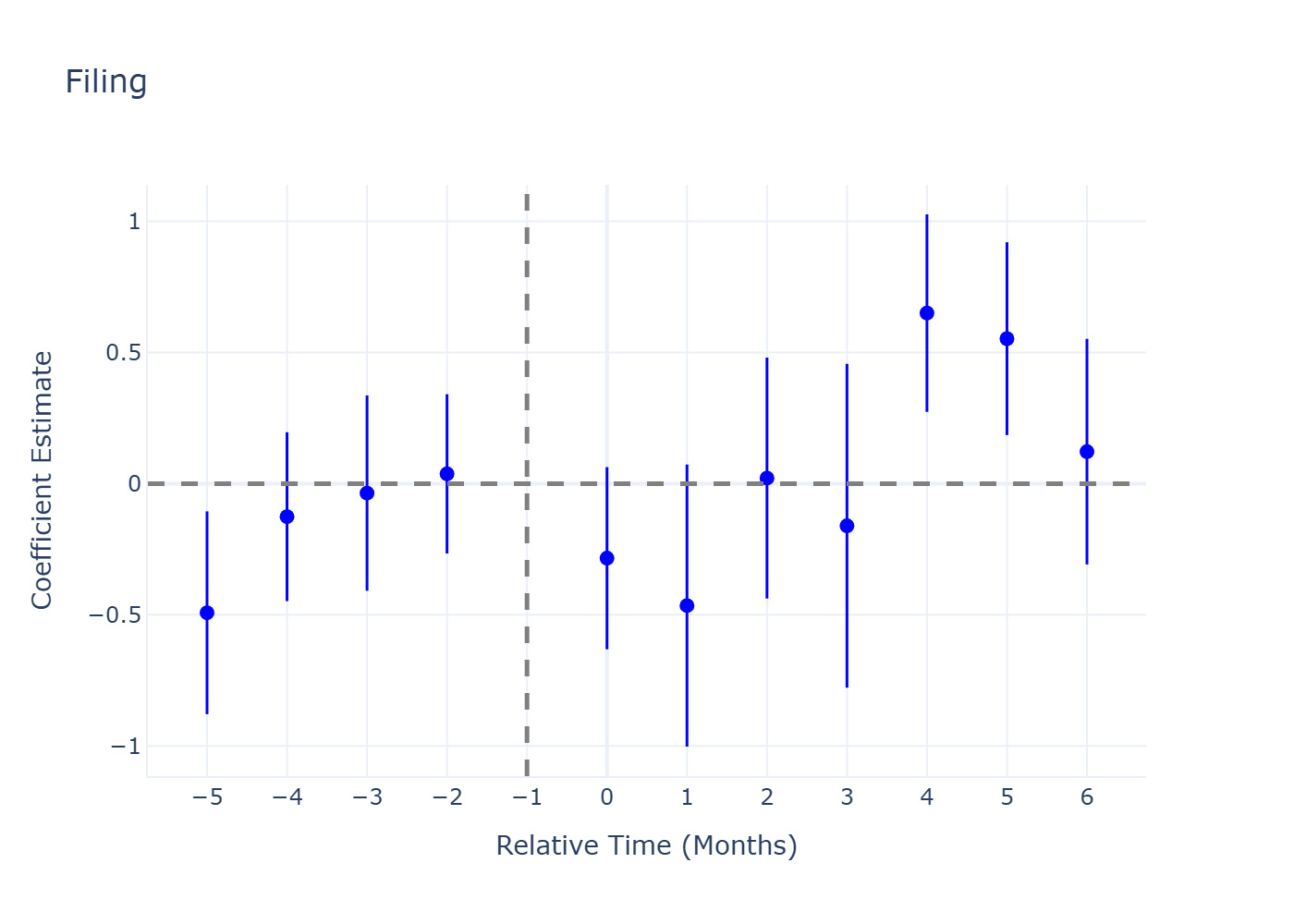}}
  \caption{Estimated coefficients from an event study model evaluating the impact of hurricane exposure on monthly eviction and filing rates at the ZIP code level with a treatment radius of 21 nautical miles. The event window spans from five months before to six months after landfall, with the month immediately preceding the hurricane (month -1) serving as the reference period. Standard errors are clustered at the ZIP code level.}
    \label{fig:evict-filing-Hurricane-21}
\end{figure}

Using our evictions dataset we find that ZIP codes that experienced hurricanes saw a maximum monthly average increase of evictions of 0.201 per month per 1000 renters\footnote{We normalize the raw number of evictions and filings using census data on the number of renters present in a ZIP code to account for heterogeneity in the percentage of renters and in the population per ZIP code.} as in \autoref{tab:DiD-radius-hurricane}. This represents a monthly increase of 10.9\% evictions per month compared to the baseline. The event study in \autoref{fig:evict-filing-Hurricane-21} shows that this effect persists up to 6 months after the initial hurricane. The event study also shows a increase around months 2-5, reflecting the nature of the eviction process that may lag from the initial missed rental payment date, to the eviction filing, to the finally court approved eviction order. 

We also show in \autoref{tab:DiD-radius-hurricane} that including ZIP codes by broadening the size of the treatment group radius (in nautical miles) results in a decreasing effect size for evictions, as one might expect given the damage from the hurricane event correlates with distance. At 100 nautical miles the effect essentially disappears. This result is important as we later show that this effect does not hold when separating into FEMA receiving and non-FEMA receiving ZIP codes.

\begin{figure}[H]
  \centering
  \subfloat[Eviction: ZIP Code - Monthly]{\includegraphics[width=.49\linewidth]{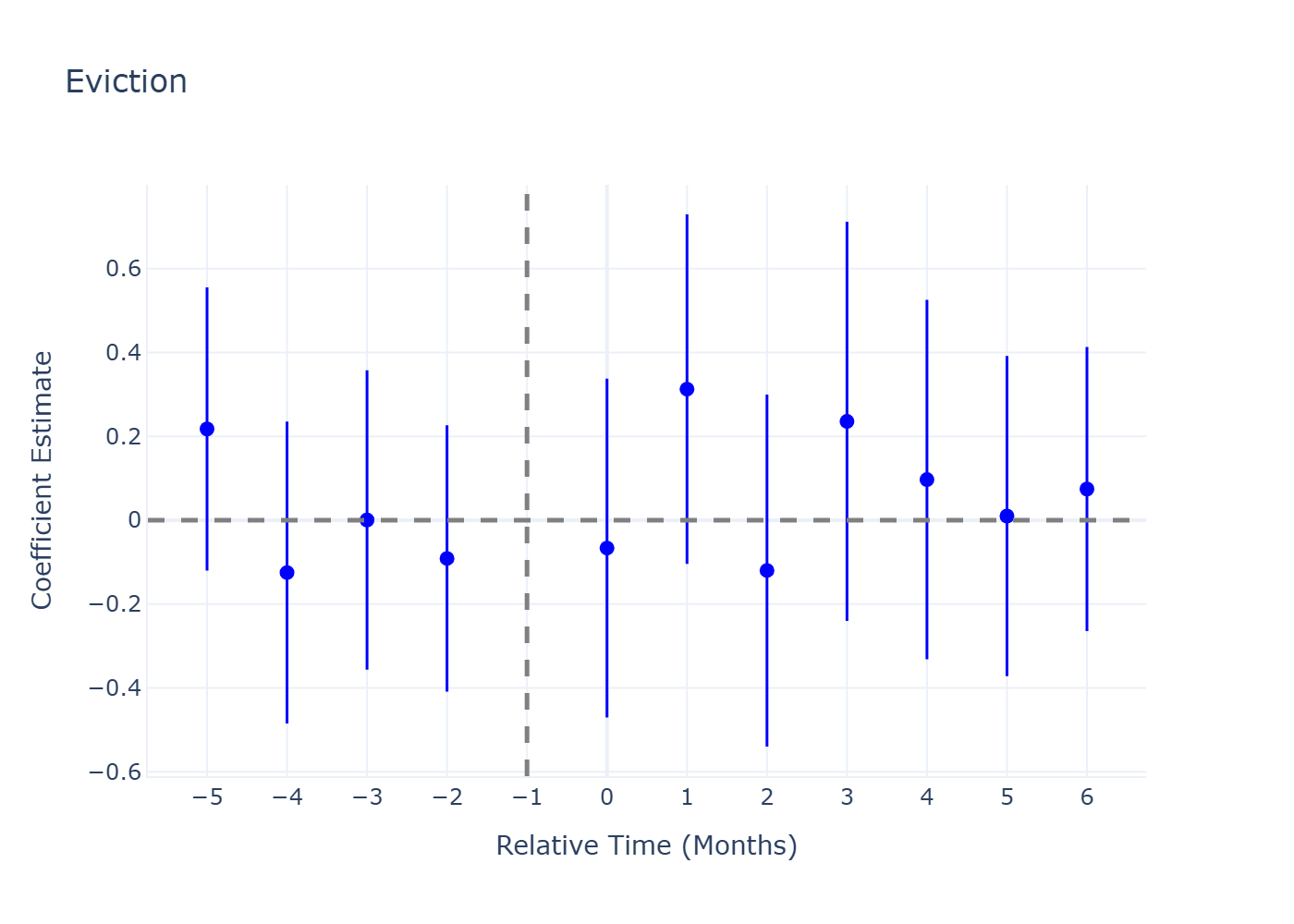}}\hfill
  \subfloat[Filing: ZIP Code - Monthly] {\includegraphics[width=.49\linewidth]{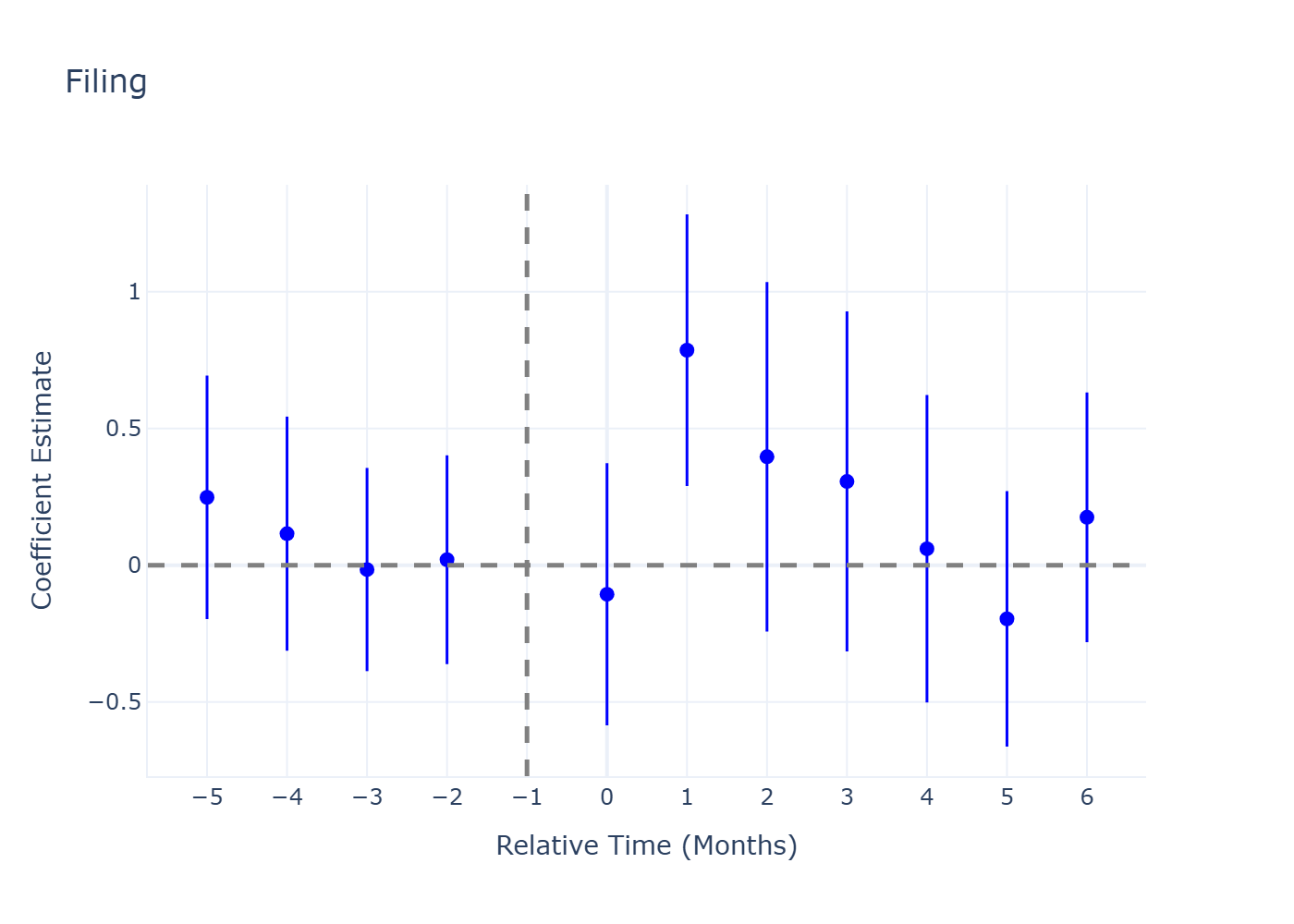}}
  \caption{Estimated coefficients from an event study model evaluating the impact of hurricane exposure on monthly eviction and filing rates at the ZIP code level with a treatment radius of 50 nautical miles. The event window spans from five months before to six months after landfall, with the month immediately preceding the hurricane (month -1) serving as the reference period. Standard errors are clustered at the ZIP code level.}
    \label{fig:evict-filing-Hurricane-50}
\end{figure}

\begin{table}[H]
\centering
\begin{tabular}{lccccc}
\hline
 & \multicolumn{5}{c}{\textbf{Treatment Radius}} \\
\cline{2-6}
 & 15 & 21 & 26 & 50 & 100 \\
\hline
\multicolumn{6}{l}{\textbf{Panel A: Eviction}} \\
Treated$\times$Post & 0.172$^{*}$ & 0.201$^{*}$ & 0.169 & 0.073 & -0.049 \\
 & [0.097] & [0.103] & [0.114] & [0.114] & [0.084] \\
\hline
\multicolumn{6}{l}{\textbf{Panel B: Filing}} \\
Treated$\times$Post & 0.164 & 0.167 & 0.142 & 0.133 & 0.193 \\
 & [0.126] & [0.125] & [0.137] & [0.150] & [0.149] \\
\hline
Event-specific ZIP Code FE & Y & Y & Y & Y & Y  \\
Event-specific Time FE & Y & Y & Y & Y & Y \\
Observations & 21,617 & 20,383 & 19,360 & 16,939 & 15,894 \\
\hline
\end{tabular}
\caption{Difference-in-Differences results for monthly eviction and filing outcomes at the ZIP code level by treatment radius. The estimation window spans 5 months before the event to 6 months after the event, capturing both pre-event and post-event outcomes. 
Standard errors clustered at the ZIP code level are reported in square brackets. ***, **, and * indicate statistical significance at the 1\%, 5\%, and 10\% levels, respectively.}
\label{tab:DiD-radius-hurricane}
\end{table}

\begin{table}[H]
\centering
\begin{tabular}{lccccc}
\hline
 & \multicolumn{5}{c}{\textbf{Treatment Radius}} \\
\cline{2-6}
 & 15 & 21 & 26 & 50 & 100 \\
\hline
\multicolumn{6}{l}{\textbf{Panel A: Evictions}} \\
Control & 2.112 & 2.189 & 2.239 & 2.302 & 2.201 \\
Treated & 1.974 & 1.842 & 1.782 & 1.782 & 1.824 \\
\hline
\multicolumn{6}{l}{\textbf{Panel B: Filings}} \\
Control & 3.164 & 3.291 & 3.372 & 3.515 & 3.535 \\
Treated & 3.041 & 2.811 & 2.740 & 2.746 & 2.693 \\
\hline
\end{tabular}
\caption{Baseline averages by group (pre-hurricane period) for each treatment radius. Control and treated ZIP codes are compared across radii of 15, 21, 26, 50, and 100 miles.}
\label{tab:baseline-averages-radius-horizontal}
\end{table}

\begin{figure}[H]
    \centering
    \includegraphics[width=0.6\textwidth]{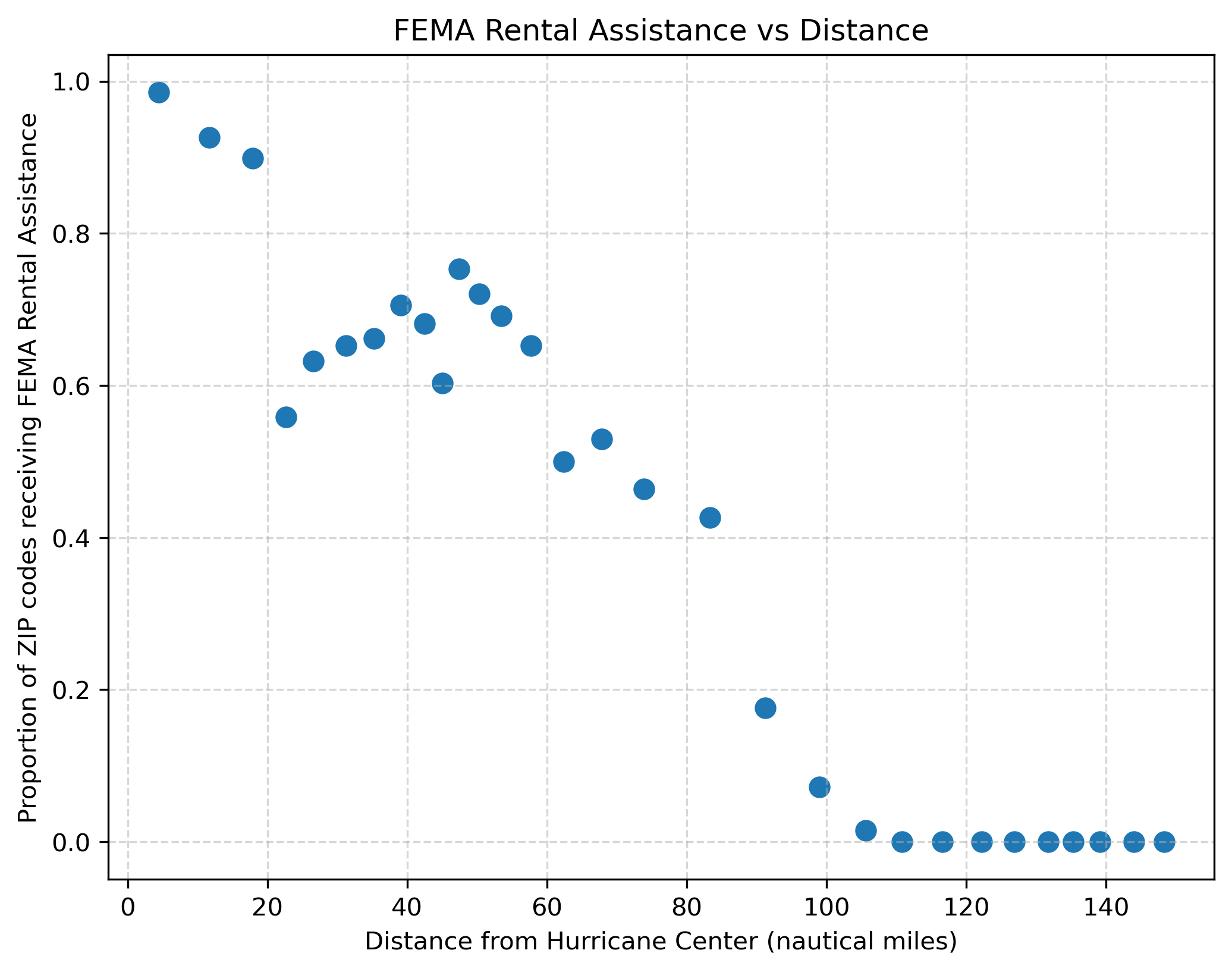}
    \caption{A bin scatter plot of the proportion of ZIP codes at increasing distances from the hurricane path that had recipients of FEMA rental assistance. The plot shows a decreasing in the proportion with a notable jump at the 20 nautical mile mark.}
    \label{fig:bin-scatter}
\end{figure}

\begin{figure}[H]
    \centering
    \includegraphics[width=0.5\textwidth]{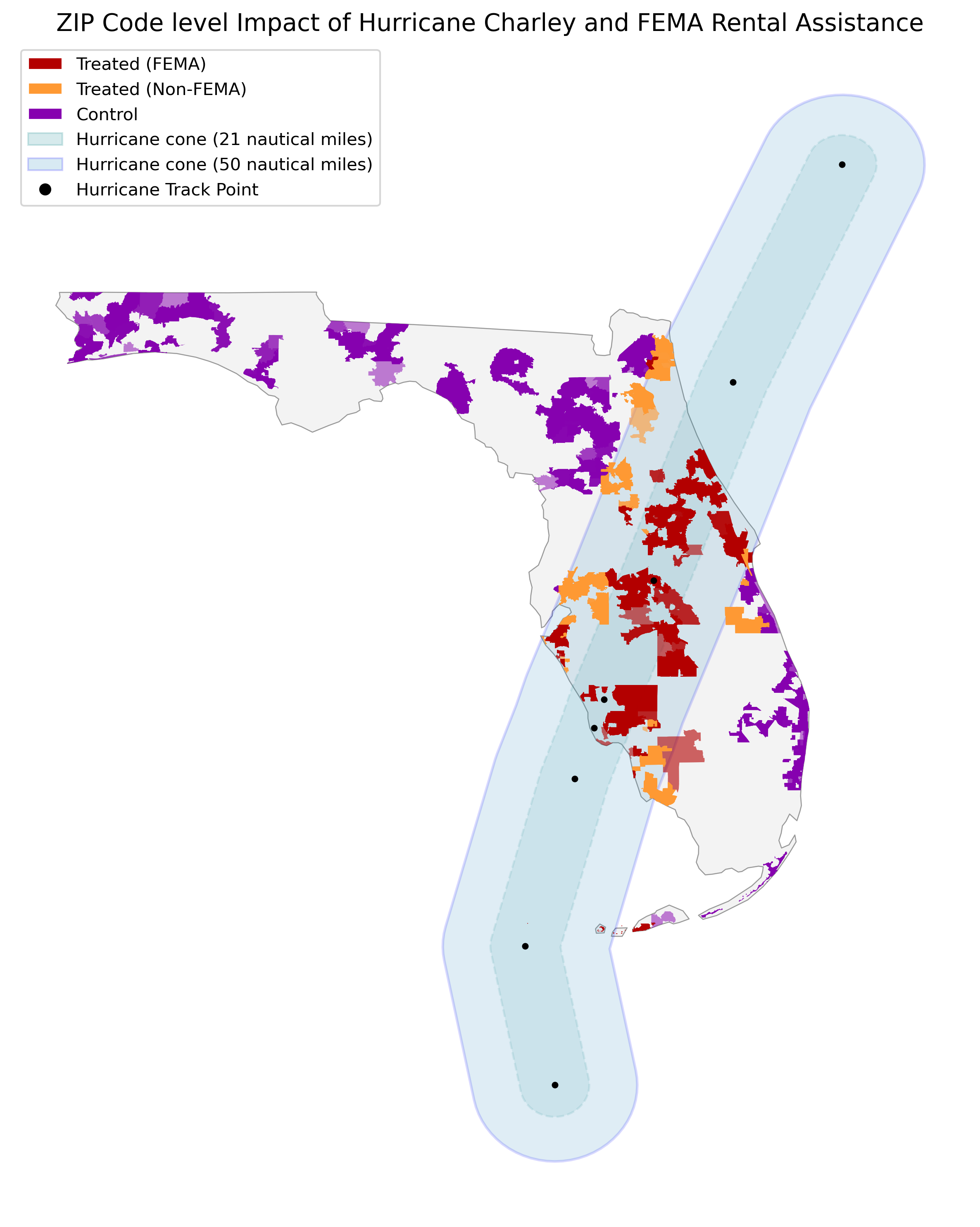}
    \caption{A map of ZIP codes that received FEMA rental assistance in red during Hurricane Charley. The orange non-FEMA treated ZIP codes are those that fall within the hurricane cone but did not receive FEMA rental assistance. The map displays two cone sizes, 21 nautical miles and 50 nautical miles, illustrating the sensitivity of the proportion of ZIP codes receiving FEMA rental assistance at different distances.}
    \label{fig:fema-map}
\end{figure}

We then investigate the role of FEMA, in particular hurricane events in which FEMA made renter assistance available in impacted ZIP codes. Eligibility for assistance through FEMA depends on a geography being declared FEMA eligible\footnote{FEMA provides a web page for individuals to lookup whether their ZIP code lies in a FEMA disaster declared county https://www.fema.gov/node/how-do-i-know-if-my-area-eligible-assistance?}, as well as individual assessments of damage and hardship which correlate with proximity to the path of the hurricane event. \autoref{fig:fema-map} shows the ZIP codes that received assistance via FEMA and their distance to the hurricane event. We observe that the intersection of ZIP code boundaries, storm damage intensity and receiving FEMA assistance correlate and vary in relation to distance. \autoref{fig:bin-scatter} shows the proportion of ZIP codes with individuals receiving FEMA assistance and the distance from the path of the hurricane during FEMA declared hurricanes.

Building on our baseline eviction and filings results we observe a bifurcation in the resultant filings and evictions in the period following the hurricane event that varies with FEMA assistance. Both the event-study and difference-in-differences regression results illuminate the decrease in the rate of filings and evictions for hurricane events where FEMA assistance is made available. In particular, from \autoref{tab:triple-did-radius-combined} we find a maximal effect size in the monthly average evictions with an increase of 0.646 in non-FEMA receiving ZIP codes and a 0.638 increase in filings at the 50 nautical mile radius of inclusion of ZIP codes. This result is largely reversed by FEMA eligibility, as the coefficient of the interaction is statistically significant with a 0.571 decrease for evictions and 0.536 decrease for filings. Our instrumental variable results match the direction and are of similar magnitude, whilst also remaining statistically significant for many of the same treatment radii as in the non-instrumented results.

\begin{figure}[H]
  \centering
  \subfloat[Eviction: FEMA]{\includegraphics[width=.49\linewidth]{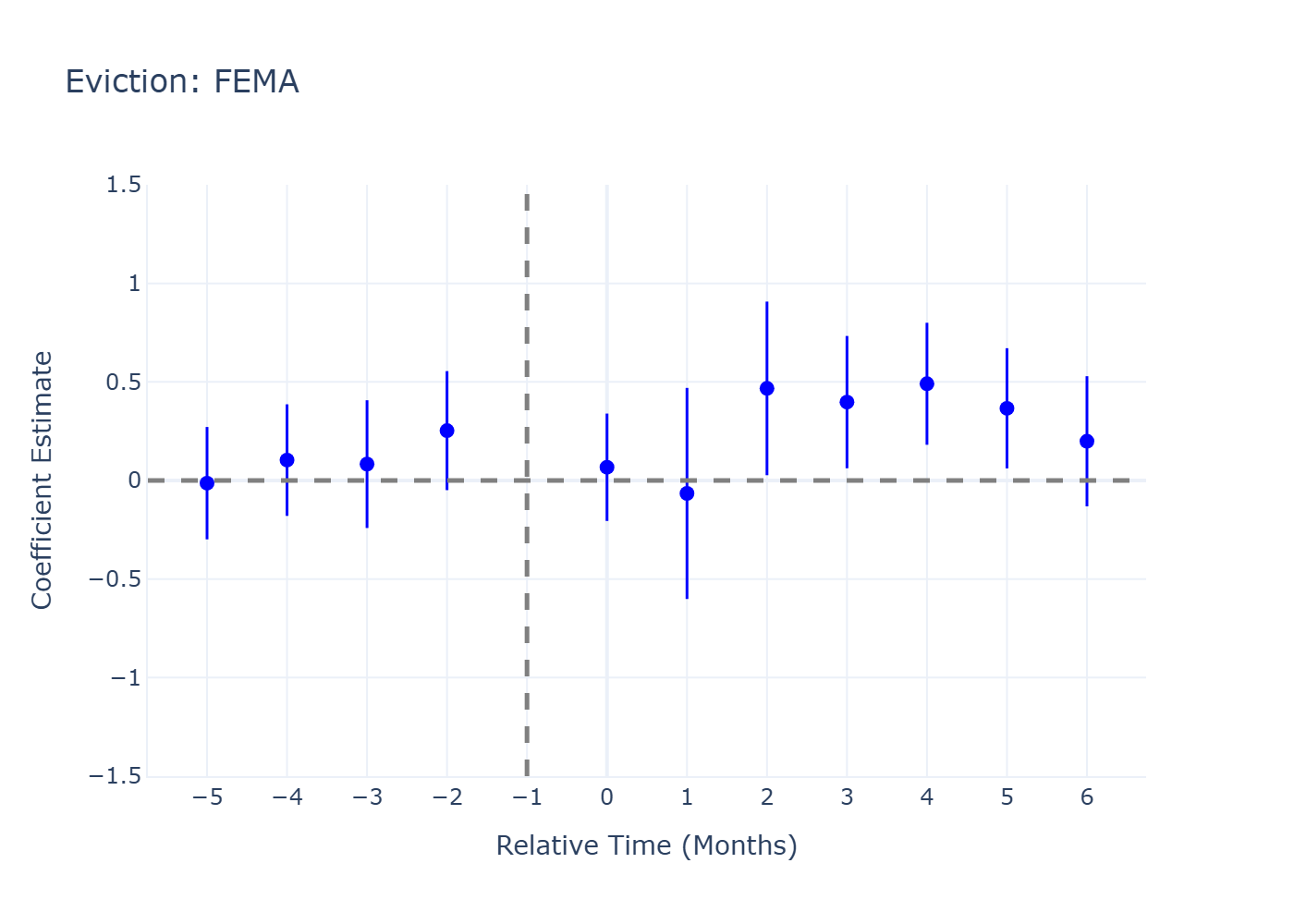}}\hfill
  \subfloat[Eviction: Non-FEMA]{\includegraphics[width=.49\linewidth]{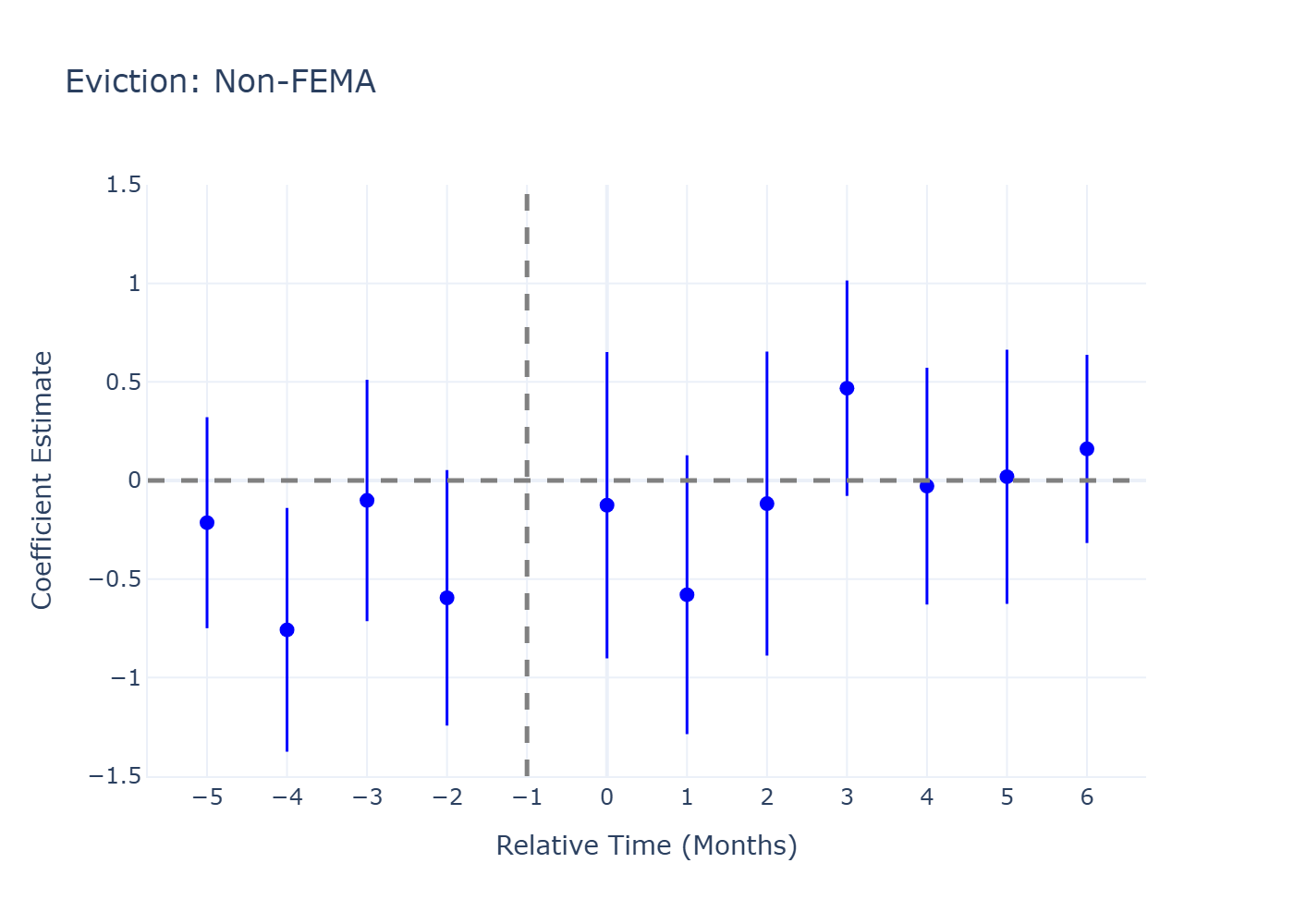}}\\[1ex]
  \subfloat[Filing: FEMA]{\includegraphics[width=.49\linewidth]{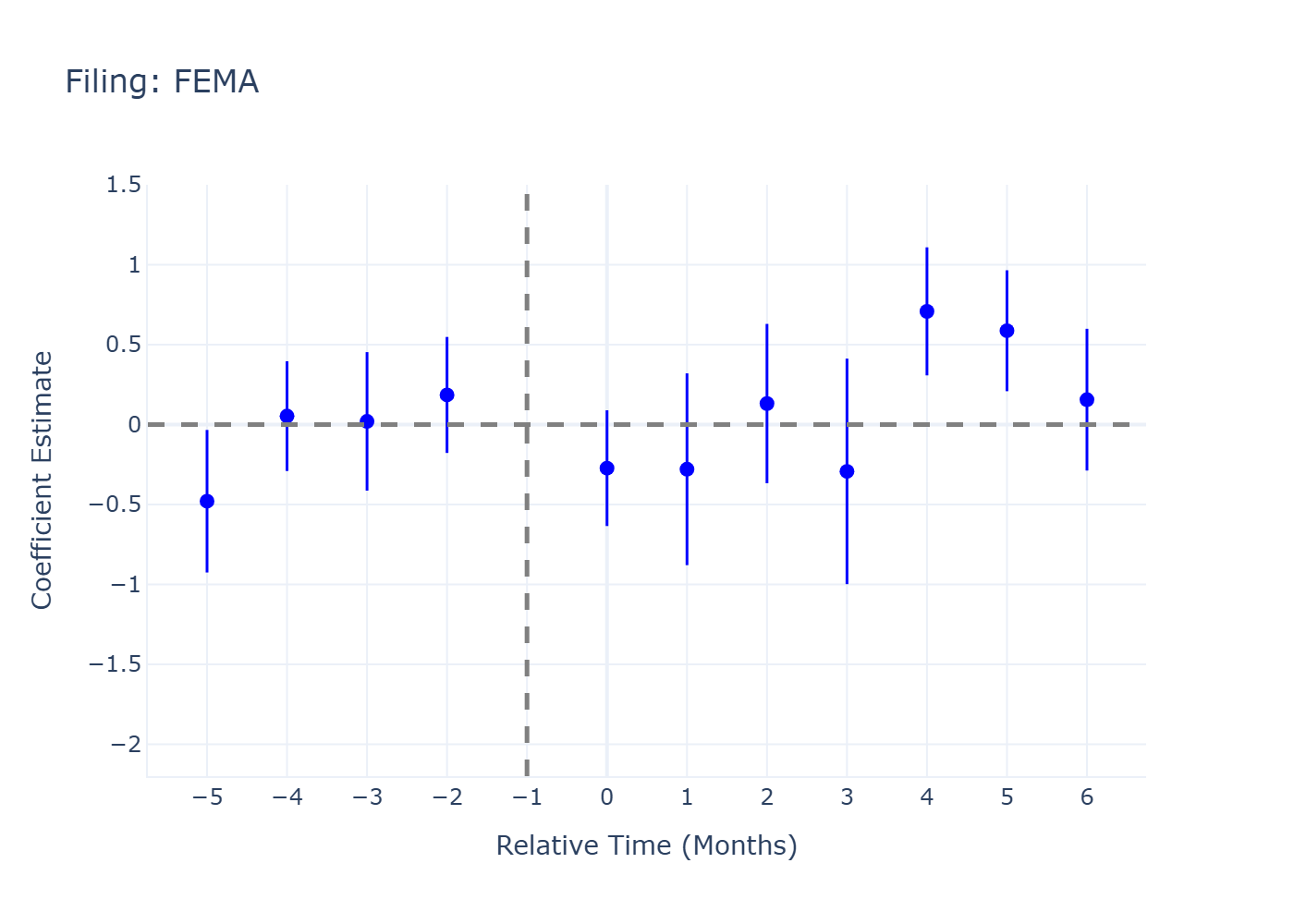}}\hfill
  \subfloat[Filing: Non-FEMA]{\includegraphics[width=.49\linewidth]{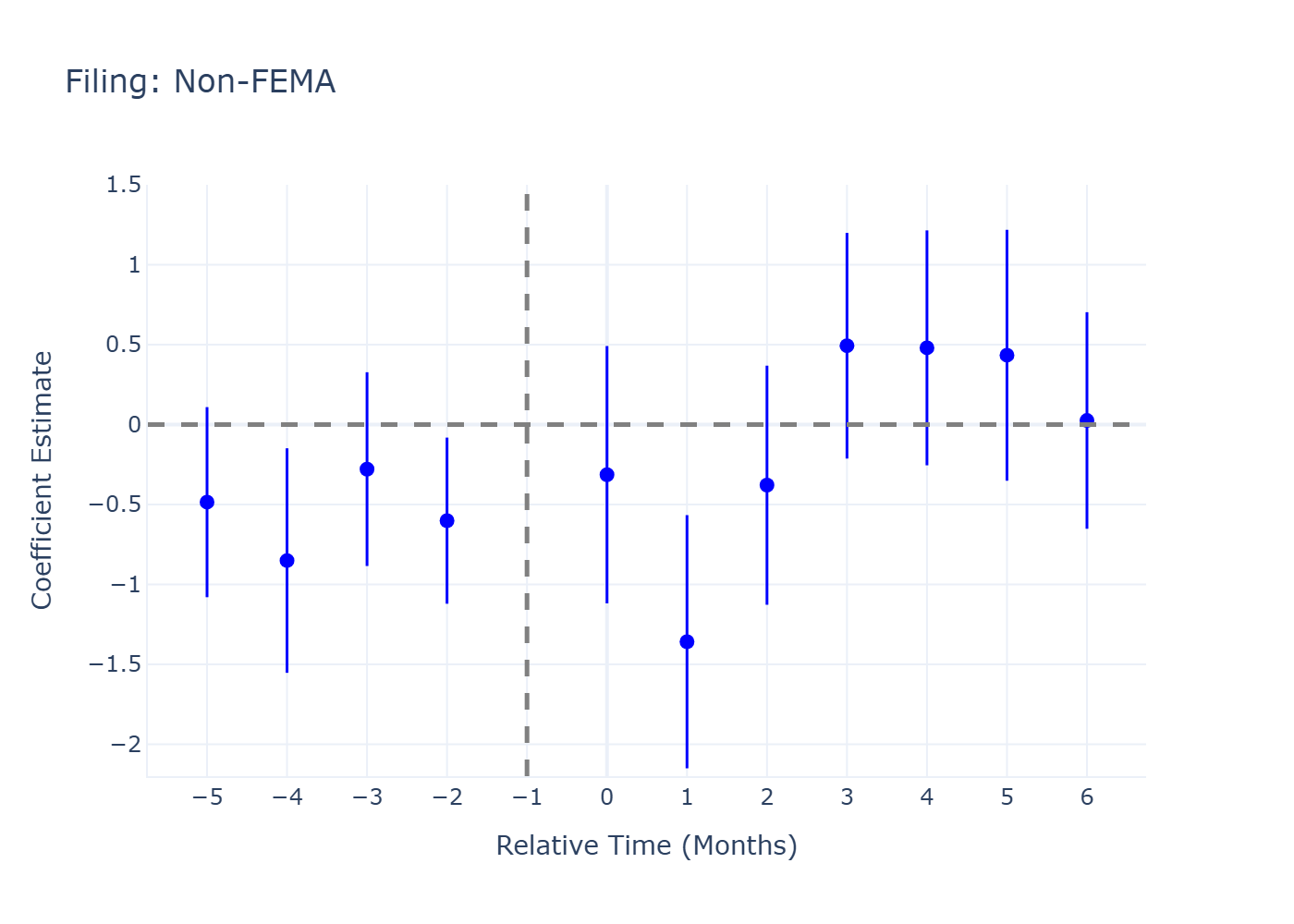}}
  \caption{Estimated coefficients from an event study model  evaluating the impact of hurricane exposure on monthly eviction rates at the ZIP code level with a treatment radius of 21 nautical miles. The event window spans from five months before to six months after landfall, with the month immediately preceding the hurricane (month -1) serving as the reference period. The treated group is divided based on FEMA rental assistance availability. Standard errors are clustered at the ZIP code level.}
        \label{fig:evict_fema_comparison_21}
\end{figure}

\begin{figure}[H]
  \centering
  \subfloat[Eviction: FEMA]{\includegraphics[width=.49\linewidth]{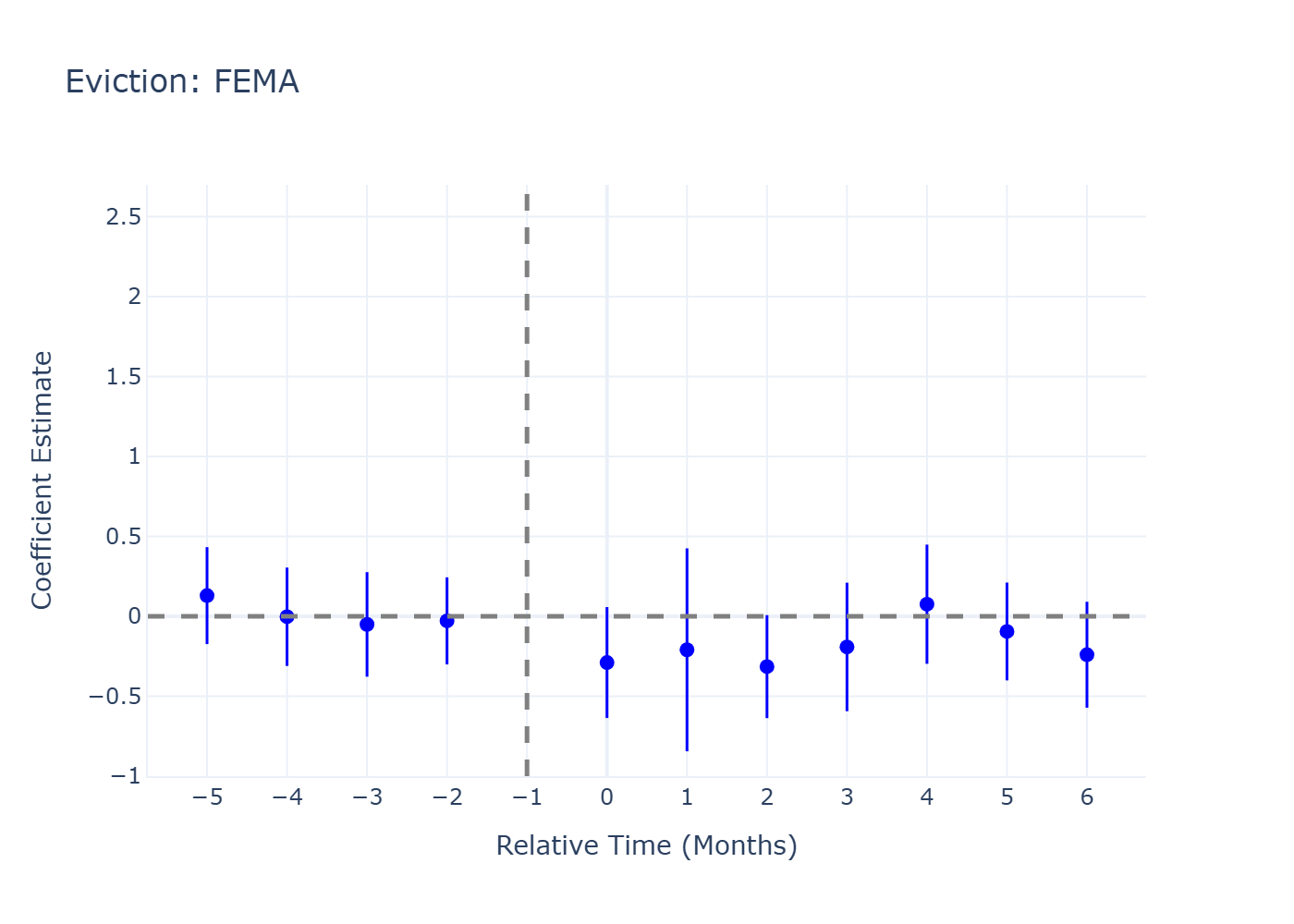}}\hfill
  \subfloat[Eviction: Non-FEMA]{\includegraphics[width=.49\linewidth]{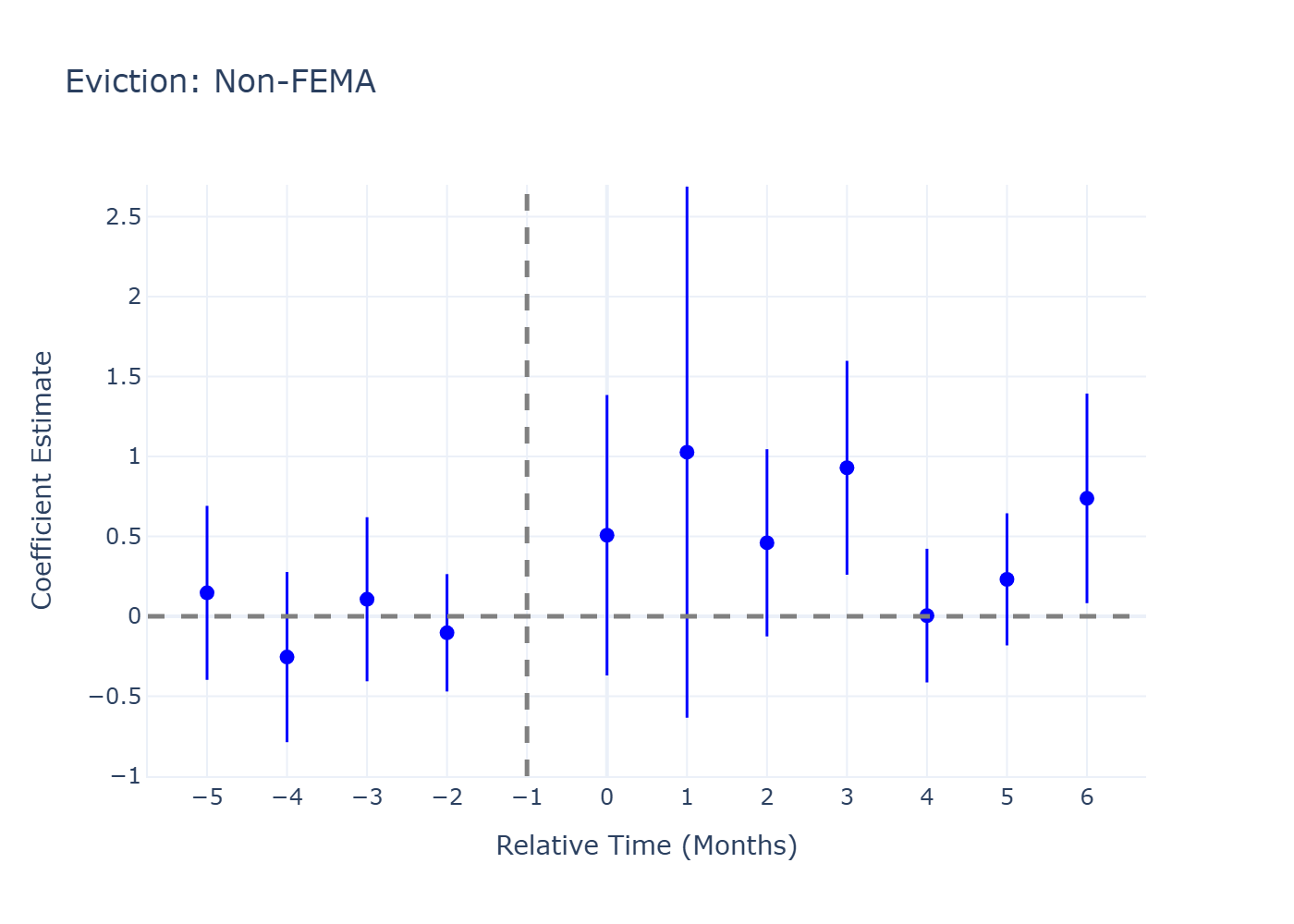}}\\[1ex]
  \subfloat[Filing: FEMA]{\includegraphics[width=.49\linewidth]{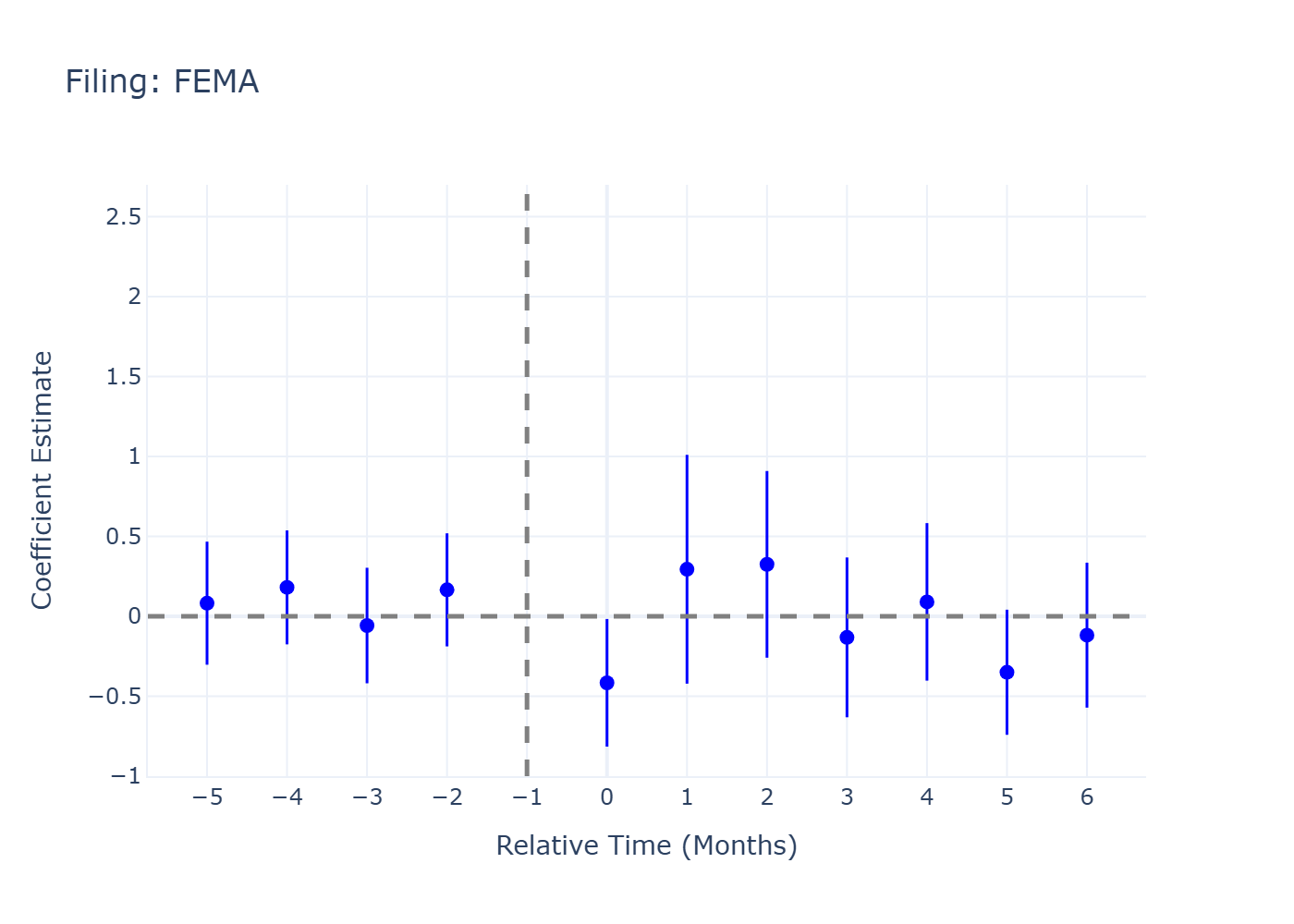}}\hfill
  \subfloat[Filing: Non-FEMA]{\includegraphics[width=.49\linewidth]{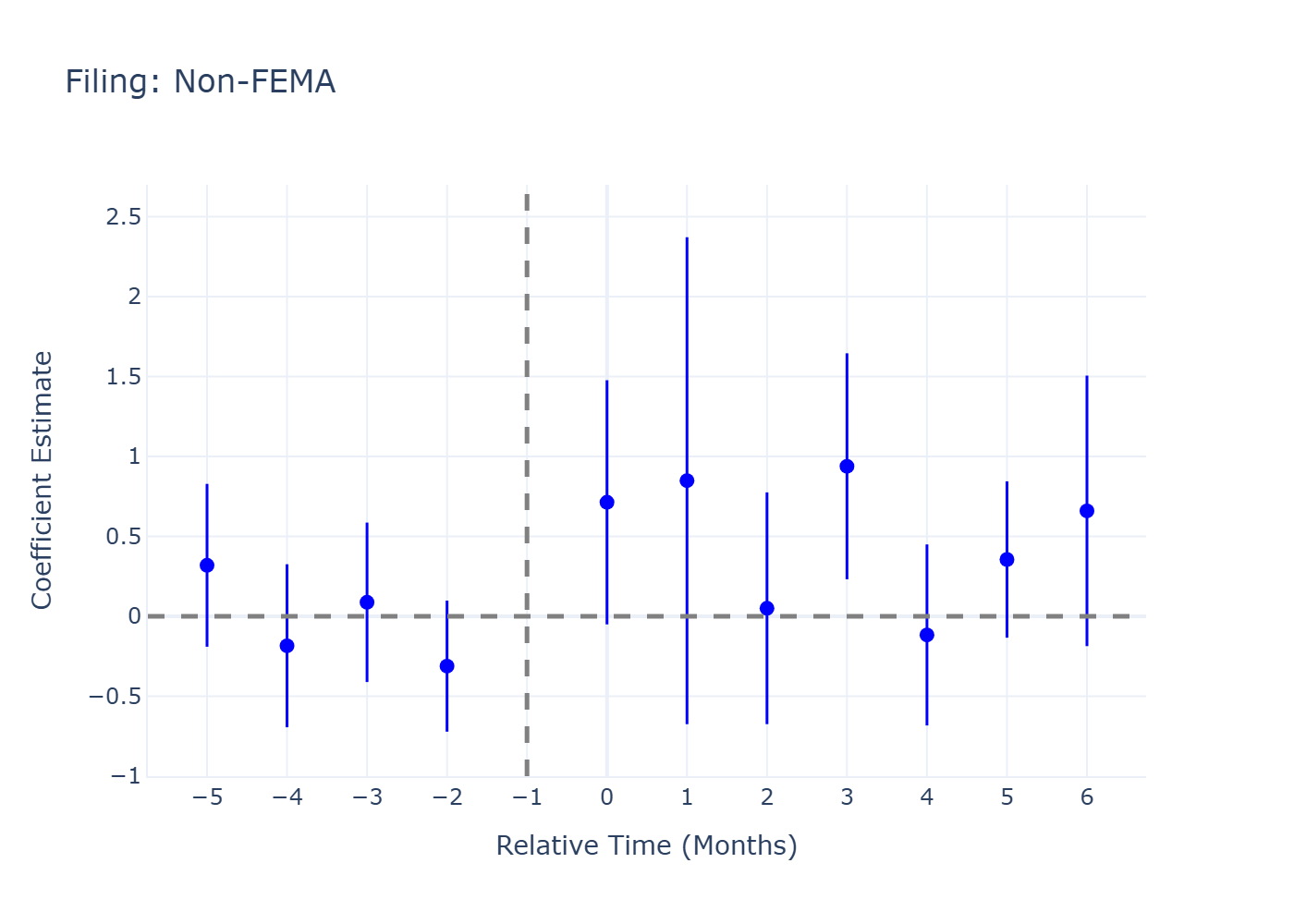}}
  \caption{Estimated coefficients from an event study model  evaluating the impact of hurricane exposure on monthly eviction rates at the ZIP code level with a treatment radius of 50 nautical miles. The event window spans from five months before to six months after landfall, with the month immediately preceding the hurricane (month -1) serving as the reference period. The treated group is divided based on FEMA rental assistance availability. Standard errors are clustered at the ZIP code level.}
        \label{fig:evict_fema_comparison_50} 
\end{figure}

\begin{table}[H]
\centering
\begin{tabular}{lccccc}
\hline
 & \multicolumn{5}{c}{\textbf{Treatment Radius}} \\
\cline{2-6}
 & 15 & 21 & 26 & 50 & 100 \\
\hline
\multicolumn{6}{l}{\textbf{Panel A: Eviction (OLS)}} \\

Treatment$\times$Post
 & -0.355 & 0.088 & 0.339$^{**}$ & 0.646$^{***}$ & 0.178 \\
 & [0.313] & [0.224] & [0.151] & [0.218] & [0.126] \\

Treatment$\times$Post$\times$FEMA
 & 0.902$^{***}$ & 0.470$^{**}$ & 0.128 & -0.571$^{***}$ & -0.248 \\
 & [0.325] & [0.233] & [0.143] & [0.208] & [0.194] \\

\hline
\multicolumn{6}{l}{\textbf{Panel B: Filing (OLS)}} \\

Treatment$\times$Post
 & -0.633 & 0.097 & 0.443$^{**}$ & 0.638$^{***}$ & 0.257 \\
 & [0.502] & [0.295] & [0.180] & [0.247] & [0.180] \\

Treatment$\times$Post$\times$FEMA
 & 1.164$^{**}$ & 0.407 & -0.055 & -0.536$^{*}$ & -0.865$^{**}$ \\
 & [0.528] & [0.314] & [0.164] & [0.274] & [0.369] \\

\hline\hline
\multicolumn{6}{l}{\textbf{Panel C: Eviction (IV)}} \\

Treatment$\times$Post
 & -0.268 & 0.200 & 0.319$^{**}$ & 0.410$^{***}$ & 0.059 \\
 & [0.401] & [0.211] & [0.136] & [0.156] & [0.104] \\

Treatment$\times$Post$\times$FEMA
 & 0.761$^{*}$ & 0.291 & 0.112 & -0.265 & -0.392 \\
 & [0.423] & [0.242] & [0.173] & [0.194] & [0.273] \\

\hline
\multicolumn{6}{l}{\textbf{Panel D: Filing (IV)}} \\

Treatment$\times$Post
 & -0.265 & 0.406$^{**}$ & 0.567$^{***}$ & 0.691$^{***}$ & 0.317$^{**}$ \\
 & [0.375] & [0.193] & [0.172] & [0.195] & [0.158] \\

Treatment$\times$Post$\times$FEMA
 & 0.680$^{*}$ & -0.057 & -0.332$^{*}$ & -0.949$^{***}$ & -2.711$^{***}$ \\
 & [0.367] & [0.224] & [0.181] & [0.283] & [0.533] \\

\hline
Event-specific ZIP Code FE & Y & Y & Y & Y & Y \\
Event-specific Time FE & Y & Y & Y & Y & Y \\
Observations
 & 21,617 & 20,383 & 19,360 & 16,939 & 15,894 \\
\hline
\end{tabular}
\caption{Triple-differences estimates of the impact of storm exposure and FEMA assistance on eviction and filing outcomes across alternative treatment radii. Panels A and B report OLS estimates, while Panels C and D report instrumental-variables estimates where FEMA assistance is instrumented. All specifications include event-specific ZIP code and time fixed effects. Standard errors clustered at the ZIP code level are reported in brackets. ***, **, and * denote statistical significance at the 1\%, 5\%, and 10\% levels, respectively.}
\label{tab:triple-did-radius-combined}
\end{table}

\subsubsection{Multiple and Tropical Storms}

We expand our baseline results to include the effects of multiple hurricanes in one season, as well as the impact of tropical storms to study the household response to damage intensity across two margins, accumulative damage intensity and immediate damage intensity. We view this as a way to gauge the extent that potential damage severity may be driving our results, suggesting that the sensitivity to the impact of availability for FEMA assistance has important heterogeneity. 

\begin{figure}[H]
  \centering
  \subfloat[Eviction: 21 nautical miles]{\includegraphics[width=.49\linewidth]{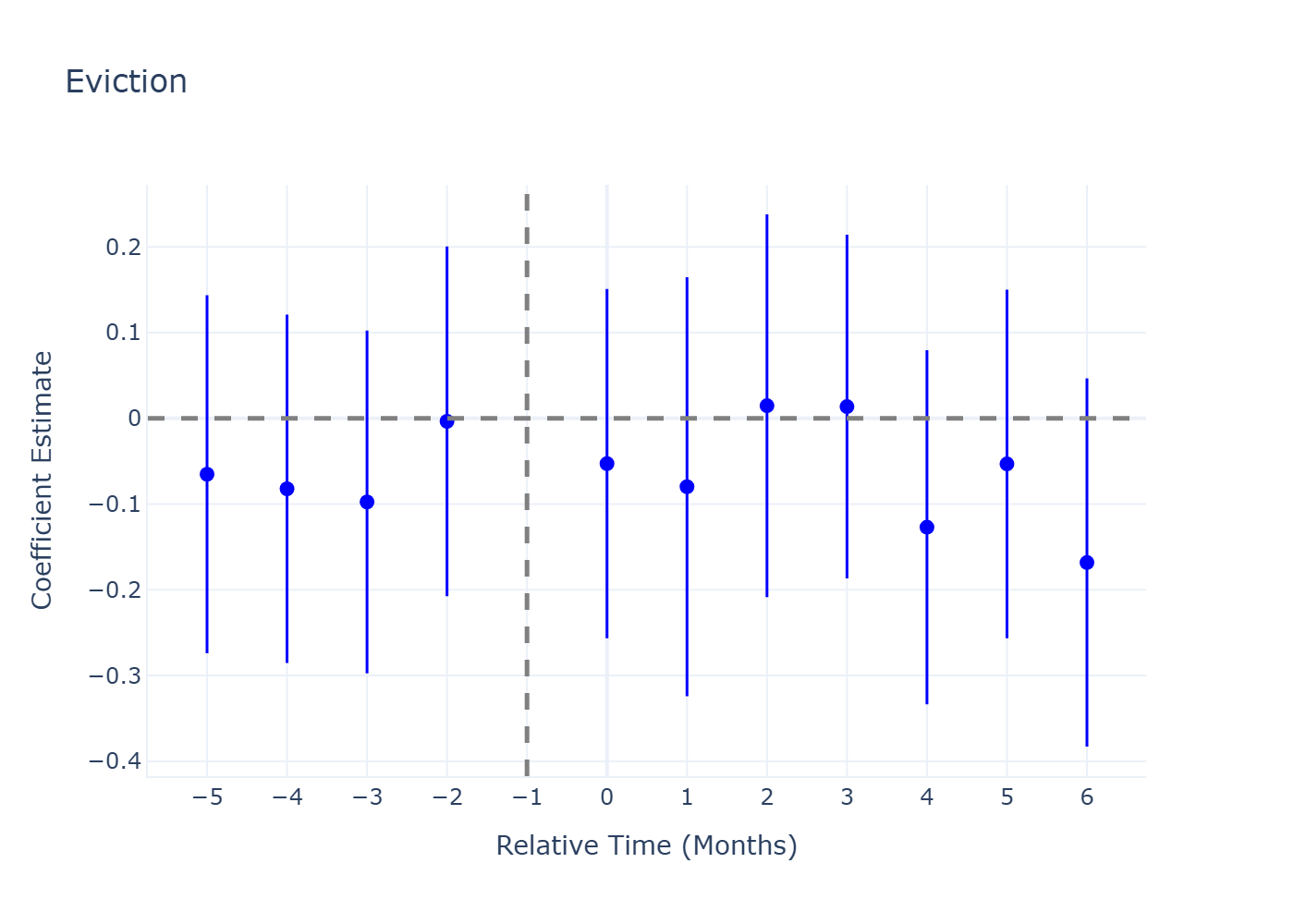}}\hfill
  \subfloat[Filing: 21 nautical miles]{\includegraphics[width=.49\linewidth]{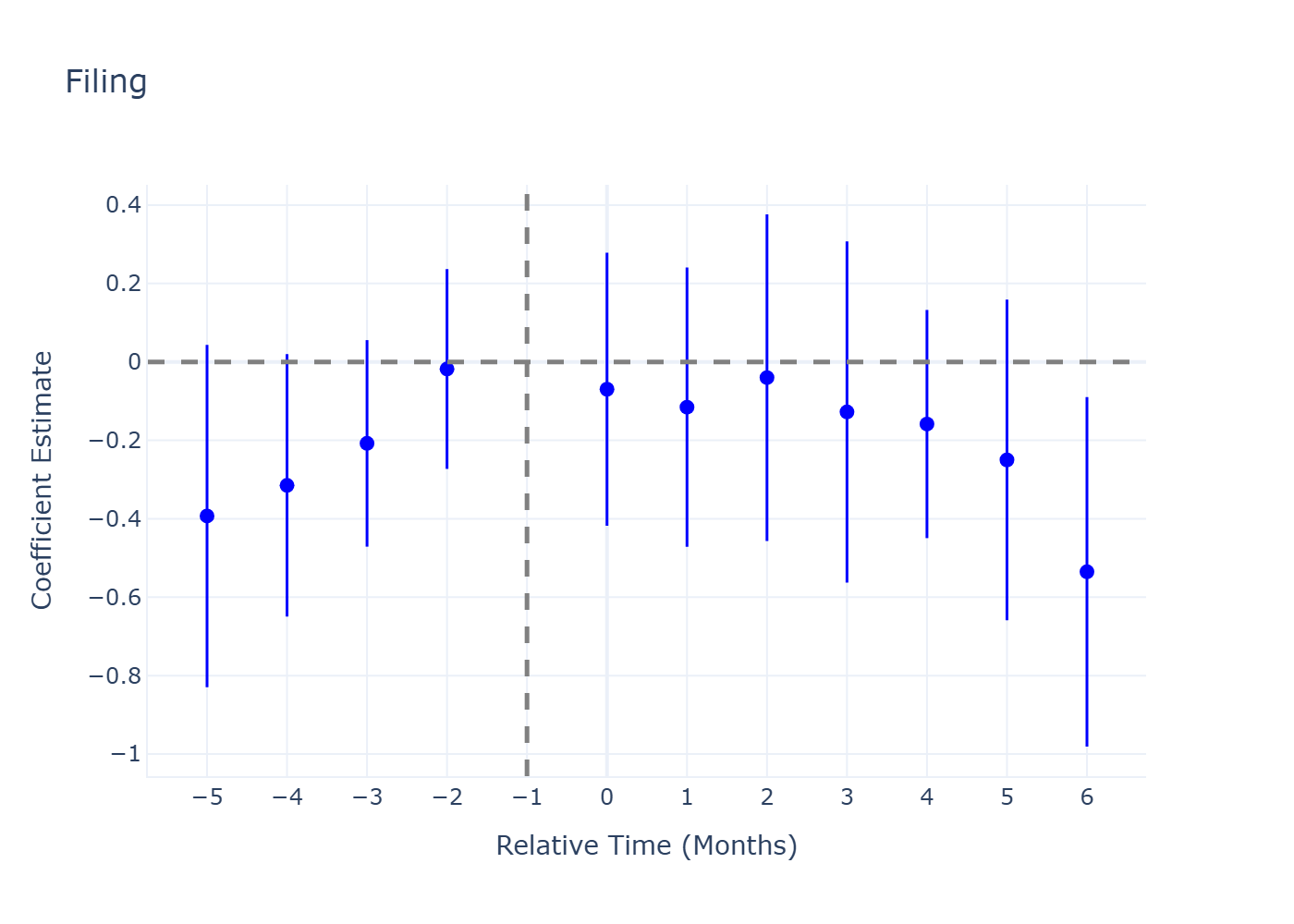}}\\[1ex]
  \subfloat[Eviction: 50 nautical miles]{\includegraphics[width=.49\linewidth]{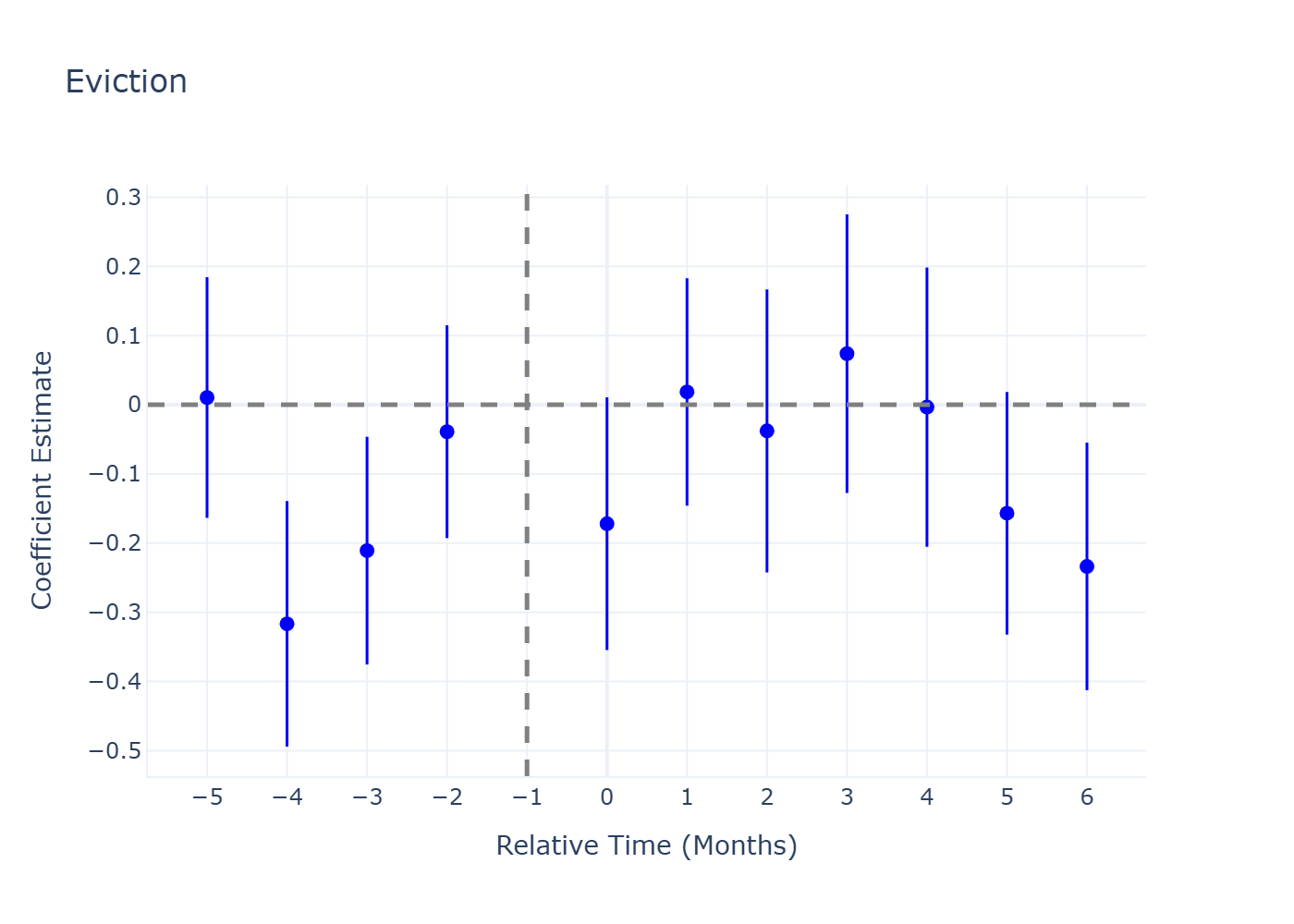}}\hfill
  \subfloat[Filing: 50 nautical miles]{\includegraphics[width=.49\linewidth]{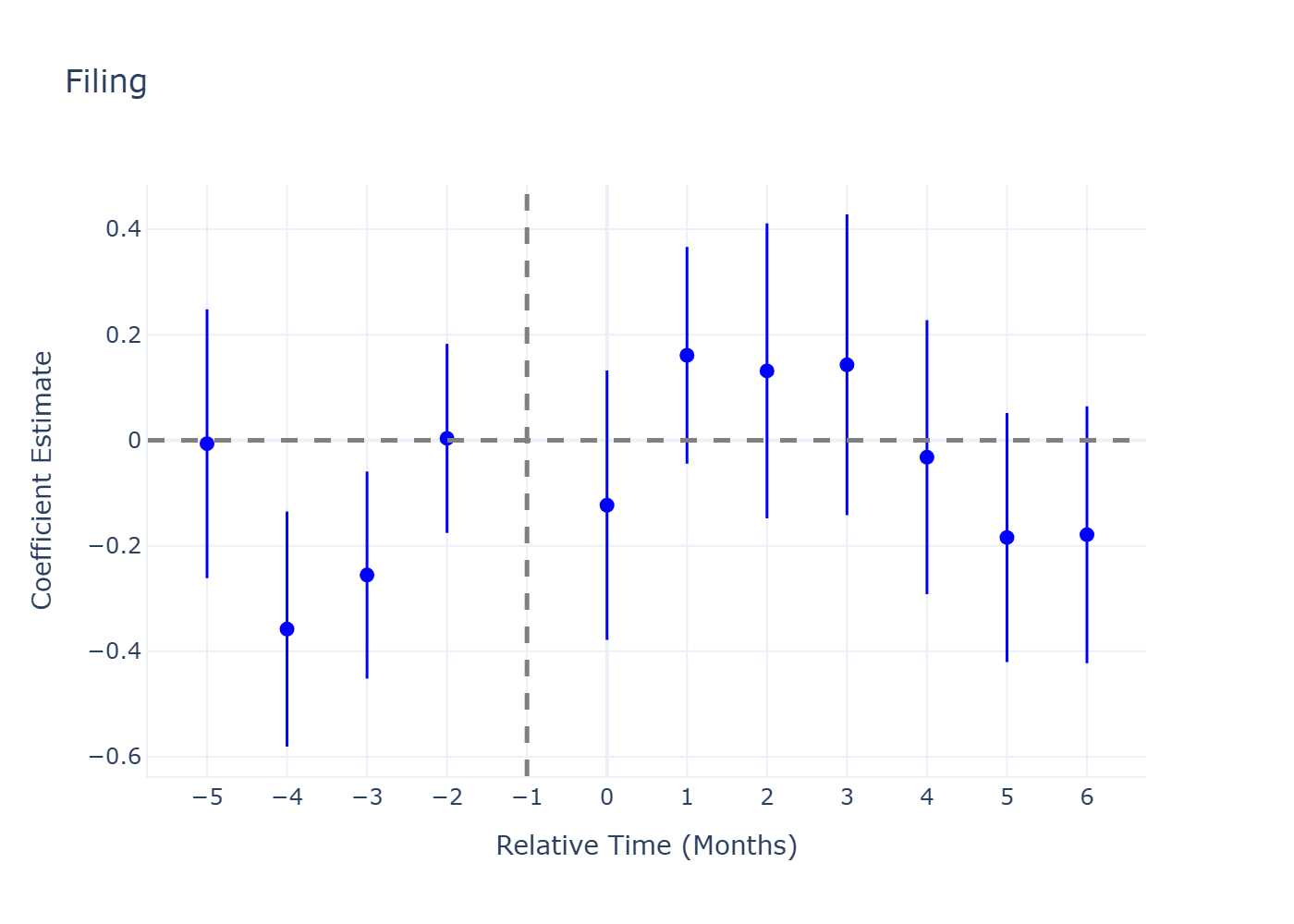}}
  \caption{The figure provides the estimated coefficients from an event study examining the combined effect of storm exposure, including hurricanes and tropical storms on monthly eviction-related outcomes at the ZIP code level. The analysis covers a window from five months before to six months after the storm event, with the month immediately preceding the event (month -1) serving as the reference period. Panel (a) displays estimates for monthly eviction rates, while Panel (b) shows estimates for monthly filing rates at the 21 nautical mile treatment radius, whereas Panel (c) displays estimates for monthly eviction rates, while Panel (d) shows estimates for monthly filing rates at the 50 nautical mile treatment radius. Vertical lines indicate 95\% confidence intervals, constructed using standard errors clustered at the ZIP code level.}
        \label{fig:evict-filing-allstorms} 
\end{figure}

We also view including the more frequent lower intensity storms as an additional test to possible seasonality or region specific factors above and beyond the fixed effect controls. Specifically, our result is likely to be robust from other temporal patterns that may have coincided in our analysis with the comparatively fewer hurricane events in our sample. Adding additional storm events help provide a degree of variability in damage intensity in between the control and major hurricane treated events. The results show that an increase in evictions and filings in the post-event period are not driven by tropical storm events, increasing our confidence in our results as stemming from coinciding with intense hurricane events and the additional absence of FEMA assistance for some geographies. In particular, \autoref{fig:evict-filing-allstorms} and \autoref{tab:DiD-radius-hurricane-storm-combined} show the baseline result when including the greater number of tropical storm events no longer holds (note the three times increase in sample size). Separating the two storm types, shows even a statistically significant decline in evictions along the path of tropical storms for treatment radii under 26 nautical miles.

Furthermore, we find that multiple storm events show that geographies on the periphery of FEMA eligibility appear to have a greater likelihood for evictions and filings. Specifically, as we increase the size of our treatment radius, the coefficient on the interaction with multiply treated ZIP codes significantly increases at the 50 nautical mile radius, as in \autoref{tab:multi-interaction}. This suggests, that the accumulative impact of hurricane damage may also be additionally impacting ZIP codes that are also more likely to be non-FEMA eligible. That is, the impact of increased damage intensity via accumulated events, crucially effects peripheral but still damage-sensitive geographies.

\begin{table}[H]
\centering
\begin{tabular}{lccccc}
\hline
 & \multicolumn{5}{c}{\textbf{Treatment Radius}} \\
\cline{2-6}
 & 15 & 21 & 26 & 50 & 100 \\
\hline
\multicolumn{6}{l}{\textbf{Hurricanes and Tropical Storms}} \\
\multicolumn{6}{l}{\textbf{Panel A: Eviction}} \\
Treated$\times$Post & -0.042 & -0.014 & 0.013 & 0.038 & -0.005 \\
 & [0.059] & [0.054] & [0.057] & [0.051] & [0.043] \\
\hline
\multicolumn{6}{l}{\textbf{Panel B: Filing}} \\
Treated$\times$Post & -0.018 & 0.005 & 0.048 & 0.113 & 0.188$^{***}$ \\
 & [0.099] & [0.090] & [0.079] & [0.072] & [0.058] \\
\hline
\multicolumn{6}{l}{\textbf{Separate Effects: Hurricane and Tropical Storms}} \\
\multicolumn{6}{l}{\textbf{Panel C: Eviction}} \\
Post$\times$Hurricane & 0.172$^{*}$ & 0.201$^{*}$ & 0.169 & 0.073 & -0.049 \\
 & [0.097] & [0.103] & [0.114] & [0.114] & [0.084] \\
Post$\times$Tropical Storm & -0.157$^{**}$ & -0.121$^{**}$ & -0.061 & 0.028 & 0.001 \\
 & [0.071] & [0.060] & [0.060] & [0.052] & [0.048] \\
\hline
\multicolumn{6}{l}{\textbf{Panel D: Filing}} \\
Post$\times$Hurricane & 0.164 & 0.167 & 0.142 & 0.133 & 0.193 \\
 & [0.126] & [0.125] & [0.137] & [0.150] & [0.149] \\
Post$\times$Tropical Storm & -0.116 & -0.076 & 0.003 & 0.107 & 0.187$^{***}$ \\
 & [0.110] & [0.095] & [0.078] & [0.069] & [0.062] \\
\hline
Event-specific ZIP Code FE & Y & Y & Y & Y & Y \\
Event-specific Time FE & Y & Y & Y & Y & Y \\
Observations & 83,704 & 81,921 & 80,495 & 75,703 & 71,206 \\
\hline
\end{tabular}
\caption{Difference-in-Differences estimates of eviction and filing outcomes following hurricanes and tropical storms. The top panel reports combined treatment effects for all events, while the bottom panels separate effects for hurricanes and tropical storms. Standard errors clustered at the ZIP code level are in brackets. ***, **, and * indicate significance at the 1\%, 5\%, and 10\% levels, respectively.}
\label{tab:DiD-radius-hurricane-storm-combined}
\end{table}

\begin{table}[H]
\centering
\begin{tabular}{lccccc}
\hline
 & \multicolumn{5}{c}{\textbf{Treatment Radius}} \\
\cline{2-6}
 & 15 & 21 & 26 & 50 & 100 \\
\hline
\multicolumn{6}{l}{\textbf{Panel A: Eviction}} \\

Treatment$\times$Post
 & 0.307$^{***}$ & 0.283$^{***}$ & 0.237$^{**}$ & 0.024 & -0.115$^{*}$ \\
 & [0.115] & [0.109] & [0.110] & [0.089] & [0.068] \\

Treatment$\times$Post$\times$Multiple
 & 0.066 & 0.158 & 0.134 & 0.316$^{***}$ & -0.052 \\
 & [0.141] & [0.115] & [0.111] & [0.118] & [0.114] \\

\hline
\multicolumn{6}{l}{\textbf{Panel B: Filing}} \\

Treatment$\times$Post
 & 0.363$^{**}$ & 0.327$^{**}$ & 0.274$^{*}$ & 0.031 & -0.005 \\
 & [0.183] & [0.158] & [0.145] & [0.104] & [0.091] \\

Treatment$\times$Post$\times$Multiple
 & 0.096 & 0.142 & 0.112 & 0.466$^{***}$ & 0.033 \\
 & [0.223] & [0.154] & [0.135] & [0.142] & [0.149] \\

\hline
Event-specific ZIP Code FE & Y & Y & Y & Y & Y \\
Event-specific Time FE & Y & Y & Y & Y & Y \\
Observations
 & 27,536 & 27,553 & 27,548 & 27,531 & 27,540 \\
\hline
\end{tabular}
\caption{Triple-differences estimates of the impact of major storm exposure and multiple-event exposure within the same year on eviction and filing outcomes across alternative treatment radii. Panel A presents results for eviction outcomes, and Panel B for eviction filings. The indicator \emph{Multiple} equals one for ZIP codes exposed to more than one hurricane event within a given year. All specifications include event-specific ZIP code and time fixed effects. Standard errors clustered at the ZIP code level are reported in brackets. ***, **, and * denote statistical significance at the 1\%, 5\%, and 10\% levels, respectively.}
\label{tab:multi-interaction}
\end{table}

\subsection{Payday Loans}

\subsubsection{Hurricanes}

\begin{figure}[H]
  \centering
  \subfloat[Transaction Volume: Weekly]{\includegraphics[width=.49\linewidth]{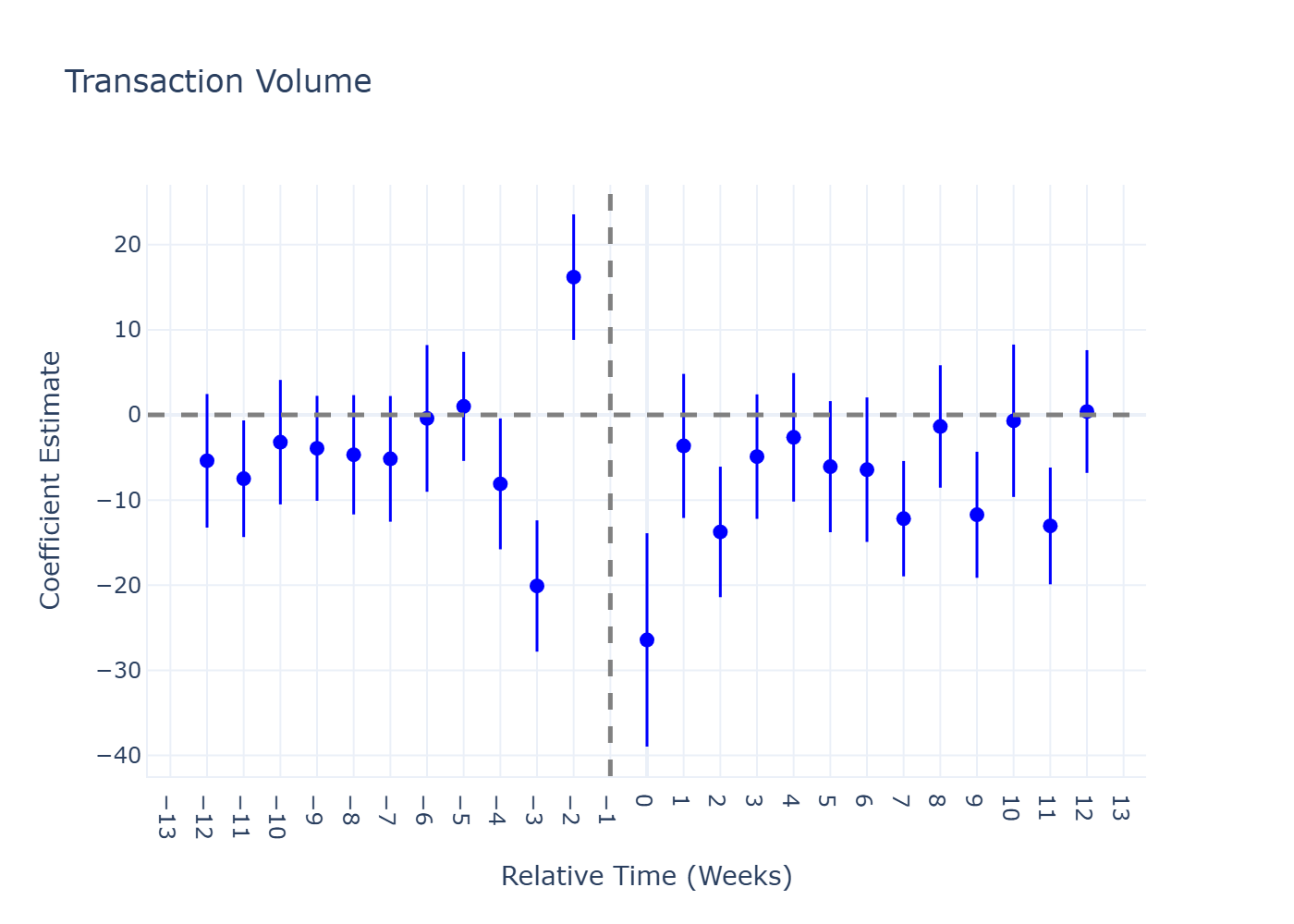}}\hfill
  \subfloat[Default: Weekly] {\includegraphics[width=.49\linewidth]{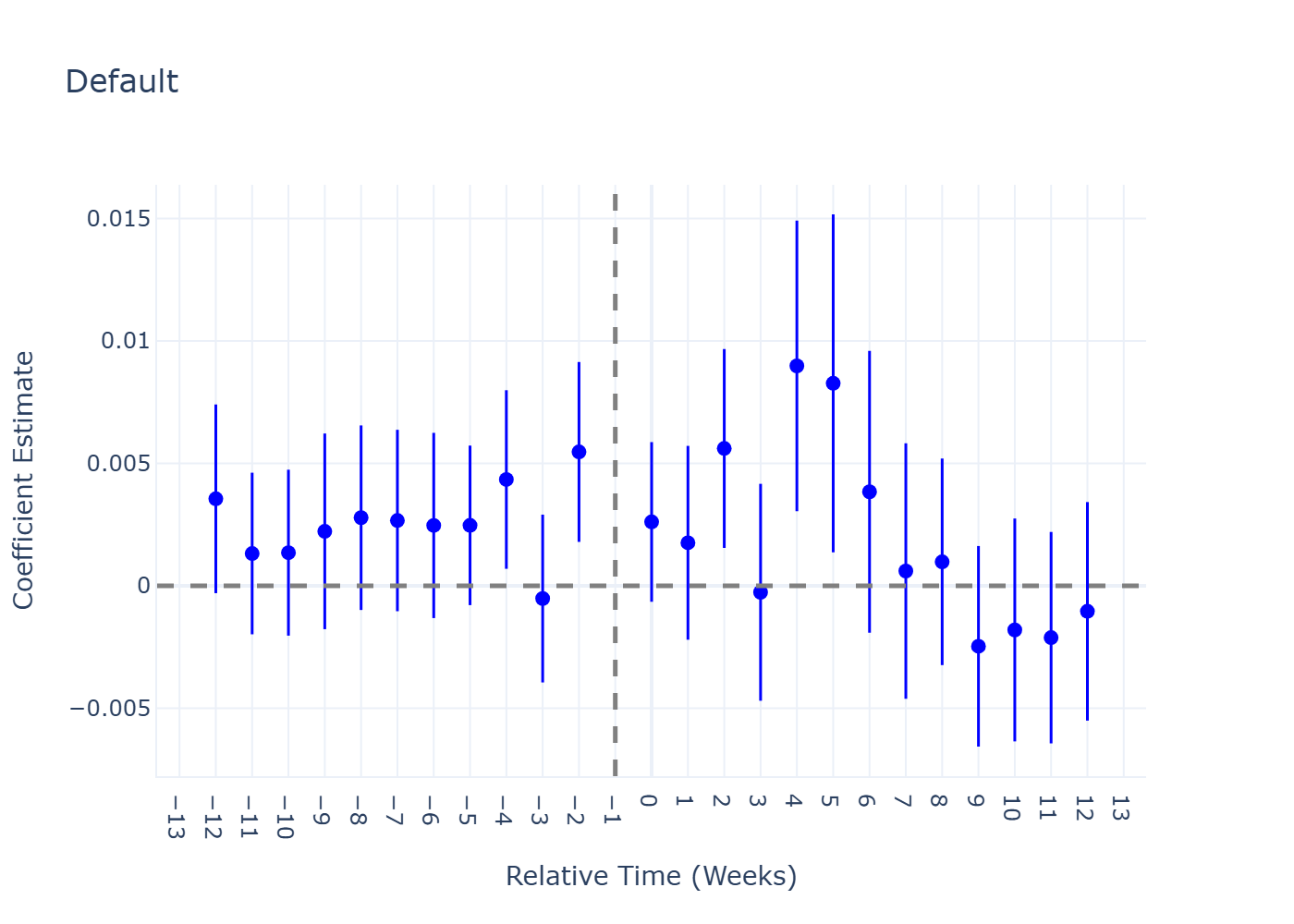}}
  \caption{The figure provides the estimated coefficients from an event study examining the impact of hurricane exposure on weekly payday loan-related outcomes at the ZIP code level with a treatment radius of 21 nautical miles. The analysis spans a window from five weeks before to twelve weeks after the hurricane event, with the week immediately preceding the event (week -1) serving as the reference period. Panel (a) reports estimates for weekly transaction volume, while Panel (b) shows estimates for weekly default rates. Vertical lines represent 95\% confidence intervals, calculated using standard errors clustered at the ZIP code level.}
 \label{fig: trans-default}
\end{figure}

We further compare the usage of high cost credit during hurricane events with the availability of FEMA rental assistance. In particular, as the individual rental assistance provided by FEMA is of similar magnitudes to the average loan amount available to households via payday loans\footnote{The maximum amount in the state of Florida is \$500 and often households borrow at the maximum}. We find that transaction volumes decreases whereas default rates are higher post hurricane event and that interestingly the decrease in usage  is particularly driven by areas where FEMA assistance is available, and the opposite is true for defaults. Transactions volumes decrease in areas where FEMA assistance is available and increase when it is not, reflecting an overall decrease in demand of payday credit in areas in the hurricane path. On the other hand, \autoref{fig: trans-default} we observe a decrease in defaults in defaults in areas not receiving in FEMA assistance and interestingly an increase where it is available. 

Specifically, in \autoref{tab:DiD-radius-payday} we show that the volume of transactions reduce by a maximal amount of roughly 25 loans per month, 8.6\% per week relative to their pre-period baseline and an increase the proportion of defaulting loans by 0.3\% percentage points per week on average per ZIP code, which reflects an overall 6.5\% increase in the percentage of defaulting loans relative to the pre-period baseline. 

Comparatively, in \autoref{tab:triple-did-payday}, when FEMA rental assistance is available, we find transactions volumes decrease by 6.9\% per week relative to non-FEMA ZIP codes which see an increase of 4.9\% per week increase at the most maximal, with similarly statistically significant increases in non-FEMA ZIP codes and decreases in FEMA ZIP codes under our instrumental variable implementation. Furthermore, we find FEMA receiving ZIP codes result in an increase in defaults by 0.014 percentage points per week at the most maximal and that non-FEMA ZIP codes see a decrease of 0.012 percentage points at their most maximal, reflecting up to 30\% swings in the percentage of defaults in either direction for FEMA and non-FEMA receiving ZIP codes.

We interpret the opposing effects of defaults between FEMA receiving and non-FEMA receiving ZIP codes as possibly interacting with the composition of borrowers. Given the observed reductions in loan volumes in FEMA receiving ZIP codes, and the additional liquidity from FEMA assistance, the incentive to rollover loans may differ for individuals receiving versus not receiving FEMA assistance.

We should suggestive evidence of this compositional change by comparing the share of new customers using Payday loans. In \autoref{tab:DiD-new-costumers} we show a 0.8\% overall decline in the share of new customers in levels. This represents, a 7.7\% decline relative to the baseline 10.3\% share of new customers. Hence, this suggests the composition of repeat borrowers increases. Furthermore, \autoref{tab:triple-did-new-costumers} shows that this effect comes directly from those receiving FEMA rental assistance, with a null effect in areas not receiving FEMA rental assistance. Together with the increase in defaults and decrease in transaction volume, this suggests areas receiving FEMA rental assistance see a potential compositional change of payday borrowers of only those most dependent on renewed access to payday loans.

\begin{figure}[H]
  \centering
  \subfloat[Transaction Volume: Weekly]{\includegraphics[width=.49\linewidth]{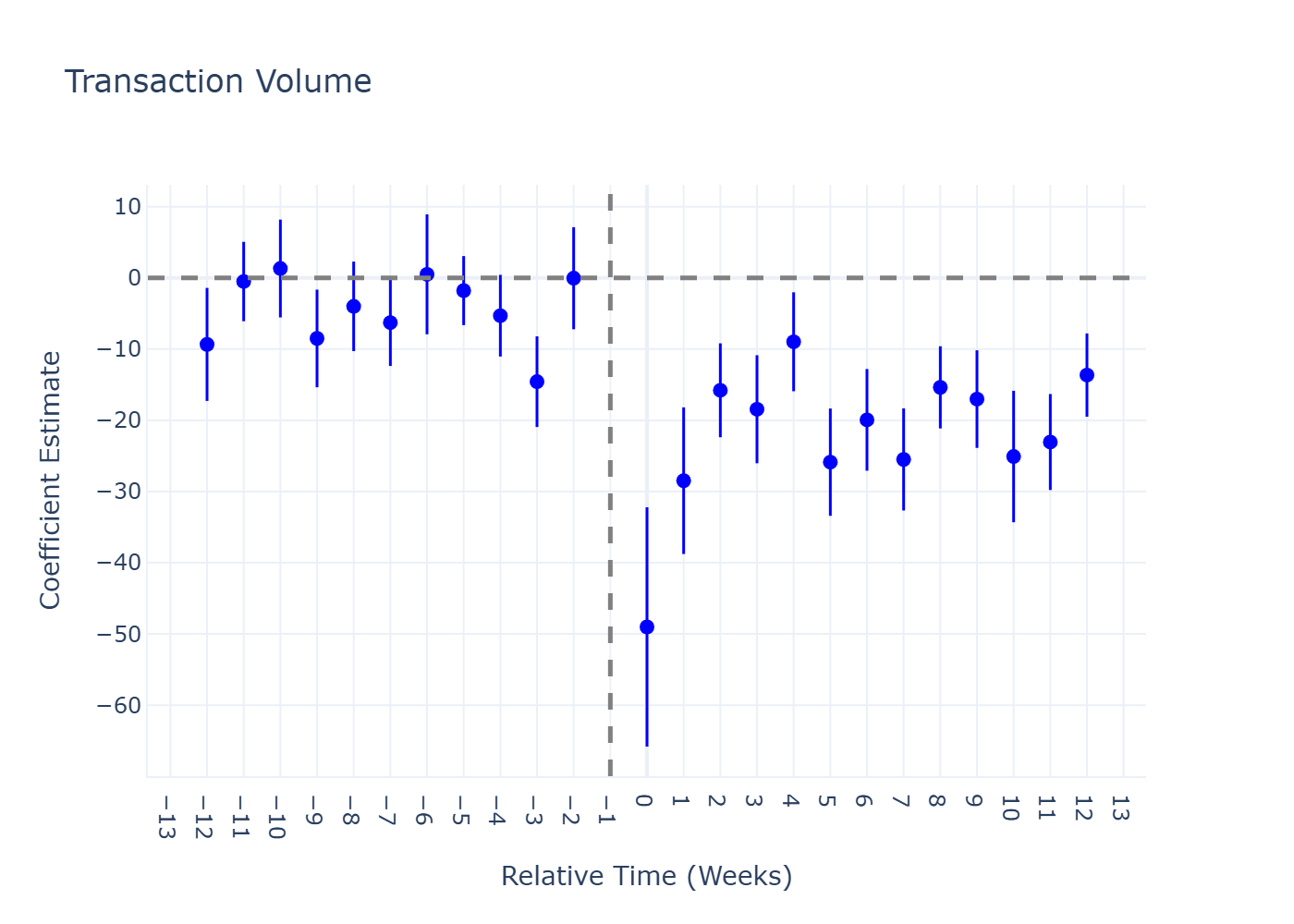}}\hfill
  \subfloat[Default: Weekly] {\includegraphics[width=.49\linewidth]{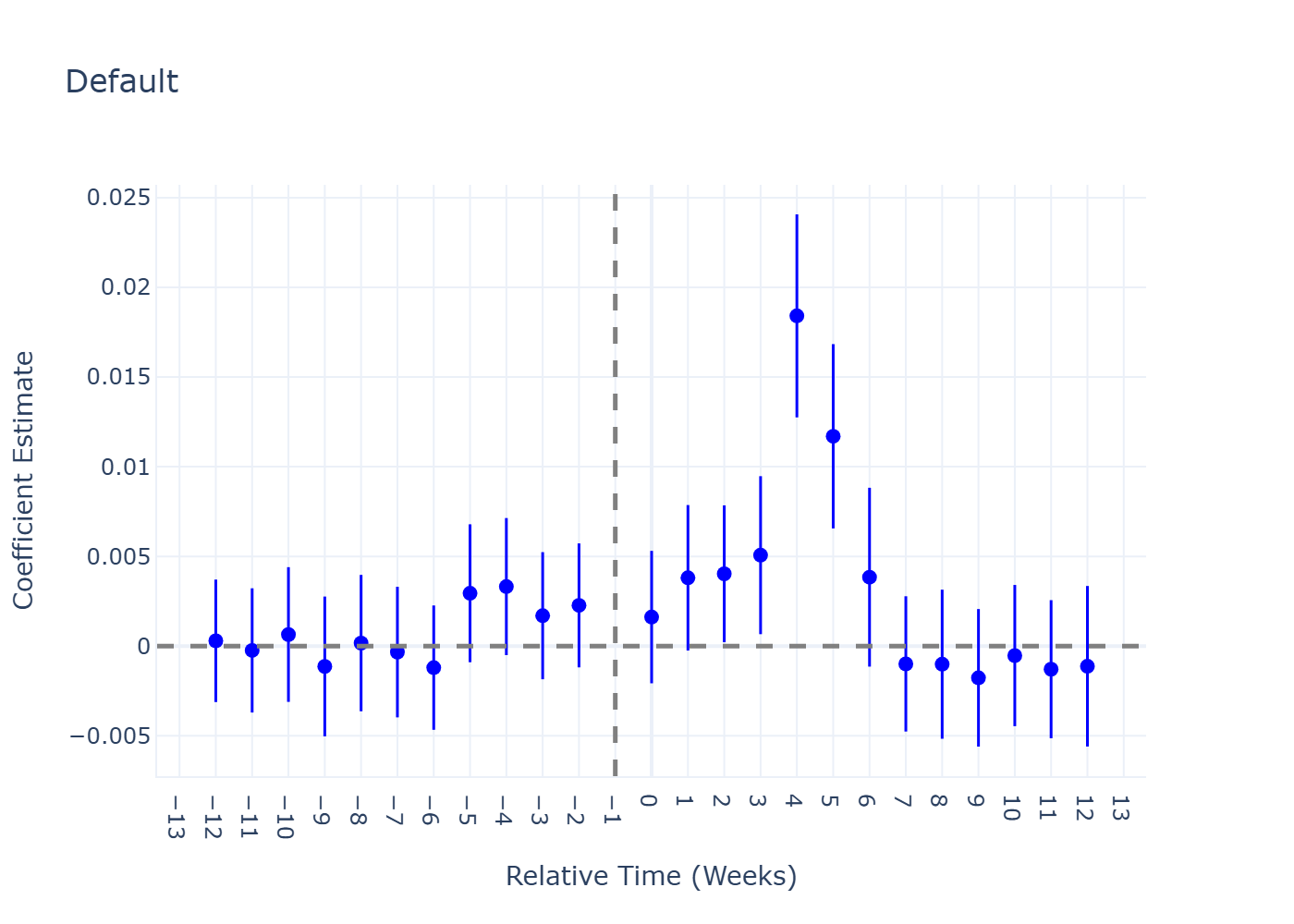}}
  \caption{The figure provides the estimated coefficients from an event study examining the impact of hurricane exposure on weekly payday loan-related outcomes at the ZIP code level with a treatment radius of 50 nautical miles. The analysis spans a window from five weeks before to twelve weeks after the hurricane event, with the week immediately preceding the event (week -1) serving as the reference period. Panel (a) reports estimates for weekly transaction volume, while Panel (b) shows estimates for weekly default rates. Vertical lines represent 95\% confidence intervals, calculated using standard errors clustered at the ZIP code level.}
 \label{fig: trans-default-50}
\end{figure}

\begin{table}[H]
\centering
\begin{tabular}{lccccc}
\hline
 & \multicolumn{5}{c}{\textbf{Treatment Radius}} \\
\cline{2-6}
 & 15 & 21 & 26 & 50 & 100 \\
\hline

\multicolumn{6}{l}{\textbf{Panel A: Transaction Volume}} \\

Treated$\times$Post 
& -4.951$^{*}$ & -4.355$^{*}$ & -5.123$^{**}$ & -17.661$^{***}$ & -25.146$^{***}$ \\
& [2.636] & [2.538] & [2.375] & [2.190] & [2.550] \\

\hline
\multicolumn{6}{l}{\textbf{Panel B: Default}} \\

Treated$\times$Post 
& -0.000 & -0.000 & -0.000 & 0.003$^{*}$ & 0.003$^{***}$ \\
& [0.001] & [0.001] & [0.001] & [0.001] & [0.001] \\

\hline
Event-specific ZIP Code FE & Y & Y & Y & Y & Y \\
Event-specific Time FE & Y & Y & Y & Y & Y \\

\multicolumn{6}{l}{\textbf{Observations}} \\
Panel A & 55,254 & 51,212 & 47,991 & 41,510 & 38,297 \\
Panel B & 50,855 & 47,124 & 44,496 & 38,335 & 35,204 \\

\hline
\end{tabular}
\caption{Difference-in-Differences estimates of post-event impacts on payday loan transaction volume and default outcomes across treatment radii. Panel A presents results for transaction volume, and Panel B for default outcomes. Standard errors clustered at the ZIP code level are reported in brackets. ***, **, and * denote statistical significance at the 1\%, 5\%, and 10\% levels, respectively.}
\label{tab:DiD-radius-payday}
\end{table}

\begin{table}[H]
\centering
\begin{tabular}{lccccc}
\hline
 & \multicolumn{5}{c}{\textbf{Treatment Radius}} \\
\cline{2-6}
 & 15 & 21 & 26 & 50 & 100 \\
\hline

\multicolumn{6}{l}{\textbf{Panel A: Transaction Volume}} \\

Control 
& 258.014 & 261.330 & 268.798 & 286.881 & 293.635 \\
Treated 
& 234.561 & 235.955 & 224.797 & 212.251 & 218.100 \\

\hline
\multicolumn{6}{l}{\textbf{Panel B: Default}} \\

Control 
& 0.035 & 0.034 & 0.033 & 0.030 & 0.027 \\
Treated 
& 0.046 & 0.046 & 0.045 & 0.045 & 0.046 \\

\hline
\end{tabular}
\caption{Baseline averages by group (pre-hurricane period) across treatment radii. Panel A reports transaction volume, and Panel B reports default rates for control and treated ZIP codes at radii of 15, 21, 26, 50, and 100 nautical miles.}
\label{tab:baseline-averages-payday}
\end{table}

\begin{figure}[H]
  \centering
  \subfloat[Transaction Volume: FEMA]{\includegraphics[width=.49\linewidth]{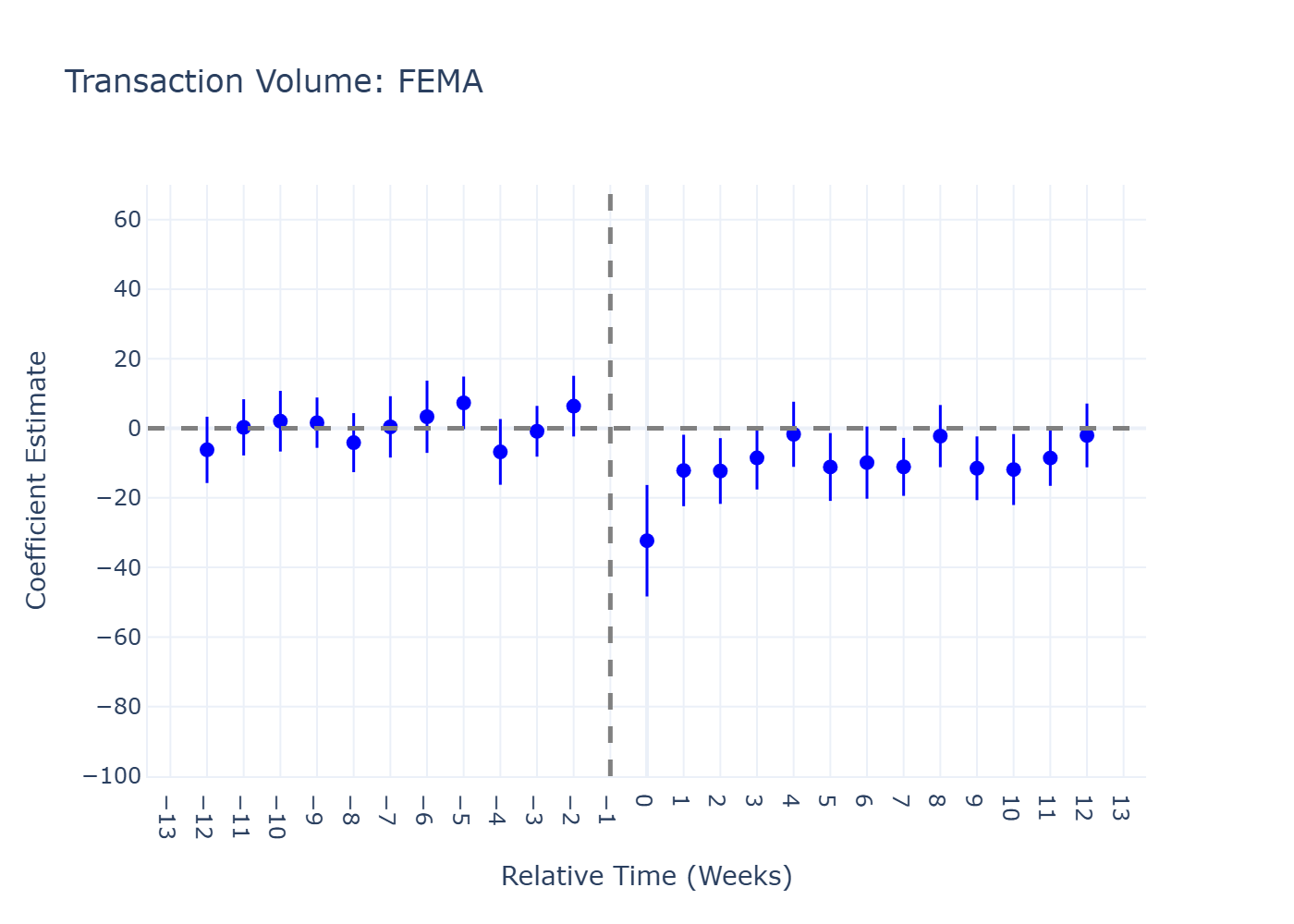}}\hfill
  \subfloat[Transaction Volume: Non-FEMA]{\includegraphics[width=.49\linewidth]{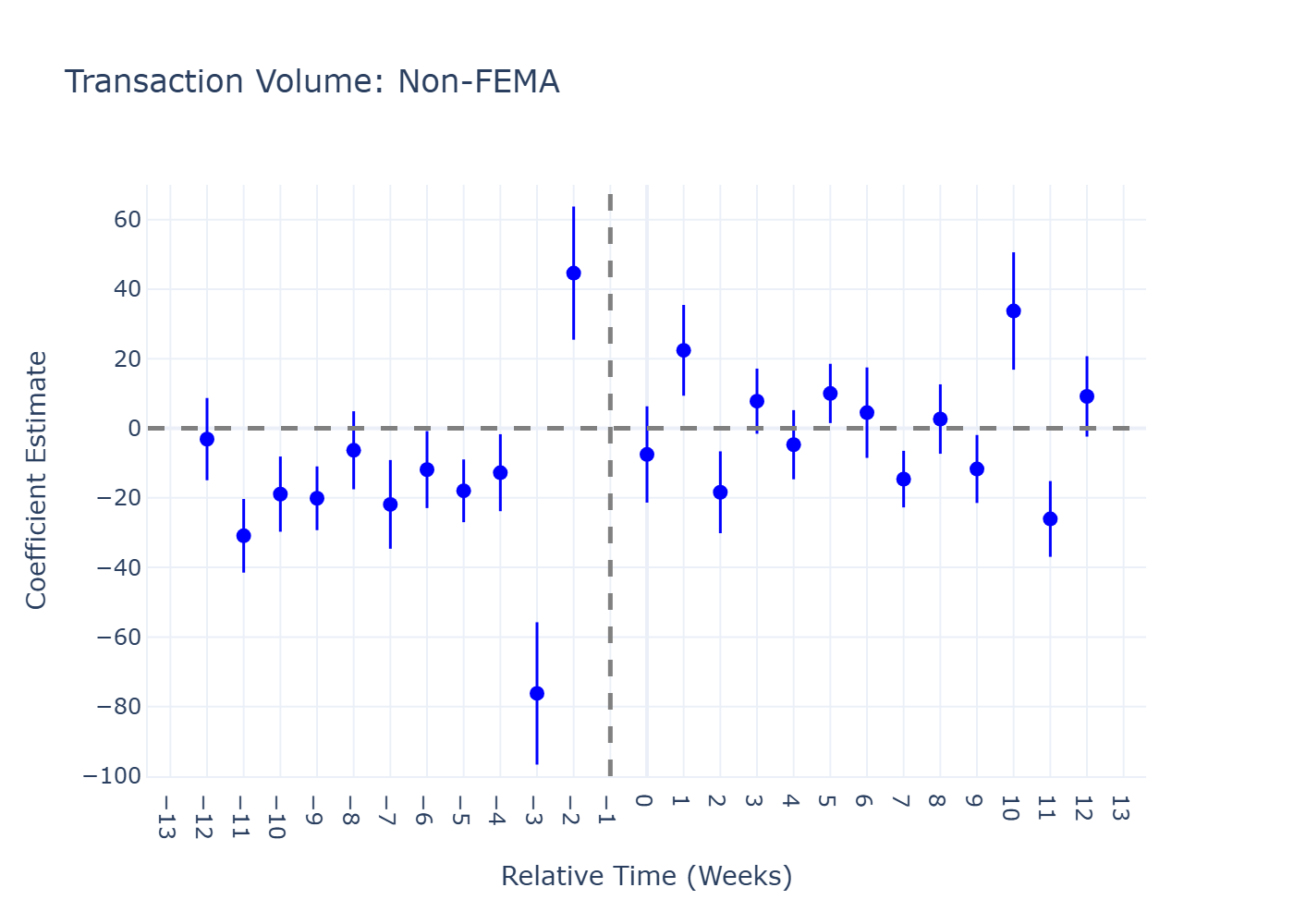}}
  \caption{Estimated coefficients from an event study model evaluating the impact of hurricane exposure on weekly payday loan transaction volume at the ZIP code level with a treatment radius of 21 nautical miles. The event window spans from twelve weeks before to twelve weeks after landfall, with the week immediately preceding the hurricane (week -1) serving as the reference period. The treated group is divided based on FEMA rental assistance availability. Standard errors are clustered at the ZIP code level.}
        \label{fig:transaction_fema_21}
\end{figure}

\begin{figure}[H]
  \centering
  \subfloat[Transaction Volume: FEMA]{\includegraphics[width=.49\linewidth]{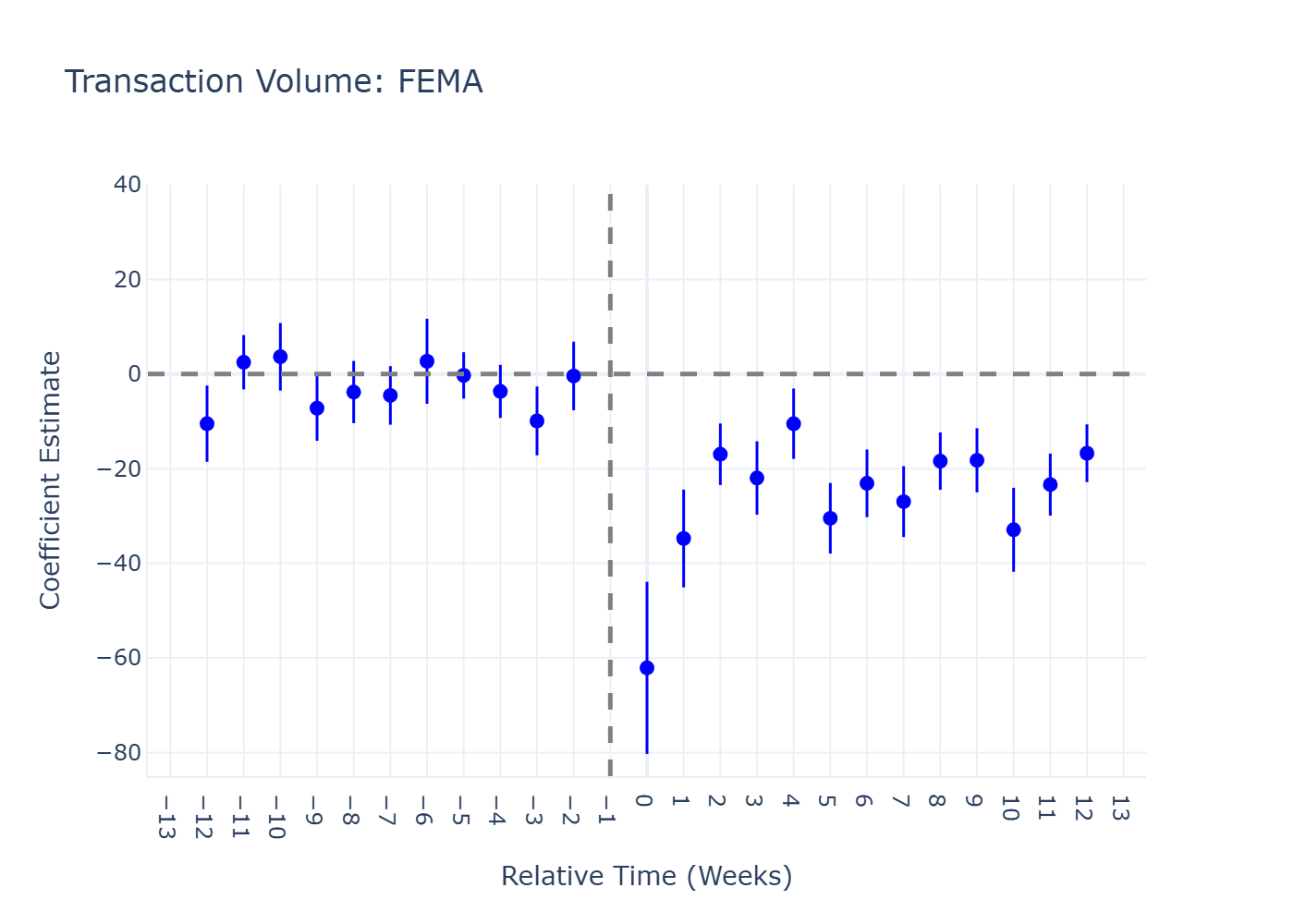}}\hfill
  \subfloat[Transaction Volume: Non-FEMA]{\includegraphics[width=.49\linewidth]{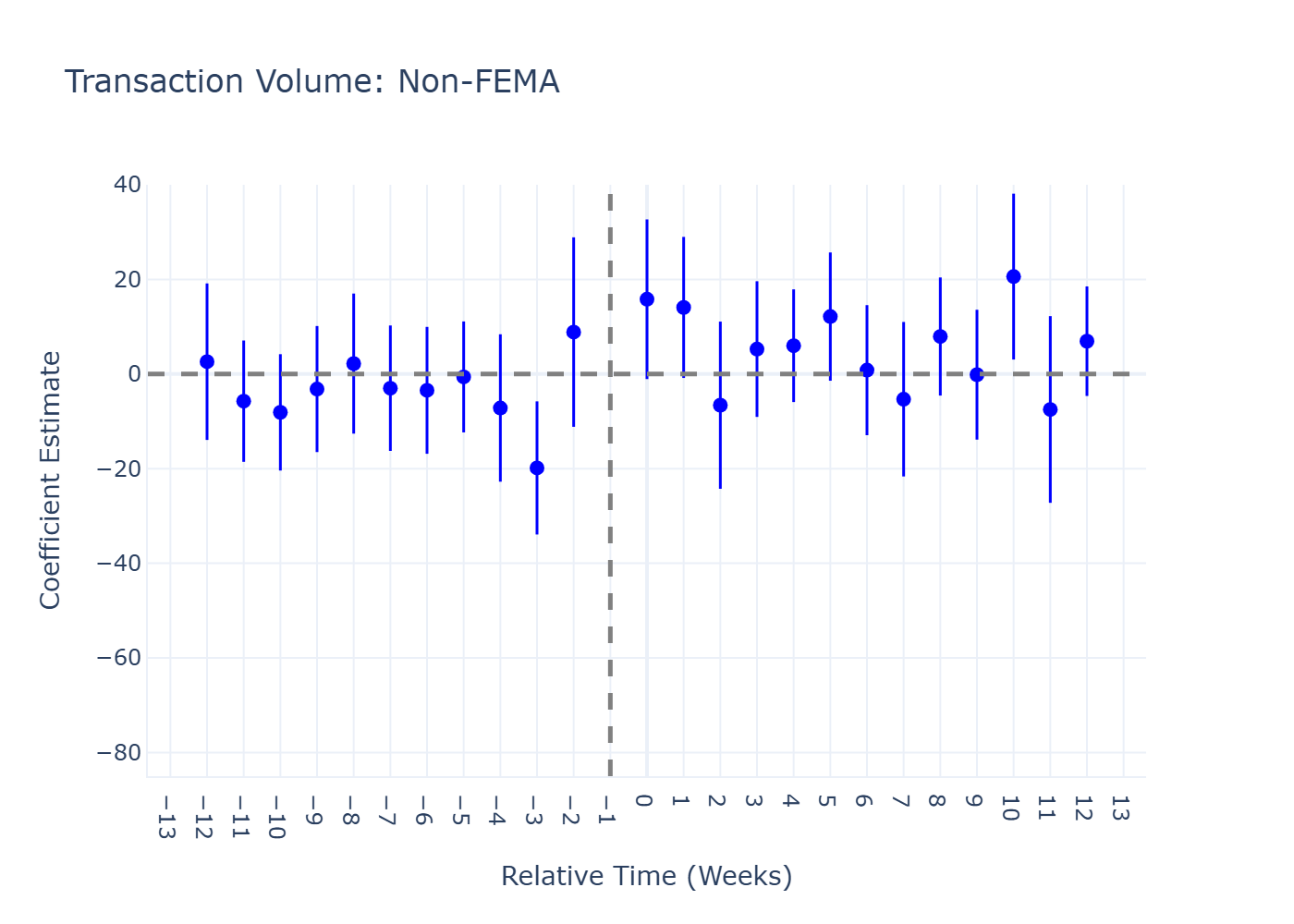}}
  \caption{Estimated coefficients from an event study model evaluating the impact of hurricane exposure on weekly payday loan transaction volume at the ZIP code level with a treatment radius of 50 nautical miles. The event window spans from twelve weeks before to twelve weeks after landfall, with the week immediately preceding the hurricane (week -1) serving as the reference period. The treated group is divided based on FEMA rental assistance availability. Standard errors are clustered at the ZIP code level.}
        \label{fig:transaction_fema_50}
\end{figure}

\begin{figure}[H]
  \centering
  \subfloat[Default: FEMA]{\includegraphics[width=.49\linewidth]{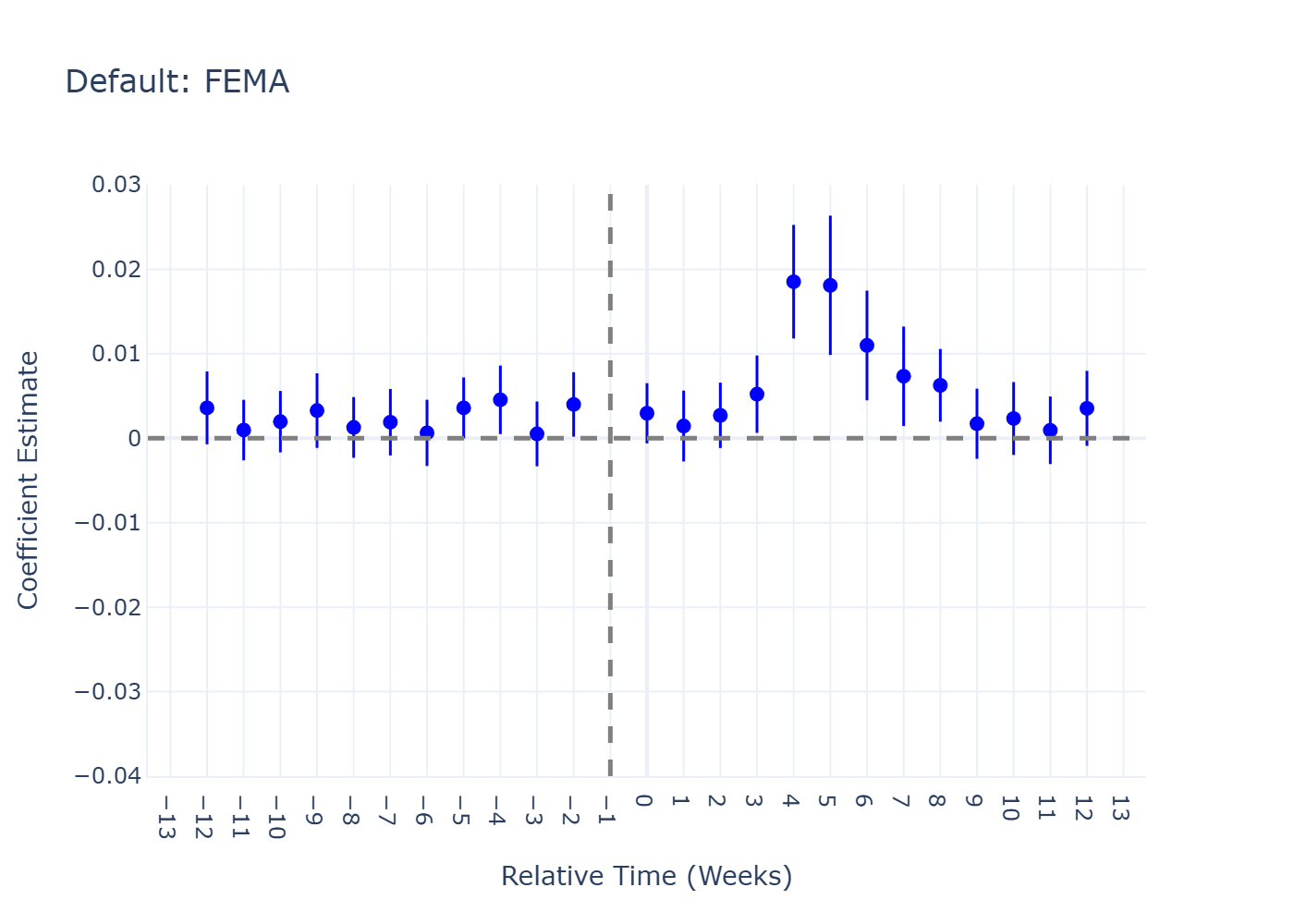}}\hfill
  \subfloat[Default: Non-FEMA]{\includegraphics[width=.49\linewidth]{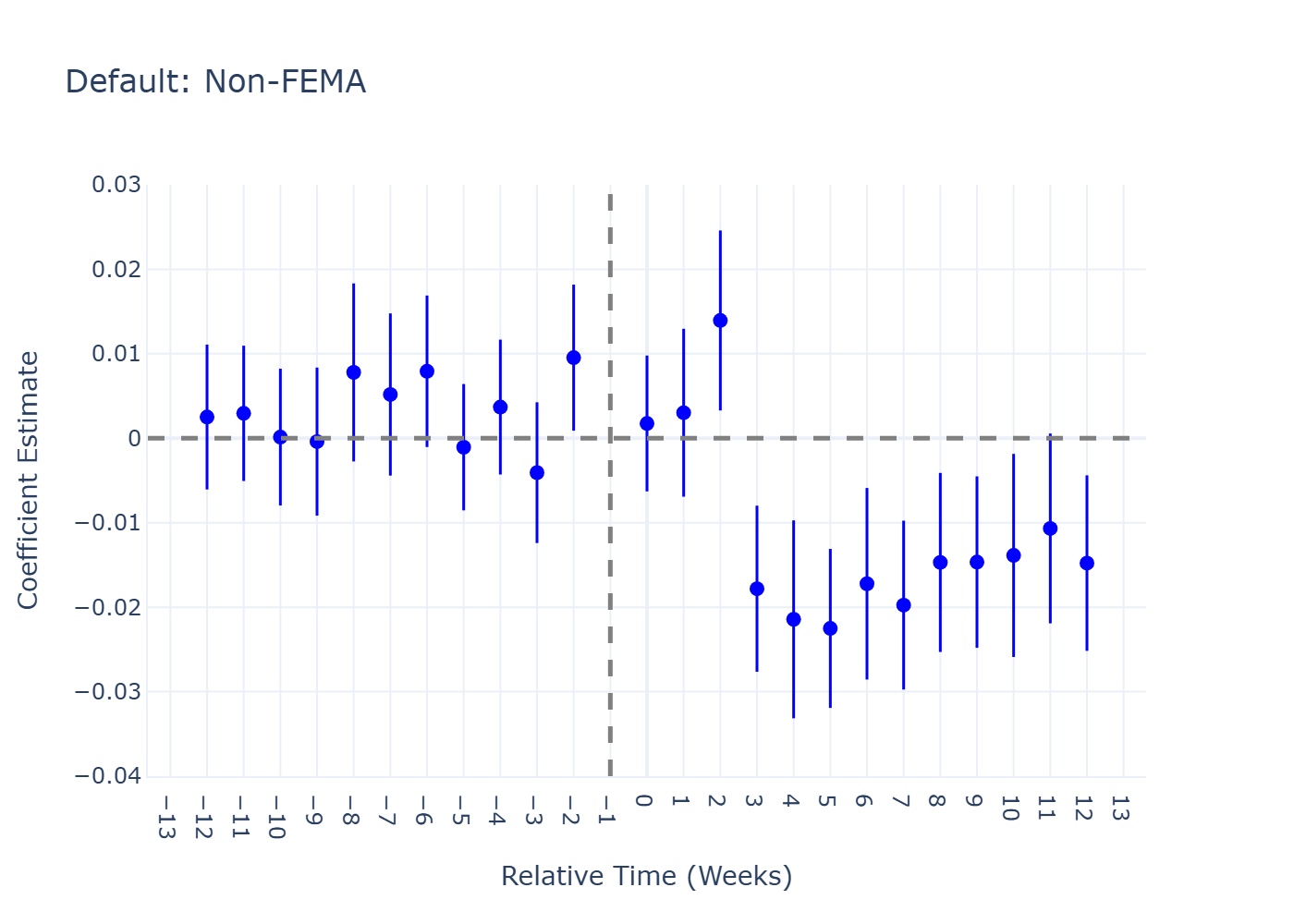}}
  \caption{Estimated coefficients from an event study model evaluating the impact of hurricane exposure on weekly payday loan default at the ZIP code level with a treatment radius of 21 nautical miles. The event window spans from twelve weeks before to twelve weeks after landfall, with the week immediately preceding the hurricane (week -1) serving as the reference period. The treated group is divided based on FEMA rental assistance availability. Standard errors are clustered at the ZIP code level.}
        \label{fig:default_fema_21}
\end{figure}

\begin{figure}[H]
  \centering
  \subfloat[Default: FEMA]{\includegraphics[width=.49\linewidth]{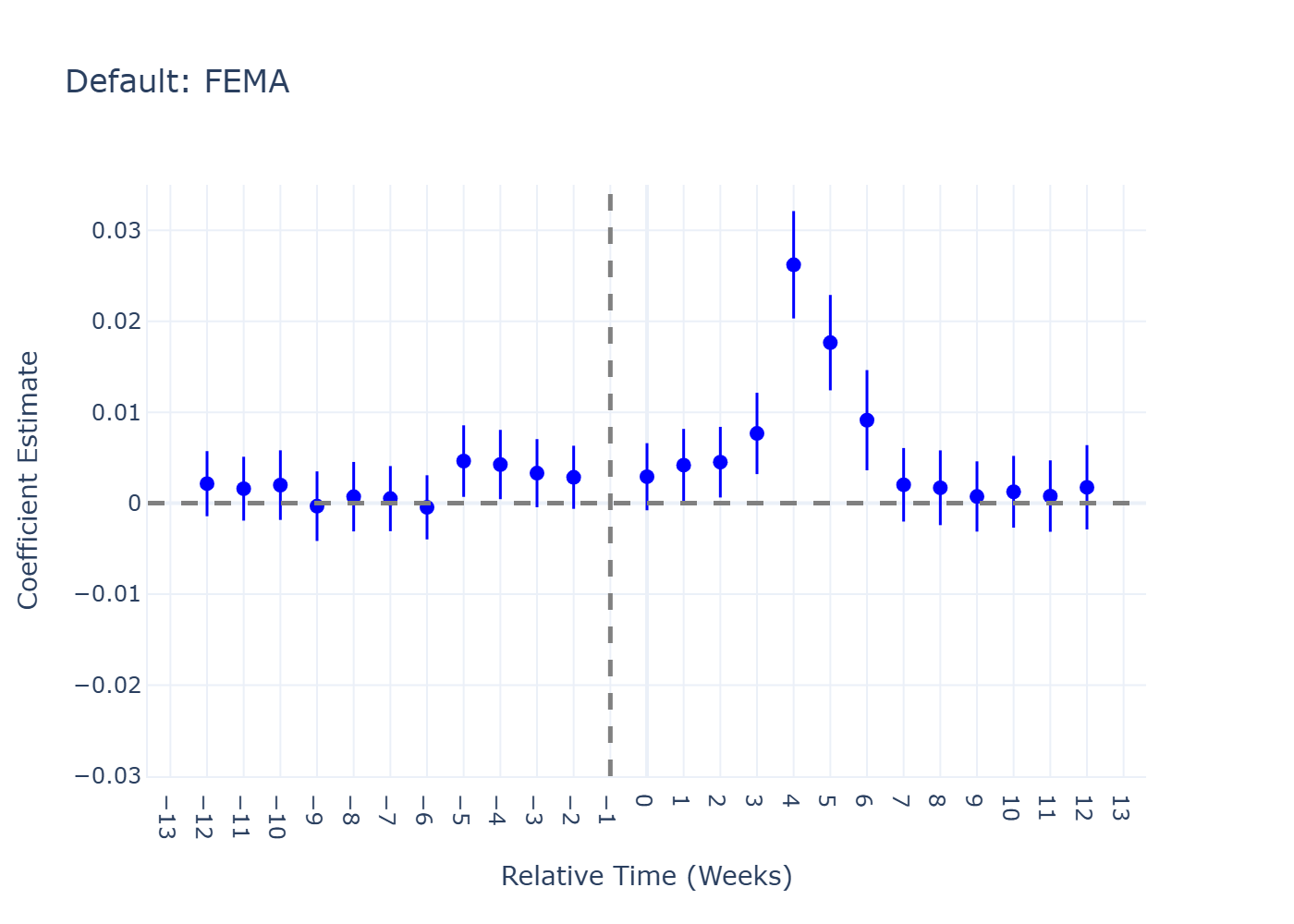}}\hfill
  \subfloat[Default: Non-FEMA]{\includegraphics[width=.49\linewidth]{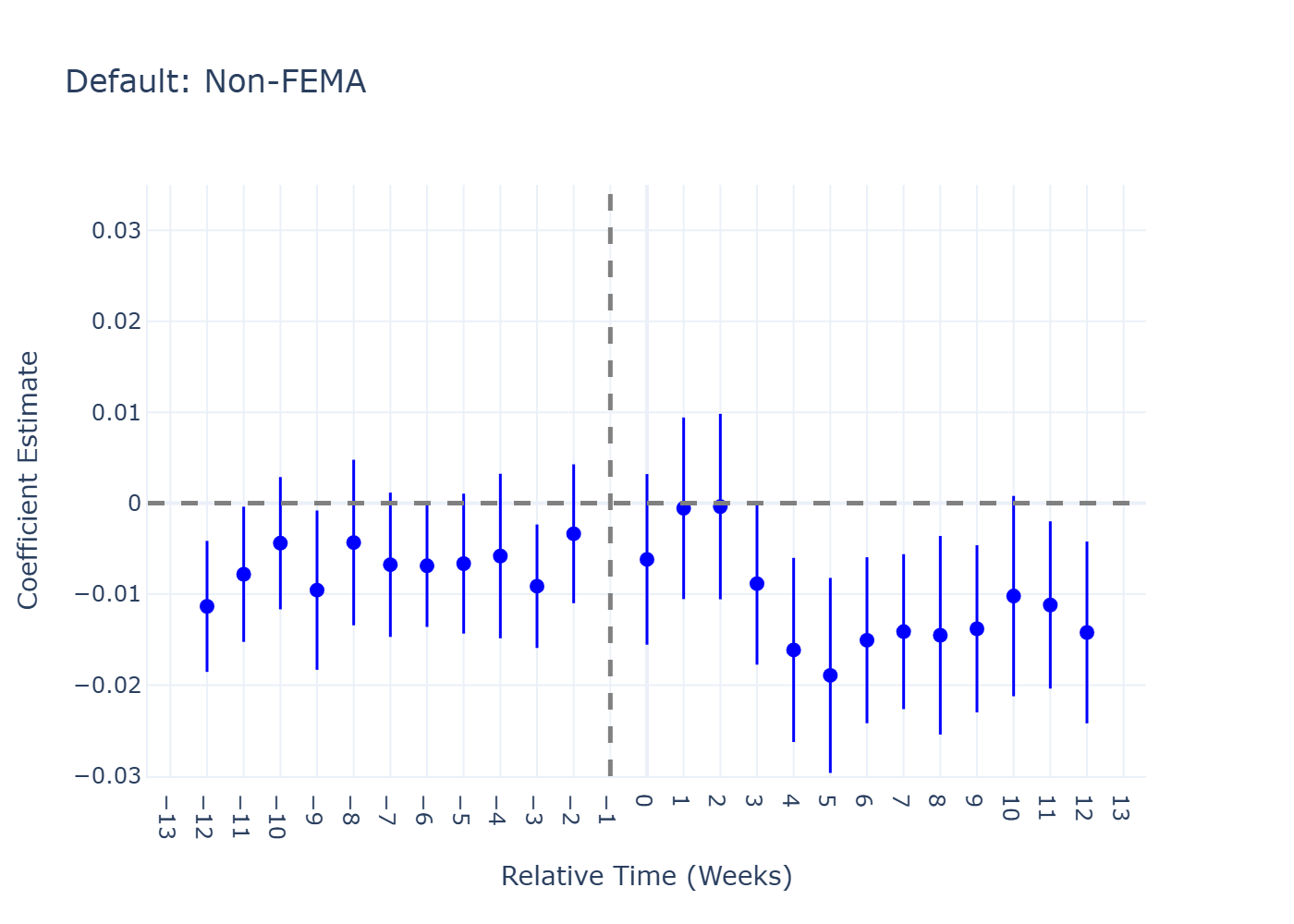}}
  \caption{Estimated coefficients from an event study model evaluating the impact of hurricane exposure on weekly payday loan default at the ZIP code level with a treatment radius of 50 nautical miles. The event window spans from twelve weeks before to twelve weeks after landfall, with the week immediately preceding the hurricane (week -1) serving as the reference period. The treated group is divided based on FEMA rental assistance availability. Standard errors are clustered at the ZIP code level.}
        \label{fig:default_fema_50}
\end{figure}

\begin{table}[H]
\centering
\begin{tabular}{lccccc}
\hline
 & \multicolumn{5}{c}{\textbf{Treatment Radius}} \\
\cline{2-6}
 & 15 & 21 & 26 & 50 & 100 \\
\hline

\multicolumn{6}{l}{\textbf{Panel A: Transaction Volume (OLS)}} \\

Treatment$\times$Post
 & 9.461$^{**}$ & 12.982$^{***}$ & 11.570$^{***}$ & 3.558 & -6.309$^{*}$ \\
 & [3.710] & [2.844] & [3.014] & [3.333] & [3.481] \\

Treatment$\times$Post$\times$FEMA
 & -16.917$^{***}$ & -18.002$^{***}$ & -15.741$^{***}$ & -11.661$^{***}$ & -11.558$^{**}$ \\
 & [5.878] & [4.731] & [4.121] & [4.204] & [4.898] \\

\hline
\multicolumn{6}{l}{\textbf{Panel B: Default (OLS)}} \\

Treatment$\times$Post
 & -0.010$^{*}$ & -0.004 & -0.010$^{***}$ & -0.005$^{**}$ & -0.005$^{***}$ \\
 & [0.005] & [0.003] & [0.002] & [0.002] & [0.002] \\

Treatment$\times$Post$\times$FEMA
 & 0.014$^{**}$ & 0.008$^{**}$ & 0.014$^{***}$ & 0.008$^{***}$ & 0.008$^{***}$ \\
 & [0.006] & [0.004] & [0.003] & [0.002] & [0.002] \\

\hline
\hline
\multicolumn{6}{l}{\textbf{Panel C: Transaction Volume (IV)}} \\

Treatment$\times$Post
 & 7.105$^{**}$ & 11.911$^{***}$ & 10.584$^{***}$ & 4.387 & -4.600 \\
 & [3.342] & [2.883] & [3.268] & [3.626] & [3.049] \\

Treatment$\times$Post$\times$FEMA
 & -12.373$^{**}$ & -15.408$^{***}$ & -13.155$^{***}$ & -10.016$^{*}$ & -16.928$^{***}$ \\
 & [5.745] & [4.939] & [4.542] & [5.158] & [6.208] \\

\hline
\multicolumn{6}{l}{\textbf{Panel D: Default (IV)}} \\

Treatment$\times$Post
 & -0.012$^{**}$ & -0.005 & -0.011$^{***}$ & -0.005$^{**}$ & -0.003$^{**}$ \\
 & [0.006] & [0.004] & [0.003] & [0.002] & [0.001] \\

Treatment$\times$Post$\times$FEMA
 & 0.015$^{**}$ & 0.008$^{**}$ & 0.015$^{***}$ & 0.007$^{***}$ & 0.006$^{***}$ \\
 & [0.006] & [0.004] & [0.003] & [0.003] & [0.002] \\

\hline
Event-specific ZIP Code FE & Y & Y & Y & Y & Y \\
Event-specific Time FE & Y & Y & Y & Y & Y \\

\multicolumn{6}{l}{\textbf{Observations}} \\
Panels A \& C & 55,254 & 51,212 & 47,991 & 41,510 & 38,297 \\
Panels B \& D & 50,855 & 47,124 & 44,496 & 38,335 & 35,204 \\

\hline
\end{tabular}
\caption{Triple-differences estimates of the impact of storm exposure and FEMA assistance across alternative treatment radii. Panels A and B report OLS estimates for transaction volume and default, respectively, while Panels C and D report corresponding IV estimates. All specifications include event-specific ZIP code and time fixed effects. Standard errors clustered at the ZIP code level are reported in brackets. ***, **, and * denote statistical significance at the 1\%, 5\%, and 10\% levels, respectively.}
\label{tab:triple-did-payday}
\end{table}

\begin{table}[H]
\centering
\begin{tabular}{lcc}
\hline
 & \multicolumn{2}{c}{\textbf{Treatment Radius}} \\
\cline{2-3}
 & 26 & 50 \\
\hline
Treated$\times$Post 
& -0.008$^{***}$ & -0.006$^{**}$ \\
& [0.003] & [0.003] \\
\hline
Event-specific ZIP Code FE & Y & Y \\
Event-specific Time FE & Y & Y \\
Observations 
& 47,886 & 41,389 \\
\hline
\end{tabular}
\caption{Difference-in-differences estimates of post-event impacts on the share of new customers across treatment radii. The outcome variable is the share of new customers, defined as the proportion of loans in a ZIP code and week that are taken by first-time borrowers—i.e., individuals with no prior payday loan activity observed within the one-year event window at the time of the loan. Standard errors clustered at the ZIP code level are reported in brackets. ***, **, and * denote statistical significance at the 1\%, 5\%, and 10\% levels, respectively.}
\label{tab:DiD-new-costumers}
\end{table}

\begin{table}[H]
\centering
\begin{tabular}{lcc}
\hline
 & \multicolumn{2}{c}{\textbf{Treatment Radius}} \\
\cline{2-3}
 & 26 & 50 \\
\hline
\multicolumn{3}{l}{\textbf{Panel A: New Customers Share (OLS)}} \\
Treatment$\times$Post
 & 0.001 & -0.001 \\
 & [0.004] & [0.004] \\
Treatment$\times$Post$\times$FEMA
 & -0.010$^{**}$ & -0.004 \\
 & [0.005] & [0.005] \\
\hline\hline
\multicolumn{3}{l}{\textbf{Panel B: New Customers Share (IV)}} \\
Treatment$\times$Post
 & 0.000 & 0.001 \\
 & [0.004] & [0.005] \\
Treatment$\times$Post$\times$FEMA
 & -0.008 & -0.006 \\
 & [0.005] & [0.007] \\
\hline
Event-specific ZIP Code FE & Y & Y \\
Event-specific Time FE & Y & Y \\
Observations
 & 47,886 & 41,389 \\
\hline
\end{tabular}
\caption{Triple-differences estimates of the impact of storm exposure and FEMA assistance on the share of new customers across alternative treatment radii. Panel A reports OLS estimates, while Panel B reports instrumental-variables estimates where FEMA assistance is instrumented. All specifications include event-specific ZIP code and time fixed effects. Standard errors clustered at the ZIP code level are reported in brackets. ***, **, and * denote statistical significance at the 1\%, 5\%, and 10\% levels, respectively.}
\label{tab:triple-did-new-costumers}
\end{table}

\begin{table}[H]
\centering
\caption{Baseline averages by group (pre-hurricane period)}
\label{tab:baseline-new-costumers}
\begin{tabular}{lcc}
\hline
& \multicolumn{2}{c}{\textbf{Treatment Radius}} \\
\cline{2-3}
& 26 & 50 \\
\hline
\multicolumn{3}{l}{\textbf{New Customers Share}} \\

Control
& 0.103649 & 0.105360 \\
Treated
& 0.112572 & 0.109821 \\
\hline
\end{tabular}
\end{table}

\begin{figure}[H]
  \centering
  \subfloat[New Customers Share: FEMA]{\includegraphics[width=.49\linewidth]{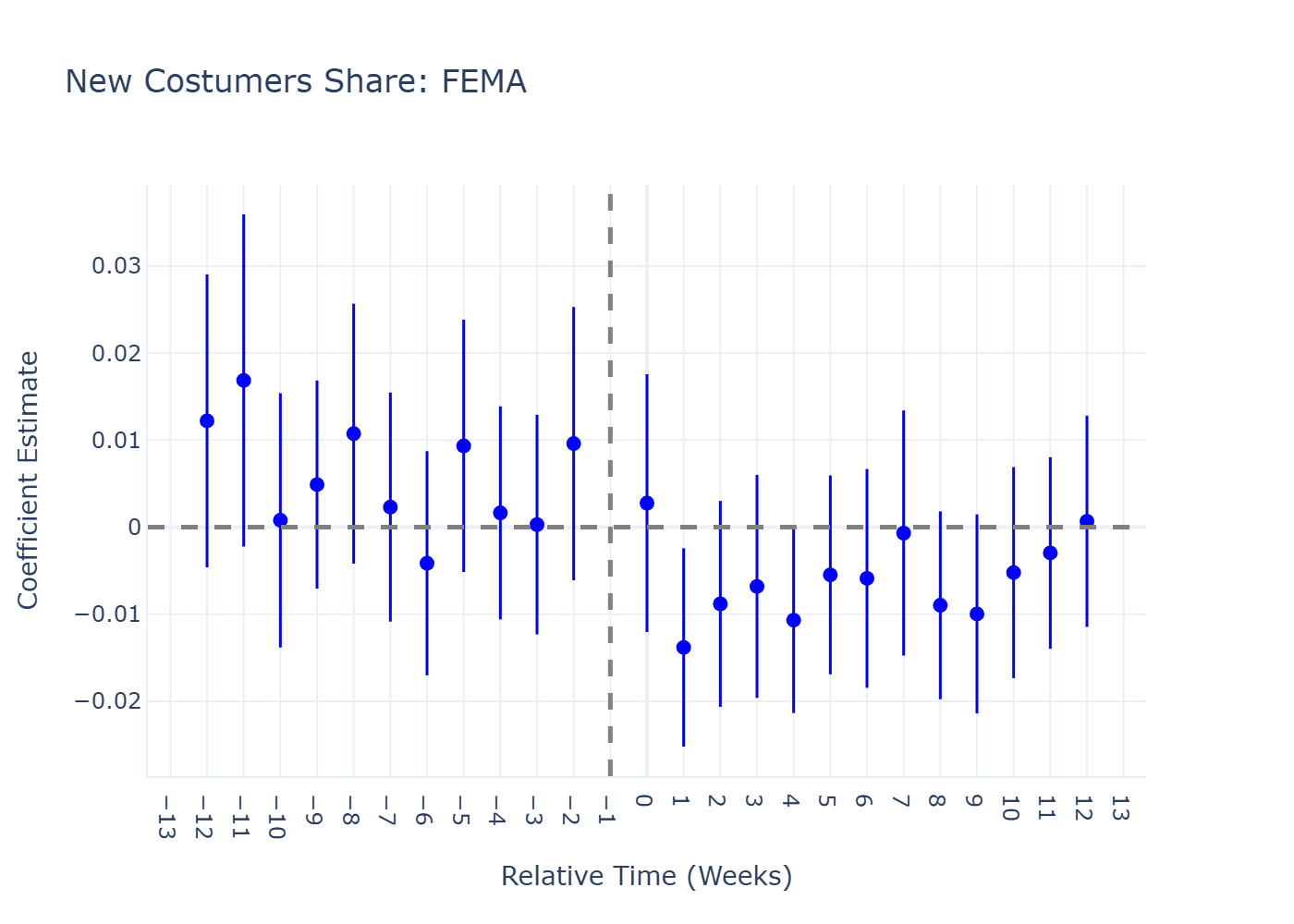}}\hfill
  \subfloat[New Customers Share: Non-FEMA]{\includegraphics[width=.49\linewidth]{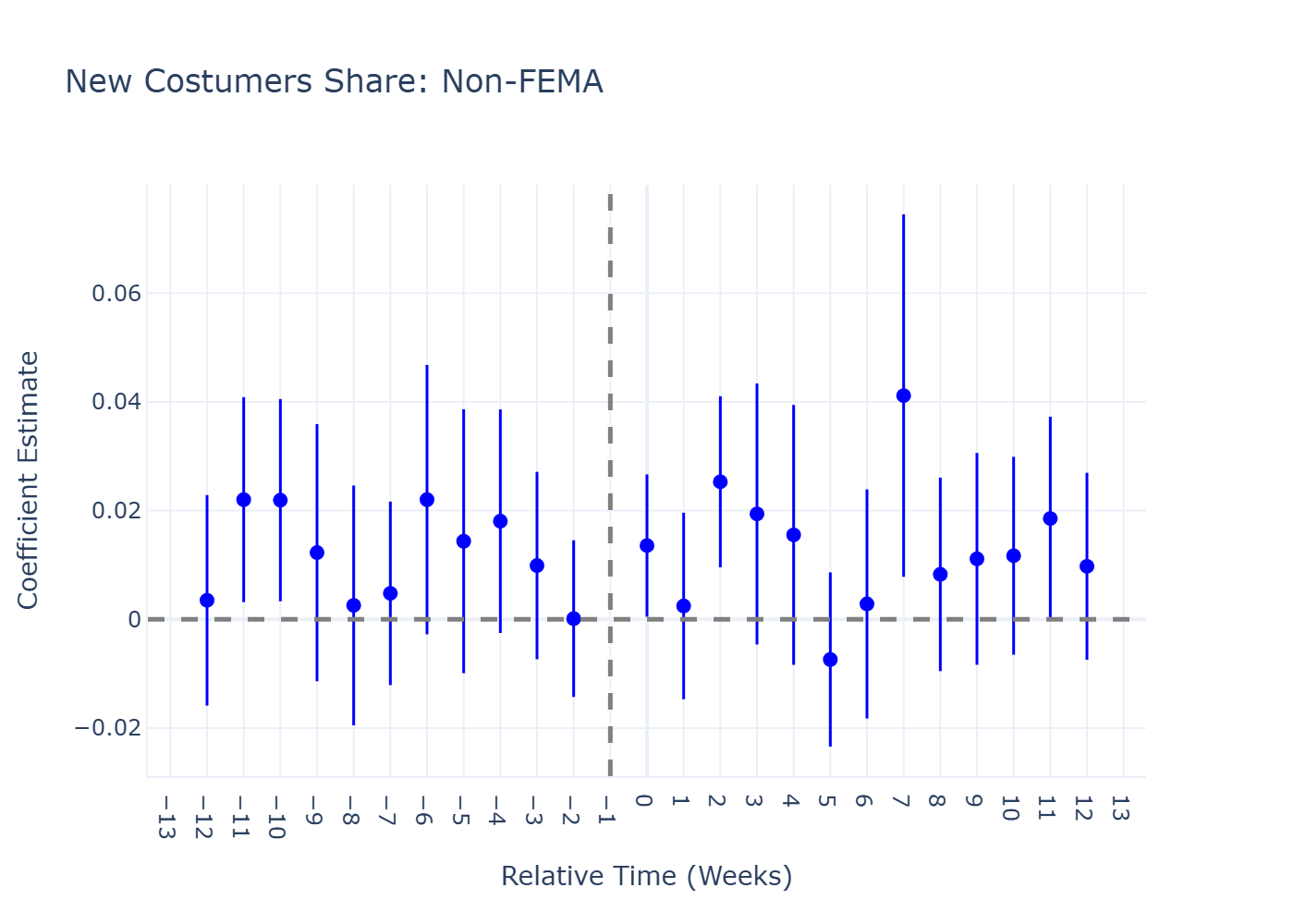}}
  \caption{Estimated coefficients from an event study model evaluating the impact of hurricane exposure on the weekly share of new customers at the ZIP code level, using a treatment radius of 26 nautical miles. The event window spans from 12 weeks before to 12 weeks after landfall, with the week immediately preceding the hurricane (week -1) serving as the reference period. The treated group is divided based on FEMA rental assistance availability. Standard errors are clustered at the ZIP code level.}
        \label{fig:event-study-newcostumer-26}
\end{figure}

\begin{figure}[H]
  \centering
  \subfloat[New Customers Share: FEMA]{\includegraphics[width=.49\linewidth]{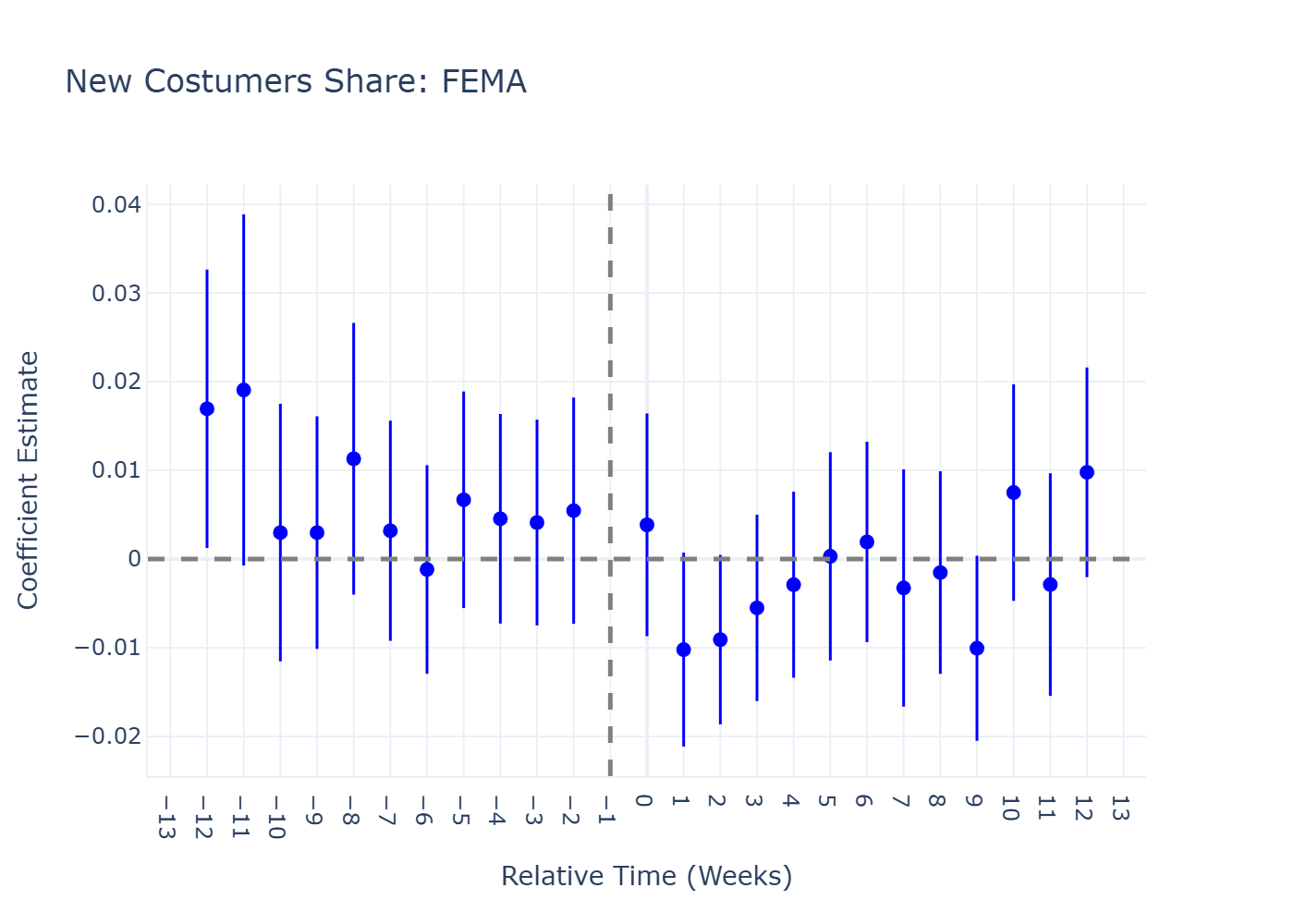}}\hfill
  \subfloat[New Customers Share: Non-FEMA]{\includegraphics[width=.49\linewidth]{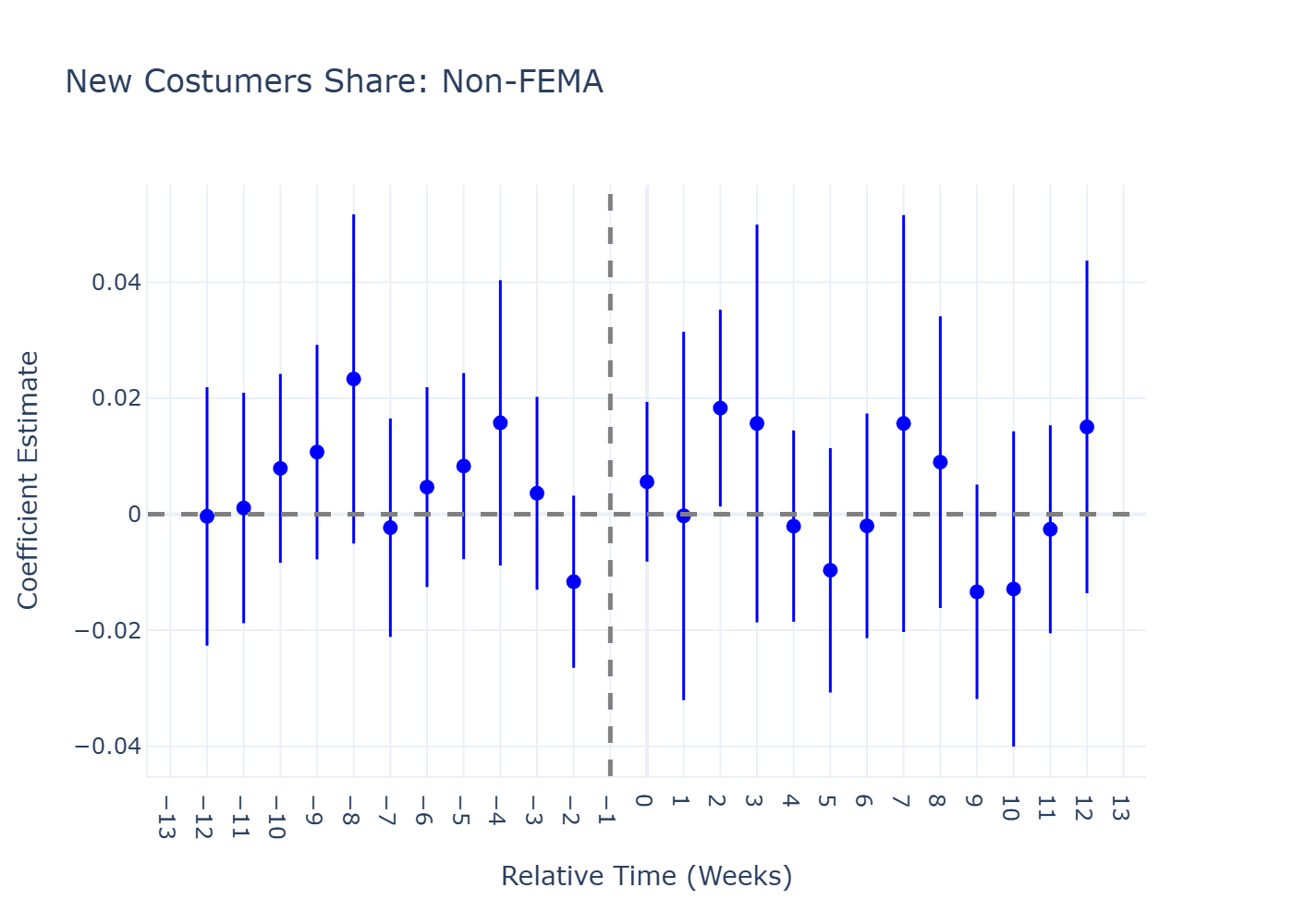}}
  \caption{Estimated coefficients from an event study model evaluating the impact of hurricane exposure on the weekly share of new customers at the ZIP code level, using a treatment radius of 50 nautical miles. The event window spans from 12 weeks before to 12 weeks after landfall, with the week immediately preceding the hurricane (week -1) serving as the reference period. The treated group is divided based on FEMA rental assistance availability. Standard errors are clustered at the ZIP code level.}
        \label{fig:event-study-newcostumer-50}
\end{figure}

\subsubsection{Multiple and Tropical Storms}

Adding lower intensity tropical storms for payday loans show a dampened but statistically significant effect for transaction volumes and defaults, unlike the case with evictions where the effect was non-significant. In this case, the results remain significant but the magnitudes decrease with treatment radii, which is in line with the a decrease in potential damage from the storm and the absence of FEMA intervention that tropical storms represent.

\begin{figure}[H]
  \centering
  \subfloat[Transaction Volume: 21 nautical miles]{\includegraphics[width=.49\linewidth]{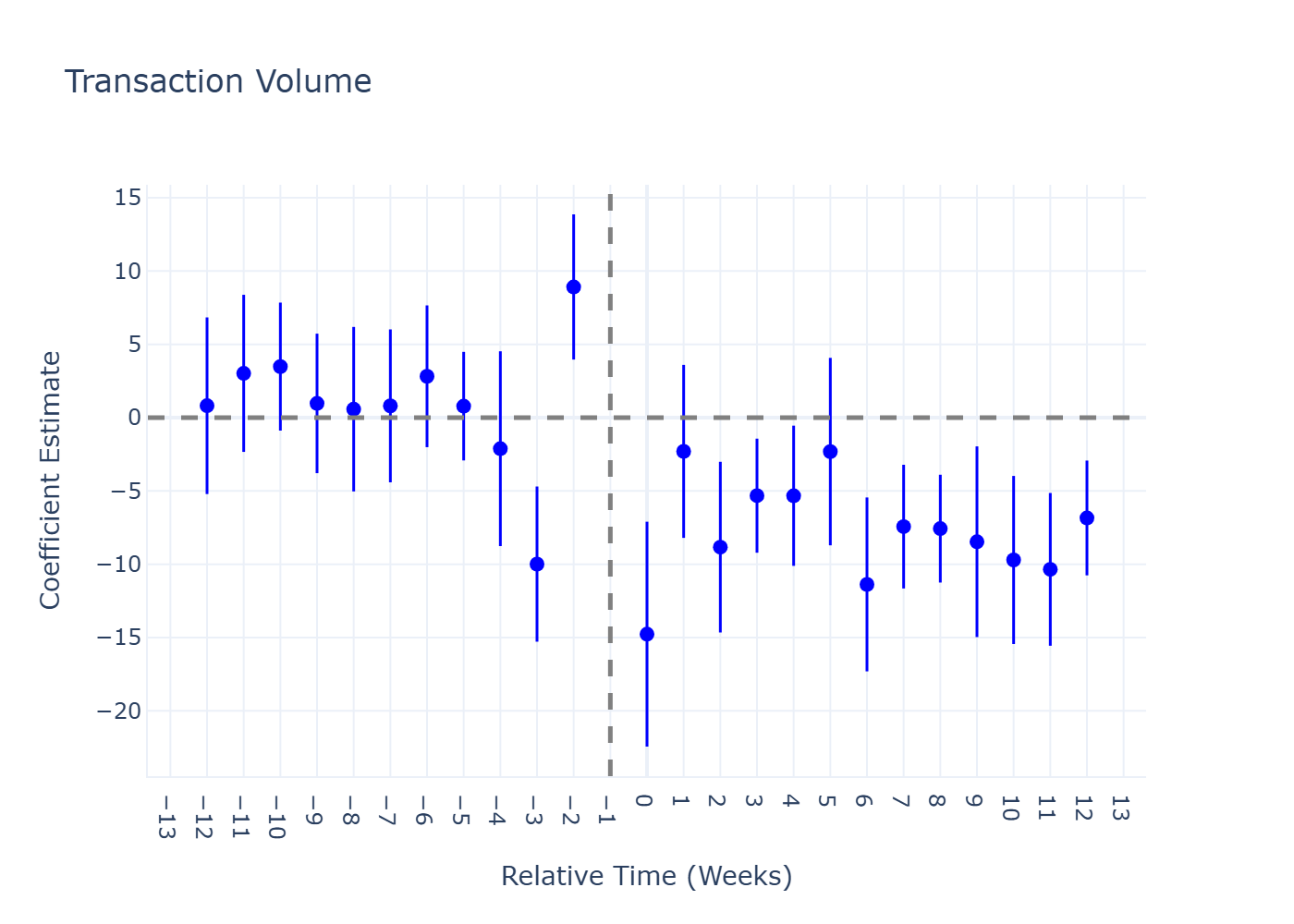}}\hfill
  \subfloat[Default: 21 nautical miles]{\includegraphics[width=.49\linewidth]{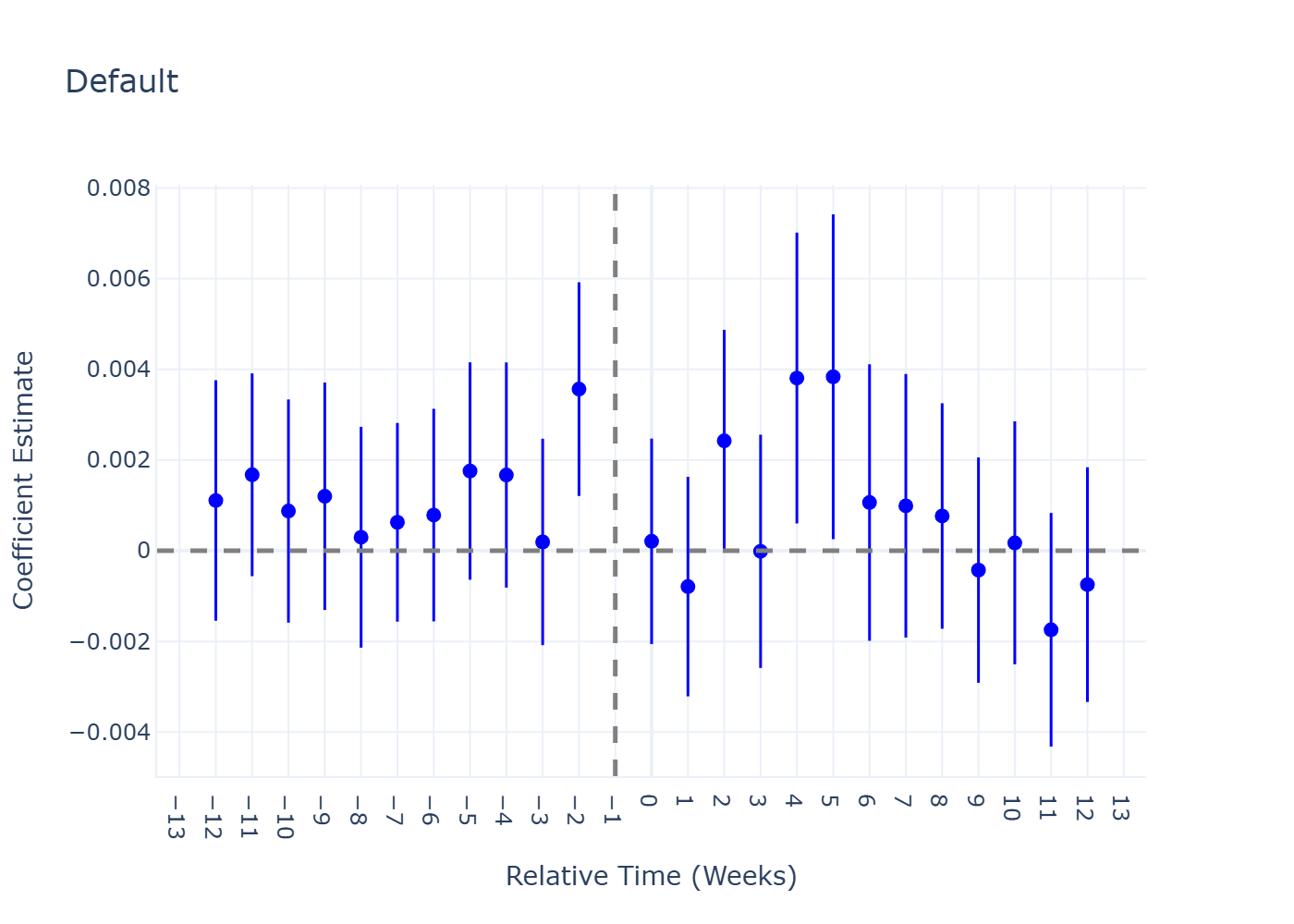}}
  \caption{The figure provides the estimated coefficients from an event study examining the impact of combined hurricane and tropical storm exposure on weekly payday loan-related outcomes at the ZIP code level. The analysis spans a window from five weeks before to twelve weeks after the hurricane event, with the week immediately preceding the event (week -1) serving as the reference period. Panel (a) reports estimates for weekly transaction volume, while Panel (b) shows estimates for weekly default rates at 21 nautical miles. Panel (c) reports estimates for weekly transaction volume, while Panel (d) shows estimates for weekly default rates at 50 nautical miles. Vertical lines represent 95\% confidence intervals, calculated using standard errors clustered at the ZIP code level.}
        \label{fig:transaction-default-allstorms} 
\end{figure}

\autoref{fig:transaction-default-allstorms} and \autoref{tab:DiD-radius-payday-allstorm} show that overall transaction volumes decrease by a maximal amount of 10 loans per ZIP code, a 3.5\% decrease relative to the pre-period average and that defaults show no statistically significant change for tropical storms. Interestingly, we also see that there is a greater bifurcation between hurricane and tropical storm events for larger radii, where transaction volumes decrease by a larger amount for hurricanes compared to tropical storms. A result, which is in line with the previous observation that greater intensity storms have a larger distance from the storm path where effects are observable, and in particular, the larger distance also interacts with a greater proportion of ZIP codes that become ineligible for FEMA assistance.

Similarly, we find a statistically significant reversion of the baseline observed effect of decreasing transaction volumes and an overall increase in defaults for ZIP codes that see multiple storms within a calendar year in \autoref{tab:multi-interaction-radius-payday}. This result in particular holds for larger treatment radii for transaction volumes. Suggesting that the increase in cumulative damage may undo the decrease in reliance on high cost credit that FEMA eligibility provides. Once again aligning with previous results via the channel that an increasing treatment radii correlates with FEMA ineligibility.

\begin{table}[H]
\centering
\begin{tabular}{lccccc}
\hline
 & \multicolumn{5}{c}{\textbf{Treatment Radius}} \\
\cline{2-6}
 & 15 & 21 & 26 & 50 & 100 \\
\hline

\multicolumn{6}{l}{\textbf{Hurricanes and Tropical Storms}} \\

\multicolumn{6}{l}{\textbf{Panel A: Transaction Volume}} \\
Treated$\times$Post 
& -8.731$^{***}$ & -8.560$^{***}$ & -8.264$^{***}$ & -10.059$^{***}$ & -8.951$^{***}$ \\
& [1.555] & [1.445] & [1.308] & [1.132] & [1.508] \\

\multicolumn{6}{l}{\textbf{Panel B: Default}} \\
Treated$\times$Post 
& -0.000 & -0.000 & -0.000 & 0.000 & 0.000 \\
& [0.001] & [0.001] & [0.001] & [0.000] & [0.000] \\

\hline
\multicolumn{6}{l}{\textbf{Separate Effects: Hurricane and Tropical Storms}} \\

\multicolumn{6}{l}{\textbf{Panel A: Transaction Volume}} \\
Post$\times$Hurricane 
& -4.951$^{*}$ & -4.355$^{*}$ & -5.123$^{**}$ & -17.661$^{***}$ & -25.146$^{***}$ \\
& [2.636] & [2.538] & [2.375] & [2.190] & [2.550] \\

Post$\times$Tropical Storm 
& -12.202$^{***}$ & -11.822$^{***}$ & -10.396$^{***}$ & -7.551$^{***}$ & -5.896$^{***}$ \\
& [1.999] & [1.784] & [1.656] & [1.401] & [1.591] \\

\multicolumn{6}{l}{\textbf{Panel B: Default}} \\
Post$\times$Hurricane 
& -0.000 & -0.000 & -0.000 & 0.003$^{*}$ & 0.003$^{***}$ \\
& [0.001] & [0.001] & [0.001] & [0.001] & [0.001] \\

Post$\times$Tropical Storm 
& -0.001$^{**}$ & -0.000 & -0.000 & -0.000 & -0.000 \\
& [0.000] & [0.000] & [0.000] & [0.000] & [0.000] \\

\hline
Event-specific ZIP Code FE & Y & Y & Y & Y & Y \\
Event-specific Time FE & Y & Y & Y & Y & Y \\

\multicolumn{6}{l}{\textbf{Observations}} \\
Panels A & 165,659 & 171,106 & 167,319 & 159,961 & 150,169 \\
Panels B & 154,134 & 159,124 & 155,997 & 146,450 & 134,922 \\

\hline
\end{tabular}
\caption{Difference-in-Differences estimates of payday loan outcomes following hurricanes and tropical storms. Panel A reports transaction volume, and Panel B reports default outcomes. The top section reports combined treatment effects, while the bottom section separates effects by hurricane and tropical storm exposure. Standard errors clustered at the ZIP code level are reported in brackets. ***, **, and * denote statistical significance at the 1\%, 5\%, and 10\% levels, respectively.}
\label{tab:DiD-radius-payday-allstorm}
\end{table}

\begin{table}[H]
\centering
\begin{tabular}{lccccc}
\hline
 & \multicolumn{5}{c}{\textbf{Treatment Radius}} \\
\cline{2-6}
 & 15 & 21 & 26 & 50 & 100 \\
\hline

\multicolumn{6}{l}{\textbf{Panel A: Transaction Volume}} \\
Treatment$\times$Post
 & -9.158$^{**}$ & -4.286 & -3.503 & -12.665$^{***}$ & -16.602$^{***}$ \\
 & [4.237] & [3.304] & [2.541] & [1.770] & [1.670] \\

Treatment$\times$Post$\times$Multiple
 & -2.851 & -4.635 & -5.049 & 9.906$^{***}$ & 9.889$^{***}$ \\
 & [7.655] & [5.039] & [4.099] & [2.539] & [2.732] \\

\hline
\multicolumn{6}{l}{\textbf{Panel B: Default}} \\

Treatment$\times$Post
 & 0.002 & 0.001 & -0.000 & 0.000 & -0.000 \\
 & [0.002] & [0.001] & [0.001] & [0.001] & [0.001] \\

Treatment$\times$Post$\times$Multiple
 & 0.003 & 0.005$^{**}$ & 0.005$^{**}$ & 0.001 & 0.001 \\
 & [0.003] & [0.002] & [0.002] & [0.001] & [0.001] \\

\hline
Event-specific ZIP Code FE & Y & Y & Y & Y & Y \\
Event-specific Time FE & Y & Y & Y & Y & Y \\
\multicolumn{6}{l}{\textbf{Observations}} \\
Panle A & 76,022 & 76,026 & 76,022 & 76,020 & 76,018 \\
Panel B & 69,544 & 69,544 & 69,537 & 69,536 & 69,522 \\

\hline
\end{tabular}
\caption{Triple-differences estimates of the impact of storm exposure and multiple-event exposure within the same year on transaction volume and default outcomes across alternative treatment radii. Panel A presents results for transaction volume, and Panel B for default outcomes. The indicator \emph{Multiple} equals one for ZIP codes exposed to more than one hurricane event within a given year. All specifications include event-specific ZIP code and time fixed effects. Standard errors clustered at the ZIP code level are reported in brackets. ***, **, and * denote statistical significance at the 1\%, 5\%, and 10\% levels, respectively.}
\label{tab:multi-interaction-radius-payday}
\end{table}

\section{Conclusion}

Using granular data on evictions and loan-level data on payday loans, we provide evidence of disaster induced increase in evictions and reliance on high cost credit in the state of Florida following major storms and hurricanes. We find that eligibility for federal assistance helps mitigate the rate of evictions while also suggesting that households reduce reliance on higher cost credit. Overall, this highlights that major household level impacts such as continued access to housing are sensitive to the availability of emergency liquidity, suggesting the utility for such emergency assistance programs while highlighting the need to better understand administrative cutoffs to access.

\clearpage
\bibliographystyle{ecta}
\bibliography{references}

\end{document}